\documentclass[12pt,preprint]{aastex}
\usepackage{natbib}
\usepackage{epsfig}
\def\LMC{$18.48\pm0.04$ }

\begin{document}
\title{The Carnegie Hubble Program: The Leavitt Law at 3.6 and 4.5 $\micron$ in the Milky Way}

\author{\bf Andrew J. Monson, Wendy L. Freedman, Barry F. Madore, S. E.  Persson, \\
 Victoria Scowcroft, Mark Seibert,}
\affil{Observatories of the Carnegie Institution of Washington \\ 813
Santa Barbara St., Pasadena, CA 91101}

\author{\bf Jane R. Rigby}
\affil{Observational Cosmology Lab, NASA Goddard Space Flight Center, Greenbelt MD 20771}

\email{
amonson@obs.carnegiescience.edu,
wendy@obs.carnegiescience.edu,
barry@obs.carnegiescience.edu,
persson@obs.carnegiescience.edu,
vs@obs.carnegiescience.edu,
mseibert@obs.carnegiescience.edu,
Jane.R.Rigby@nasa.gov}


\begin{abstract}
The Carnegie Hubble Program (CHP) is designed to calibrate the extragalactic distance scale using data from the post-cryogenic era of the \emph{Spitzer Space Telescope}.  The ultimate goal of the CHP is a systematic improvement in the distance scale leading to a determination of the Hubble Constant to within an accuracy of 2\%. This paper focuses on the measurement and calibration of the Galactic Cepheid Period-Luminosity (Leavitt) Relation using the warm Spitzer IRAC 1 and 2 bands at 3.6 and 4.5 $\mu$m.  We present photometric measurements covering the period range 4 - 70 days for 37 Galactic Cepheids. Data at 24 phase points were collected for each star.  

Three PL relations of the form $M=a(\log(P)-1)+b$ are derived. The method adopted here takes the slope $a$ to be -3.31, as determined from the Spitzer LMC data of Scowcroft et al. (2012). Using the geometric HST guide-star distances to ten Galactic Cepheids we find a calibrated 3.6 $\micron$ PL zero-point of $-5.80\pm0.03$. Together with our value for the LMC zero-point  we determine a reddening-corrected distance modulus  of \LMC mag to the LMC.

The mid-IR Period-Color diagram and the $[3.6] - [4.5]$ color variation with phase are interpreted in terms of CO absorption at 4.5 $\micron$.
This situation compromises the use of the 4.5 $\micron$ data for distance determinations.

\end{abstract}


\section{Introduction}
The Carnegie Hubble Program (CHP) is designed to reduce systematic uncertainties in the distance scale. The compelling reasons for doing so are provided in an overview by \cite{Freedman:2011}. The first phase of the CHP is a warm Spitzer legacy mission, the preliminary goal of which is to reduce the systematic uncertainties in $H_0$ to 3\% or better.  The second phase will include observations from GAIA and the \emph{James Webb Space Telescope} (JWST) where the goal will be to push this number to 2\%.  

The warm Spitzer phase consists of observing Cepheids in the Milky Way (MW), Large Magellanic Cloud (LMC), Small Magellanic Cloud (SMC) and other Local Group Galaxies (see \citealt{Freedman:2011} for a complete list) to calibrate the local distance scale in the mid-infrared (mid-IR).  The program extends into the Hubble Flow  by calibrating the mid-IR Tully-Fisher relation and farther to Type Ia supernova host galaxies observed as part of the Carnegie Supernova Project \citep{Folatelli:2009, Contreras:2010}.  

This paper presents warm Spitzer IRAC channel 1 (3.6 $\micron$) and channel 2 (4.5 $\micron$) light curves for 37 Galactic Cepheids. Each star was observed at 24 phase points. The data are used to derive robust mean magnitudes and colors, and, with the adoption of HST parallaxes for 10 stars, an accurate calibration in the mid-IR. The absolute magnitudes give a distance to the LMC, which we believe is currently the value carrying the lowest systematic uncertainty. The calibrations and LMC distance values for a sample of Cepheids in Galactic Clusters and for a sample with Infrared Surface Brightness data are compared to the HST-based distance. A brief discussion of the observed Period-Color relationship and its interpretation in terms of CO affecting the 4.5 $\micron$ band is also given.

\section{Warm Spitzer Observations}
\label{warm_spitzer_observations}

\subsection{Target Selection}
\label{target_selection}

The 37 Cepheids in our sample have multiple distance and extinction estimates, and span a wide range of 4 - 70 days in period. (\citealt{Fernie:1995}, \citealt*{Tammann:2003}, \citealt{Fouque:2007}). The sample includes the majority of the nearest Cepheids; these should soon have high-precision parallaxes available from the GAIA mission \citep*{Windmark:2011}. Table \ref{target} lists the target star, the adopted reddening and three measurements of distance for each star if available: direct parallax data from HST, distance moduli obtained from main-sequence fitting of their host clusters, and distances from Infrared Surface Brightness (IRSB) measurements; the latter two methods were converted to units of parallax for comparison.    Reddening estimates from photometric and spectroscopic methods are listed in Table \ref{red} as well as space reddenings determined from stars along the same line of sight.   The average reddening of these methods was used in Table \ref{target}.

\subsection{Observations}
\label{observations}

Observations were made using the \emph{Spitzer Space Telescope} as part of a two-year Exploration Science Program, PID 60010:  The Hubble Constant \citep{Freedman_prop:2008}.  The warm Spitzer Mission started in 2009 (Cycle-6) and the Galactic Cepheid observations were completed in early 2011. Each Cepheid was observed at 24 epochs, pre-selected and scheduled to fully sample the light curve (23 epochs for $\zeta$ Gem).  At each epoch a nine-point dither pattern was used to mitigate array-dependent artifacts such as bad pixels and cosmic rays.

The majority of the data were taken using the sub-array mode of the Infrared Array Camera (IRAC\footnote{The IRAC instrument handbook and ancillary data products are available at: \\http://irsa.ipac.caltech.edu/data/SPITZER/docs/irac/.}) \citep{Fazio:2004}, with the shortest available frame time of 0.02 seconds (effective exposure time of 0.01 s). The sub-array mode outputs data from only one corner of the detector, in a 32 $\times$ 32 pixel format, thus allowing the shortest possible exposure times. The observations of CF Cas were made using the full array mode (0.4 s frame time, 0.2 s effective exposure time) because it is relatively faint compared to the other program stars.

The sub-array data are provided by the Spitzer Science Center (SSC) in two forms: as an image cube of 64 frames (each 32 $\times$ 32 pixels) and as a single combined image; all further discussions to sub-array data refer to the combined form (\emph{sub2d} extensions).  All data were retrieved in the basic calibrated data ({BCD}) format, and were reduced using the most recent pipeline (S18.18.0). 

\section{Data Reduction and Photometry}
\label{data_reduction}

The stellar flux in each image was measured using the Mosaicking and Point-source Extraction (MOPEX\footnote{The MOPEX software and documentation are available at: \\ http://irsa.ipac.caltech.edu/data/SPITZER/docs/dataanalysistools/tools/mopex/}) package \citep{MOPEX:2005}.  The pipeline script \emph{apex\_user\_list\_1frame.pl}  (included with MOPEX ) was used to perform profile fitting photometry.  Uncertainty images were supplied to MOPEX using the square root of the input image frame\footnote{The input values were converted to units of electrons first and the uncertainty images were converted back to units of MJy sr$^{-1}$.}.  

Once the first-pass fluxes had been obtained we processed the data to correct for three systematic effects. These are (1) the masking of saturated or markedly non-linear pixels; (2) corrections for non-uniformity of response across individual pixels; and (3) correction for image persistence, in which a bright source will leave behind a trail of spurious flux as the telescope executes its dither pattern. The three effects are examined and the derived corrections are explained in detail in Appendix \ref{app1}. 

The final stellar flux for each epoch is determined from the mean of the 9 dithered flux measurements from MOPEX, modified by the three corrections. The random error is adopted as the dispersion of the 9 measurements and the systematic error is taken as the zero-point error adopted by the SSC, viz., 0.016  mag for both [3.6] and [4.5] \citep{Reach:2005}.  Table \ref{CD_CYG} shows a sample of the IRAC photometric data (the magnitudes are named [3.6] and [4.5]) available for the 37 stars\footnote{The full table can be accessed in the online version of this paper.}.

\section{Results}
\label{results}

\subsection{Mid-IR Light Curves for 37 Galactic Cepheids}
\label{lightcurves}

Periods from the General Catalog of Variable Stars (GCVS, \citealt{GCVS}) were assumed.  In the cases of U Car and V340 Nor, periods were computed using photometric data from \cite{Laney:1992}.   Figure \ref{LC} presents the individual light and color curves (Vega magnitudes) for each Galactic Cepheid.  Data points from \cite{Marengo:2010} are shown as open triangles for comparison, when available.  All are in good agreement except for U Car where the difference is likely due to a phase shift resulting from a period increase between the \cite{Laney:1992} data and ours. All the light curves are plotted with the same magnitude range to emphasize relative changes in signal-to-noise ratio and amplitudes.  The internal photometric precision is high, ranging from 0.004 to 0.029 mag.  Interestingly, for the longer-period Cepheids, there is strong variation in the $[3.6]-[4.5]$ color.  This is due to the temperature-dependent carbon monoxide (CO) band in the [4.5] Channel, as will be discussed in \S\ref{period_color}.

Smooth light curves were generated using a Gaussian local estimation (GLOESS) algorithm.   GLOESS is an interpolating method that uses second-order polynomials to fit the data locally throughout the cycle. The data points surrounding the point to be fit are assigned weights according to a Gaussian window function; weights depend on their distance (in phase) from the fit point.  The method has been used to conveniently obtain mean magnitudes by \cite{Persson:2004} and \cite{Scowcroft:2011} for LMC Cepheids, and by \cite{Monson:2011} for Galactic Cepheids.  
These data are uniformly sampled so the error on the mean is: $\sigma = \frac{A}{N\sqrt{12}}$, where $A$ is the peak-to-peak amplitude of the light curve and $N$ is the number of sample points;  this is discussed in the Appendix of \cite{Scowcroft:2011}.    
The final \emph{total} uncertainty in the mean magnitude turns out to be dominated by the systematic zero-point calibration of the Spitzer warm mission.   Because the uncertainties in Channels 1 and 2 are correlated with each other, uncertainties in the color were also determined using the above equation.    Table \ref{galcephs} gives the  [3.6] and [4.5] IRAC intensity-mean magnitudes and colors for the 37 Cepheids.

\subsection{Mid-IR Extinction Corrections}
\label{extinction_corrections}

Before discussing the Period-Luminosity and Period-Color Relationships, we shall need to correct for extinction. Compared to optical wavelengths, reddening and extinction corrections are relatively small at mid-IR wavelengths. They must, nevertheless,  be quantified and applied because values of A$_V$ can exceed 3 mag in our sample.  We adopted an extinction law for all stars that is applicable along average lines-of-sight through the diffuse interstellar medium. The extinction law of \cite{Ind:2005} combined with that of \cite{Cardelli:1989} best fulfill this choice. The relations: ${A_{[3.6]}/A_K = 0.56\pm0.06}$ and ${A_{[4.5]}/A_K = 0.43\pm0.08}$ \citep{Ind:2005} were derived from field stars in the Galactic Plane and are probably applicable to the Cepheids in this study\footnote{Following \cite{Ind:2005} we use $K$ to mean the $K_s$ filter of the 2MASS survey.}. 

To scale the extinctions at $K$ to the reddenings $E(B-V)$ we used the extinction law derived by \cite{Cardelli:1989}: ${A(\lambda)/A_V = a(x)+b(x)/R_V}$, where ${a=0.574x^{1.61}, b=-0.527x^{1.61}}$,  $x = 1/ \lambda$ and ${R_V}$ is the ratio of total-to-selective absorption (${R_V = A_V/E(B-V)}$).  Using a wavelength of $\lambda=2.164$ $\mu$m for the $K$ filter (actually $K_s$) as adopted by \cite{Ind:2005} and an average ${R_V=3.1}$, we find: ${A_K/A_V=0.117}$.   The combined relations yield a final total-to-selective extinction of ${{A_{[3.6]}}/{E[B-V]}}=0.203$, ${{A_{[4.5]}}/{E[B-V]}}=0.156$, and ${{E([3.6]-[4.5])}/{E[B-V]}}=0.047$. 

Use of the  \cite{Ind:2005} mid-IR extinction law might be questioned on general grounds. Three stars have $E(B-V) >  1$, above which the  corrections will begin to introduce systematic errors. For example, measured values of $A_{[3.6]}/A_K$ range from 0.41\citep{Chapman:2009} to 0.64 \citep{Flaherty:2007}\footnote{ See www.pas.rochester.edu/~emamajek/memo\_ir\_reddening.html for a summary.}. Toward the Galactic Center \cite{Nishiyama:2009} obtain 0.50 $\pm$ 0.01. For V367 Sct ($E(B-V) =1.231) $ the total range in $A_{[3.6]}$ is 0.10 mag.   
The uncertainties in $E(B-V)$ are all $\leq$ 0.03 mag except for for GY Sge, where $\sigma(E(B-V))$ = 0.17.  This value introduces uncertainties of 0.035 and 0.027 mag into the 3.6 and 4.5 $\micron$ magnitudes, respectively. Finally, if the uncertainty in ${{A_{[3.6]}}/{E[B-V]}}$ is as large as 0.05 (see above), the corresponding corrections will remain negligibly small.


\section{Period-Luminosity Relations at 3.6 and 4.5 $\micron$}
\label{secPLR}

We now present the Period-Luminosity Relations for the 37 Cepheids in our sample. Table \ref{target} shows that the sample may be divided into three subsamples, depending on the origin of their distance measurements. Henceforth we consider each subsample separately, as the three methodologies for distance determinations are quite different. In \S\ref{weighting_techniques} we discuss in detail three weighting techniques for the data points. In the following sections we have adopted unweighted fits in finding slopes and zero-points.

\subsection{The Three Subsamples}
\label{three_samples}

Ten of the 37 Cepheids have direct geometric parallaxes determined from \emph{Hubble Space Telescope} (HST) guide camera data \citep{Benedict:2007} and therefore have the most accurate distance determinations currently available. The data for the sample are provided in Table \ref{hstcephs}.   Figure \ref{plhstw} shows the data and zero-point fit using uniform weighting;  the distance uncertainties are displayed for reference as error bars. The data include the final de-reddened [3.6] and [4.5] magnitudes, the final adopted distance moduli, extinctions, and absolute magnitudes. Following \cite{Benedict:2007} we have applied Lutz-Kelker-Hanson corrections \citep[LKH]{Lutz:1973, Hanson:1979} to these parallaxes; the corrections are systematic and range from -0.02 to -0.15 mag with uncertainties of $\pm0.01$ mag.   For completeness we include in Table \ref{hstcephswo} and Figure \ref{plhstwo} the corresponding data for the HST sample without LKH corrections. 

Eighteen of the 37 Cepheids are likely to be members of star clusters or associations for which distance moduli have been estimated from Main-Sequence (MS) fitting \citep{Turner:2002, Turner:2010, Majaess:2012a, Majaess:2012b, Majaess:2011}.  Table \ref{cluscephs} lists the data and Figure \ref{plclus} shows the forced LMC-slope and zero-point fit for the uniformly weighted data. 
The Cepheids CEa and CEb Cas are presumably at the same distance as CF Cas by virtue of common membership in the cluster NGC 7790 and although separated by only $1\farcs0$ they were easily split using PRF photometry and they have been included in the sample. 

Thirty-two of the 37 Cepheids have distance determinations based on the IRSB technique \citep[and references therein]{Storm:2011}.
Table \ref{irsbcephs} contains the data and Figure \ref{plirsb} shows the fit and zero-point for uniformly weighted data; W Sgr  was rejected from the analysis due to its relatively high uncertainty.

Each of the the data subsamples were fit using a PL relationship of the form: $M = a(\log P-1) + b$. As will be shown below we find no statistically significant difference between the slope of the [3.6] PL of $-3.31\pm0.05$ for the LMC, and that of the HST parallax sample.  We thus adopt the LMC slope for the PL fit and re-determine zero-points for each subsample.  
The magnitude residuals from the PL fits are highly correlated with each other, suggesting that the widths of the PL relations are not driven by random photometric errors.   Rather, the correlated scatter is most likely some combination of deterministically correlated (unit slope) distance errors and the intrinsic (correlated) positions of these Cepheids in the instability strip (IS). If the IS is represented by a rectangular distribution, (i.e., it is uniformly filled and has hard limits at the blue and red edges) then the peak-to-peak width in the residuals can be interpreted as the width of the IS or at least an upper limit, which in the HST subsample is  $\sim0.4$ mag.

\subsection{Dependence of PL Relations  and Uncertainties on Weighting Techniques}
\label{weighting_techniques}

The PL relations were fit to each of the three subsamples using multiple weighting schemes and also by further restricting the subsamples by period cuts.

The final uncertainty in absolute magnitude for an individual Cepheid is
dominated by the uncertainty in its distance. In deriving a PL relation, however, an additional spread is caused by the finite width of the IS, and biases in the PL slope and zero-point may result depending on how the strip is filled. To investigate
these uncertainties and their effects on the derived PL fits, we applied different weighting
schemes.   In addition, for each of the data sets, we investigated different
period cuts so that the Galaxy data sets more closely matched the period range 
of the LMC sample, viz., 6 - 60 days. Finally, a fixed
slope determined from the LMC data was force fitted to the Galactic data
to determine only the Galactic zero-point.  The data were fit using a
PL relationship of the form: $M = a(\log P-1) + b$.  Table \ref{TPL}
presents the results using three different weighting methods for
the data in this analysis.

The first weighting method applies a uniform uncertainty of 0.1 mag to each Cepheid,
the purpose of which is to provide results presumably less biased by Cepheids which may have
underestimated distance uncertainties. The second method applies traditional 
weights as $\sigma^{-2}$ to the absolute magnitudes.   
The third method falls between the first two in that it assumes
an intrinsic scatter in the IS. In this case an additional uncertainty of 0.1 mag is
added in quadrature to each individual uncertainty.   The
value of 0.1 mag is adopted from the average RMS scatter of the LMC
data points around the best fit.

\subsection{PL-Slopes}

The IRSB slopes closely match the HST slopes, which is not surprising
because the most recent IRSB distances used a projection factor ($p$-factor) 
calibrated using the HST parallaxes (but without LKH corrections) as priors.  The IRSB slope is better
constrained because of sample size, but is still
dependent on the adopted  $p$-factor.
The effect of the varying the weighting method is most noticeable in
the Cluster Main-Sequence (MS) fits where the slopes differ by more
than 2$\sigma$ between the first two methods, and converges to within
1$\sigma$ of the HST, IRSB and LMC slopes using the third weighting
method.   As discussed by \cite{Turner:2010} some long-period Cepheids
populate the blue edge of the IS and can bias the
slope, so it is necessary to include an estimate of the
intrinsic width of the IS to reduce the bias.   The
benefit of the third weighting method is that the intrinsic width of the
IS is included as well as individual uncertainties
for each Cepheid. As can be seen in Table \ref{TPL} the
slopes for all three methods agree very well with each other using the
softened weights.  Since the slope from each method agrees with that 
of the LMC, we chose to adopt the better-determined LMC slope and 
to redetermine the zero-points for both the [3.6] and [4.5] PL relations.
This decision is further backed by recent studies that find near-identical PL slopes for the MW and LMC in the near-infrared \citep{Storm2:2011}.

\subsection{PL-Intercepts}
\label{pli}
As mentioned above, some of the long-period Cepheids occupy the
blue edge of the IS and although we forced a fixed (LMC)
slope, the zero-point can now be slightly biased if the entire sample
does not uniformly populate the IS.  We therefore limit ourselves to
adopt zero-points from the uniformly weighted fits. This effectively assumes the width
of the IS is the only source of uncertainty and can be treated
as equal for each Cepheid.   We also chose to make use of the entire
period range which provides a larger sample and will more uniformly
populate the IS.  With these choices in hand, we now have zero-points
of: HST (with LKH) = $-5.80\pm0.03$,  HST (without LKH) = $-5.74\pm0.03$ ,  MS
= $-5.75\pm0.05$ and IRSB = $-5.74\pm0.02$.   We notice again the
agreement between the HST (without LKH) and IRSB zero-points as must be the case (see above).  The average LKH correction is -0.06 mag, which if applied to the IRSB calibration would shift the [3.6] IRSB PL zero-point to $-5.80\pm0.03$ mag.   Because they are calibrated using the HST parallaxes the IRSB zero-point does not offer an independent baseline measurement, however it does better sample the instability strip and since the it yields the same zero-point and scatter it indicates that the HST data is not too effected by paucity.  

The Cluster Cepheids do offer an independent check of the zero-point and they appear to confirm the HST (without LKH) zero-point.   We note, however, that the outliers S Vul and TW Nor are more than 10\% discrepant compared to their IRSB distances and rejecting them would change the Cluster zero-point to -5.79 mag; in agreement with LKH.   This is in agreement with the  \cite{Ngeow:2012} results which also find a 0.04 mag relative offset between their derived Wesenheit PL distances to the \cite{Storm:2011} IRSB and \cite{Turner:2010} Cluster samples.    Based on discussions in the literature on the use of LKH (\citealt{Lutz:1973}, \citealt{Hanson:1979}, \citealt{Smith:2003}, \citealt{Loredo:2007}) we have, at present, chosen to adopt the use of the LKH correction and an uncertainty in this correction of 0.01 mag \citep{Benedict:2007}.  

We note that the scatter in the HST data is less than the average uncertainty assigned to the HST parallaxes and that we have chosen to adopt this (smaller) empirical scatter as a measure of the total zero-point uncertainty.

\section{The Distance to the LMC}
\label{dist_to_lmc}

As will be discussed in \S\ref{period_color} the 4.5 $\micron$ data are likely to be affected by CO absorption while the 3.6 $\micron$ data are not (see also \cite{Freedman:2012}). Consequently, we have adopted the absolute PL zero-point from the HST Leavitt Law at 3.6 $\mu$m ($-5.80\pm0.03$ mag)  and compare that with the apparent zero-point of the LMC PL relationship ($12.70\pm0.02$ mag),  both zero-points are defined at $\log P=1.0$ and the PL relations are parallel to each other.  We found no measurable metallicity effects for the MW and LMC Cepheids at 3.6 microns for which there are spectroscopic [Fe/H] values \citep[Figure 2.]{Freedman:2012}.   By adopting a  net extinction to the LMC of E(B-V)=0.1 \citep{Freedman:2010} and using the extinction law discussed in \S\ref{extinction_corrections} (which yields a total LMC extinction of $A_{[3.6]}=0.02$ mag),  we find a distance modulus ${(m-M)_{[3.6]}}$ for the LMC of \LMC mag.   

Alternatively, we followed a multi-wavelength approach to solve for reddening and distance modulus simultaneously \citep{Freedman:2010}.  The multi-wavelength photometry for the LMC are taken from: \citealt{Udalski:1999} (B, V and I$_C$ ),  \citealt{Persson:2004} (J, H and K$_s$) and \citealt{Scowcroft:2012} ([3.6] and [4.5]).   The Galactic Cepheid photometry were compiled from the literature \citep{Berdnikov:2008,Barnes:1997,Laney:1992,Welch:1984,Monson:2011} and average magnitudes were found in the same manner as discussed in \ref{lightcurves}; see Table \ref{nest}.   The multi-wavelength PL relations are shown in Fig \ref{LMC_1} and summarized in Table \ref{mpl}.  The slopes were found by fitting the LMC data in the period range 3.8-60 days;  a universal slope is considered.  The Galactic zero-points were found using the LMC slopes at each wavelength using only the HST parallax Cepheids; see \ref{pli} for details.  The apparent distance moduli are plotted against inverse wavelength in Figure \ref{LMC_2}.  The standard extinction law from \cite{Cardelli:1989} was fit to the data to find the true LMC distance modulus of $18.48\pm0.03$ and average LMC reddening of $E(B-V)=0.12\pm0.01$ mag.  The K$_s$ and [4.5] data were excluded from the fit due to the effect of CO in those bands.    This supersedes the value of $18.39\pm0.06$ found in \cite{Freedman:2010} and is in excellent agreement with other independent measures recently reviewed by \cite{Walker:2011} who finds a composite distance modulus of $18.48\pm0.05$ and \cite{Laney:2012} who find a red clump distance of $18.47\pm0.02$.

\section{Period-Color Relationships}
\label{period_color}

\subsection{The Period-Color Diagram}
\label{PC}

CO absorption in the 4.5 $\micron$ band for cooler stars produces a significant period-color relation that moves the mean $[3.6]-[4.5]$ color toward the blue at cooler temperatures, and longer periods.    This effect is driven, as is the case for the light curve color variations (see below),  by the temperature dependence of CO dissociation and not by the thermal color-temperature which has little effect on the slope of the continuum at these long wavelengths.  The $[3.6] - [4.5]$ Period-Color Relation is shown in Figure \ref{PLC} and the weighted least squares fit to the de-reddened data (omitting Y Oph) is:

$[3.6] - [4.5]_{MW} = -0.09(\pm0.01)(\log P-1.0) - 0.03(\pm0.01)$.  For comparison, the LMC Period-Color Relation is $[3.6] - [4.5]_{LMC} = -0.09(\pm0.01)(\log P-1.0) + 0.01(\pm0.01)$ \citep{Scowcroft:2011}. The slopes of these fits are consistent, but the zero-points differ by $0.04\pm0.01$ mag. 

\cite{Ngeow:2012a} have presented a number of models for theoretical PC diagrams in the mid-IR. Their summary tables give slopes and zero-points for several models of the PLs and $[3.6] - [4.5 ]$ color covering a range of Helium and metal abundance. Several of their model PCs are plotted with the data in Figure \ref{PLC}. Comparison of the theoretical slope of the $[3.6] - [4.5]$ PC diagram shows reasonably good agreement with the Galaxy color data for a Helium abundance of Y = 0.31. 

\subsection{Color Curves and CO Absorption Models}
\label{color_curves}

As pointed out in \S\ref{lightcurves} the $[3.6]-[4.5]$ color is characterized by systematic variations through the cycle. The color amplitude is also  closely related to period, the longer period stars having the largest variations. This effect, in complete analogy with the cause of the Period-Color relation, is again due to CO absorption in the 4.5 $\micron$ band, as discussed by \cite{Scowcroft:2011} and \cite{Marengo:2010}. The color variation extends only toward bluer colors from a baseline red color limit of $\sim$ 0.01 mag; this can be seen Figure \ref{extent}. The blue extent (blue indicating more absorption at 4.5 $\micron$) increases with period as the Cepheids reach intrinsically cooler temperatures.    The effect of CO has only recently been observed over entire Cepheid pulsation cycles (see also \citealt{Scowcroft:2011}). 

To quantify the behavior of both the overall PC Relation and the color curves, we have computed several synthetic spectra using appropriate Kurcz stellar models \citep{Kurucz_13, Castelli:2003, atlas_linux_1, atlas_linux_2}.   Figure \ref{kurucz} shows the results. They indicate that at temperatures greater than approximately 6000 K absorption due to CO is nearly non-existent.   As the temperature falls below 6000 K CO absorption in the [4.5] band sets in, leading to the diminished flux observed in the 4.5 $\micron$ light curve.  The result is that the color curves should have larger amplitudes for Cepheids with longer periods, as they reach intrinsically cooler temperatures. This is precisely the behavior exhibited in the observed color curves. For the shorter-period Cepheids the color amplitude is diminished because these Cepheids are intrinsically hotter and the CO remains dissociated over a longer portion of the pulsation cycle.

Part of the systematic offset in the PC fits (Galaxy versus LMC) is plausibly explained by the difference in metal abundance between the two galaxies. Fig \ref{kurucz} shows that this shift should amount to $\sim$ 0.02 mag offset per 0.5 dex change in metal abundance. Other effects such as rotation may also play a role, and in any case we are dealing with a very small effect.


\section{Summary}
\label{summary}

In this work we have presented the first results from the Galactic Cepheid campaign of the CHP.  Light curves created from uniformly-spaced observations with high-precision photometry yield intensity-mean magnitudes for 37 Galactic Cepheids spanning a range of periods from 4-70 days.       

Using the precise geometric parallax measurements from \cite{Benedict:2007} we have found a Galactic zero-point (set to $\log$ P=1.0) for the 3.6 $\mu$m Period-Luminosity (Leavitt) Law of $-5.80\pm0.03$ mag.  Comparing this to the LMC zero-point we find an LMC distance modulus of \LMC mag, which is confirmed using a multi-wavelength analysis.  The uncertainty represents a factor of 2 improvement over previous Key Project measurements \citep{Freedman:2010} and will be made stronger with future geometric parallaxes to the full sample from GAIA.  The implications of this revised LMC distance modulus on the Key Project distances are discussed in \cite{Freedman:2012}.

The well-sampled light curves reveal a strong color variation for Cepheids with periods longer than 10 days.  A second and related result is a clear Period-Color relation. Both correlations are caused by enhanced temperature-sensitive CO absorption at 4.67 $\micron$ in longer period, intrinsically cooler Cepheids.   

\section{Acknowledgements}
We would like to thank the staff of the Spitzer Science Center for their assistance with scheduling such a large and complex project as well as the support received to make these analyses.  This work is based [in part] on observations made with the \emph{Spitzer Space Telescope}, which is operated by the Jet Propulsion Laboratory, California Institute of Technology under a contract with NASA.  Support for this work was provided by NASA through an award issued by JPL/Caltech.  We would also like to thank the anonymous referee for constructive comments regarding the form and content of this work.

\clearpage
\appendix

\section{Corrections for Systematic Effects on the Photometry}
\label{app1}


\emph{Saturated Pixels.} The first step in the reduction is to find and mask markedly non-linear or saturated pixels. This is particularly important for stars as bright as the Cepheids in this program. Our routine works as follows: MOPEX fits the profile by masking unwanted pixels, and by assigning weights using the uncertainty image. Saturated pixels (near the center of the PSF) will tend to have lower dispersions and hence artificially high weights (in the limit of complete saturation the dispersion will be zero). We found valid upper thresholds by experiment: the dispersions for the nine dither positions of a saturated star were found as a function of threshold level and the best level chosen. The final upper threshold values were 10,000 and 12,000 DN for the 3.6 and 4.5 $\micron$ bands, respectively. These were lower than than those recommended in the IRAC Handbook.

\emph{Point Response Profiles and Pixel Phase Corrections}. Each stellar profile was fit with a \emph{Point Response Function} (PRF) profile, a procedure that minimizes the residual between the input frame and a standard PRF supplied by MOPEX. Rather than having a functional form, the PRF is a look-up table containing different representations of a point source at various pixel phases. These are the distances from the center of the stellar profile to the center of the nearest (integer valued) pixel. This complication arises because pixels do not have uniform response across them. The \emph{Pixel Phase Correction} (PPC) was included in the PRF tables provided by the SSC for the cold mission. This was not the case for the warm mission. The correction for an arbitrary profile was thus determined using all the data to find a empirical PPC as follows. For every nine-point dither pattern constituting a measurement one has an average count, and nine deviations from that count. (The deviations arise from the pixel phase variation.) The left hand side of Figure \ref{apexppc} shows all those deviations plotted as a function of pixel phase. The total number of points is $9 \times 24 \times 37$ (9 dither positions per measurement, 24 light curve points per star, 37 stars). The strong correlation represents the \emph{residual} PPC, which is easily removed to yield corrected data. The correction is of the form: $f_{ppc} = u+v \left( \frac{1}{\sqrt{2\pi}} - p \right) $, where $p={\sqrt{((x-\rm{nint}(x))^2 + (y-\rm{nint}(y))^2)}}$ is the pixel phase\footnote{This equation is taken from the IRAC Handbook. The coefficients guarantee that the average deviation is zero and thus the correction does not introduce a spurious shift in the average measurement.}. The right-hand side of Figure \ref{apexppc} shows the results of applying the above correction to the data points on the left-hand side. The results are seen to be satisfactory. 

The PRF photometry methods correct for several other systematic effects:   

(1) The fits have slightly different coefficients for saturated and unsaturated data, a systematic effect that probably arises from centroid offsets in masking central pixels. The coefficients were confirmed by masking non-saturated data.  

(2) The PRF varies across the array, but a fixed PRF was used for the sub-array data, located at column 233, row 35. For the full frame data (CF Cas) a lookup table was used to find the nearest PRF at each position.

(3) MOPEX reports fluxes at the center of pixel flux and is normalized to a radius of 10 pixels, i.e., the IRAC standard aperture. Because the flux reported is that for a pixel phase of zero, an additional correction factor ($f_{corr}$) is applied to bring the flux to the \emph{average} pixel phase, which corresponds to where the flux zero points were defined.   

Finally, the flux is placed on the standard Vega magnitude system by dividing by the photometric flux zero-points ($zp$); $280.9\pm4.1$ and $179.7\pm2.6$ [Jy], for [3.6] and [4.5], respectively \citep{Reach:2005}.  The final corrected magnitude ($m$) is found from the PRF flux ($F_{PRF}$), which is reported in $\mu$Jy, by the following relation: 
$$ m = -2.5 \log \left( \frac{F_{PRF}\cdot10^{-6}}{f_{corr}\cdot f_{ppc}\cdot zp} \right), $$
Table \ref{constants} contains the constants for each channel.

\emph{Image Persistence or Latency}. The sub-array data were taken with short exposure times on bright objects, and with short settling times between dithers.  As a consequence, the data are prone to short-term image persistence from previous dithers and observations; see Figure \ref{persist}.  A multi-stage process was undertaken to mitigate image persistence for each frame.  First, the stellar profile was fit in each of the nine dithered frames using the PRF-fitting algorithm in the MOPEX script described above.  The residual images were averaged together using a nearest neighbor weighting scheme for each dither position.  Lowest weights were given to future frames and higher weights were given to the most recent frames to create an approximate map of the image persistence at each dither position.  The persistence maps were then subtracted from the original data leaving only the source; see Figure \ref{phot}.  Note that this process simply describes creating a local background frame by combining dithered frames where the source(s) has been modeled and subtracted rather than masked.  Each background/persistence subtracted image was passed through the MOPEX pipeline to perform a second and final PRF fit.  

The latency can affect aperture photometry measurements by nearly five-percent in the short (0.02 s) sub-array data.  The effect is larger for shorter exposure times when there is less time for the latency to dissipate.  One advantage of the PRF-fitting algorithm is that it tends to ignore the latent pixels when fitting a PRF.  After the latent image subtraction the PRF photometry changes by less than 1\%.

\emph{Photometry Checks}. As a check of the photometric fidelity the standard star HD165459 was
processed in the manner described above for various exposure times in
sub-array and full-array mode resulting in non-saturated and saturated
data.   To quantify the effect of persistence both aperture and PRF
photometry were performed on the standard star both prior to, and
after, the persistence correction.  Prior to the correction, the
aperture photometry was consistently reporting 5-15\% higher flux than
expected for the shortest exposure times while after the persistence
correction the aperture photometry was typically only 0-5\% higher
than expected.  The PRF photometry was relatively unaffected by the
persistence correction, changing by less than 1\% before and after the
persistence correction.  The standard star comparisons represent the
worse case scenario since the long exposures (saturated data) were
taken just prior to the short exposures and subsequently suffered from
a relatively large amount of persistence.    The final PRF photometry
results for the non-saturated standard data are  $6.588 \pm0.007$ mag
and $6.571\pm0.008$ mag and $6.587\pm0.014$ mag and $6.564\pm0.017$
mag  for the saturated data.   Both are in good agreement with the
standard magnitudes of $6.593\pm0.029$ mag and $6.575\pm0.028$ mag
found by \cite{Reach:2005}.   For the Cepheids, the difference in PRF
photometry before and after the persistence correction was also less
than 1\%; except for $\ell$ Car, the most severely saturated star in
our sample, in which case the difference was 6\% and 3\% in Channels 1
and 2.  We report here the final PRF photometry resulting from a
persistence-subtracted image.

\clearpage

\begin{deluxetable}{rrrrrlll}
\tabletypesize{\scriptsize}
\tablecolumns{9} 
\tablewidth{0pc} 
\tablecaption{CHP selected Galactic Cepheids and adopted parallaxes. }
\tablehead{ 
\colhead{ID}                                          &  \colhead{$\log(P)$\tablenotemark{1}}  & \colhead{RA\tablenotemark{1}}              & \colhead{DEC\tablenotemark{1} }  & 
\colhead{E(B-V)\tablenotemark{2}}  & \colhead{$\pi$[HST]\tablenotemark{3}} & \colhead{$\pi$[MS]\tablenotemark{4}} & \colhead{$\pi$[IRSB]\tablenotemark{5}}    \\

\colhead{}                                              &  \colhead{[days]}                                         & \colhead{J2000}            & \colhead{J2000} &
\colhead{[mags]}                                    & \colhead{[mas]}                                           & \colhead{[mas]}                 & \colhead{[mas]}   

 }
\startdata
         S Vul & 1.835 & 19:48:23.8 & +27:17:11 & $   0.925 \pm 0.022 $ &        \nodata       &$   {0.32  \pm 0.04  }\tablenotemark{6}$&$    0.27  \pm 0.01  $  \\ 
        GY Sge & 1.708 & 19:35:13.6 & +19:12:09 & $   1.187 \pm 0.170 $ &        \nodata       &        \nodata       &$    0.35  \pm 0.01  $  \\ 
        SV Vul & 1.653 & 19:51:30.9 & +27:27:37 & $   0.508 \pm 0.013 $ &        \nodata       &$    0.56  \pm 0.03  $&$    0.53  \pm 0.01  $  \\ 
         U Car & 1.589 & 10:57:48.2 & -59:43:56 & $   0.265 \pm 0.010 $ &        \nodata       &        \nodata       &$    0.71  \pm 0.01  $  \\ 
    $\ell$ Car & 1.551 & 09:45:14.8 & -62:30:28 & $   0.154 \pm 0.011 $ &$    2.01  \pm 0.20  $&        \nodata       &$    1.93  \pm 0.02  $  \\ 
         T Mon & 1.432 & 06:25:13.0 & +07:05:09 & $   0.181 \pm 0.010 $ &        \nodata       &        \nodata       &$    0.76  \pm 0.01  $  \\ 
        WZ Sgr & 1.339 & 18:16:59.7 & -19:04:33 & $   0.501 \pm 0.007 $ &        \nodata       &$    0.56  \pm 0.03  $&$    0.56  \pm 0.01  $  \\ 
        RU Sct & 1.294 & 18:41:56.4 & -04:06:38 & $   0.929 \pm 0.010 $ &        \nodata       &$    0.60  \pm 0.03  $&$    0.53  \pm 0.01  $  \\
\enddata
\tablecomments{Three different methods of distance determination are examined: Hubble Space Telescope (HST) parallaxes, Main-Sequence (MS) fitting to candidate cluster and Infrared Surface Brightness (IRSB) method.  The latter two methods were converted to parallax measures for comparison.  }
\tablenotetext{1}{Values adopted from the General Catalog of Variable Stars \citep{GCVS}.  The periods for V340 Nor and U Car were recomputed; see text. }
\tablenotetext{2}{Values adopted from the average value in Table \ref{red}. }
\tablenotetext{3}{HST parallaxes adopted from \cite{Benedict:2007}.  The Lutz-Kelker bias correction is applied in the PL analysis; see Table \ref{hstcephs}.}
\tablenotetext{4}{Cluster parallaxes adopted from \cite{Turner:2010}.  }
\tablenotetext{5}{IRSB parallaxes adopted from \cite{Storm:2011}. }

\tablenotetext{6}{\cite{Turner:2011b}.  }
\tablenotetext{7}{\cite{Majaess:2011}.  }
\tablenotetext{8}{\cite{Majaess:2012a}.  }
\tablenotetext{9}{\cite{Majaess:2012b}.  }
\tablenotetext{10}{\cite{Turner:2002}.  }
\tablenotetext{11}{\cite{Mathews:1995}.  }

\label{target}
\end{deluxetable}


\clearpage

\begin{deluxetable}{rllll}
\tabletypesize{\scriptsize}
\tablecolumns{5} 
\tablewidth{0pc} 
\tablecaption{CHP selected Galactic Cepheids and reddenings. }
\tablehead{ 
\colhead{ID}   &  \colhead{E(B-V)$_{phot}$}  & \colhead{E(B-V)$_{spec}$}    & \colhead{E(B-V)$_{space}$}  & \colhead{E(B-V)$_{ave}$}

}
\startdata
    S   Vul & $ 0.727 \pm 0.042 $ & $ 0.940 \pm 0.051 $ & $ 1.020 \pm 0.030 $ & $ 0.925 \pm 0.022 $ \\  
   GY   Sge & $ 1.187 \pm 0.170 $ &    \nodata      &    \nodata      & $ 1.187 \pm 0.170 $ \\  
   SV   Vul & $ 0.461 \pm 0.022 $ & $ 0.510 \pm 0.020 $ & $ 0.590 \pm 0.030 $ & $ 0.508 \pm 0.013 $ \\  
    U   Car & $ 0.265 \pm 0.010 $ &    \nodata      &    \nodata      & $ 0.265 \pm 0.010 $ \\  
$\ell$   Car & $ 0.147 \pm 0.013 $ &    \nodata      & $ 0.170 \pm 0.020\tablenotemark{1} $ & $ 0.154 \pm 0.011 $ \\  
    T   Mon & $ 0.181 \pm 0.011 $ & $ 0.179 \pm 0.029 $ &    \nodata      & $ 0.181 \pm 0.010 $ \\  
   WZ   Sgr & $ 0.431 \pm 0.011 $ & $ 0.458 \pm 0.058 $ & $ 0.560 \pm 0.010 $ & $ 0.501 \pm 0.007 $ \\  
   RU   Sct & $ 0.921 \pm 0.012 $ &    \nodata      & $ 0.950 \pm 0.020 $ & $ 0.929 \pm 0.010 $ \\
\enddata
\tablecomments{Reddenings adopted from the literature include:  average photometric reddenings \citep{Fouque:2007}, spectroscopic reddenings \citep{Kovtyukh:2008} and space reddenings \citep{Turner:2010,Benedict:2007}; the last column contains the weighted average.   When no error estimate is available for the spectroscopic reddening 0.05 is used.     }
\tablenotetext{1}{Space reddenings adopted from \cite{Benedict:2007}, otherwise \cite{Turner:2010}.  }
\tablenotetext{2}{CEa \& CEb Cas are here assumed to have same excess as CF Cas.}

\label{red}
\end{deluxetable}


\clearpage

\begin{deluxetable}{rrrr}
\tabletypesize{\scriptsize}
\tablecolumns{4} 
\tablewidth{0pc} 
\tablecaption{Spitzer IRAC photometry for Galactic Cepheids}
\tablehead{ 
 \colhead{ID}  &   \colhead{HMJD\tablenotemark{1}}    & \colhead{{[3.6]}}    &  \colhead{{[4.5]}}   \\
 \colhead{}     &   \colhead{[days]} & \colhead{[mag]}       & \colhead{[mag]}                       
 }
\startdata
         CD Cyg &    55147.1401 &  5.618 $\pm$  0.010 &  5.733 $\pm$  0.004 \\ 
         CD Cyg &    55147.8152 &  5.659 $\pm$  0.012 &  5.740 $\pm$  0.007 \\ 
         CD Cyg &    55148.3628 &  5.675 $\pm$  0.011 &  5.728 $\pm$  0.007 \\ 
         CD Cyg &    55149.3595 &  5.723 $\pm$  0.015 &  5.730 $\pm$  0.004 \\ 
         CD Cyg &    55150.2981 &  5.713 $\pm$  0.017 &  5.696 $\pm$  0.008 \\ 
         CD Cyg &    55150.8732 &  5.587 $\pm$  0.020 &  5.555 $\pm$  0.007 \\ 
         CD Cyg &    55151.7063 &  5.543 $\pm$  0.018 &  5.514 $\pm$  0.008 \\ 
         CD Cyg &    55152.0117 &  5.528 $\pm$  0.010 &  5.505 $\pm$  0.007 \\
\enddata
\tablenotetext{1}{The heliocentric modified julian date (HMJD) was determined from averaging the 18 ``HMJD'' header keywords at each epoch (9 exposures in each Channel).  Note: HMJD = HJD-2,400,000.5}
\tablecomments{The quoted errors represent the random photometric errors determined from the variance of the dithered data.  The systematic zero-point errors are 0.016 for both the [3.6] and [4.5] bands.  The full table is available with the online version of this paper. }
\label{CD_CYG}
\end{deluxetable}


\clearpage

\begin{deluxetable}{rcrrr}
\tabletypesize{\scriptsize}
\tablecolumns{5} 
\tablewidth{0pc} 
\tablecaption{Intensity averaged [3.6] and [4.5] apparent magnitudes for Galactic Cepheids}
\tablehead{ 
\colhead{ID}    &  \colhead{$\log(P)$}  & \colhead{$\langle$[3.6]$\rangle$\tablenotemark{1}}   &  \colhead{$\langle$[4.5]$\rangle$\tablenotemark{1}}  & \colhead{$\langle$[3.6]-[4.5]$\rangle$\tablenotemark{2}} \\
\colhead{}        &  \colhead{[days]} & \colhead{[mags]} & \colhead{[mags]} & \colhead{[mags]}  
 }
\startdata
               S Vul &   1.835 & $    4.358 \pm 0.003 $ & $    4.394 \pm 0.003 $ & $   -0.036 \pm 0.002 $  \\ 
              GY Sge &   1.708 & $    4.311 \pm 0.003 $ & $    4.345 \pm 0.004 $ & $   -0.034 \pm 0.003 $  \\ 
              SV Vul &   1.653 & $    3.711 \pm 0.005 $ & $    3.788 \pm 0.005 $ & $   -0.077 \pm 0.003 $  \\ 
               U Car &   1.589 & $    3.357 \pm 0.005 $ & $    3.415 \pm 0.006 $ & $   -0.058 \pm 0.003 $  \\ 
          $\ell$ Car &   1.551 & $    0.925 \pm 0.004 $ & $    1.047 \pm 0.004 $ & $   -0.122 \pm 0.003 $  \\ 
               T Mon &   1.432 & $    3.359 \pm 0.006 $ & $    3.425 \pm 0.005 $ & $   -0.066 \pm 0.003 $  \\ 
              WZ Sgr &   1.339 & $    4.364 \pm 0.006 $ & $    4.443 \pm 0.005 $ & $   -0.079 \pm 0.003 $  \\ 
              RU Sct &   1.294 & $    4.856 \pm 0.005 $ & $    4.873 \pm 0.005 $ & $   -0.017 \pm 0.002 $  \\
\enddata
\tablenotetext{1}{Shown with the expected random errors from the averaging algorithm; see text.  The systematic errors are 0.016 mags for both the [3.6] and [4.5] data.}
\tablenotetext{2}{Because the errors in Channel 1 and 2 are correlated, the color error was calculated independently by using the error algorithm described in the text.  }
\label{galcephs}
\end{deluxetable}


\clearpage

\begin{deluxetable}{lllllrllll}
\rotate
\tabletypesize{\scriptsize}
\tablecolumns{10} 
\tablewidth{0pc} 
\tablecaption{HST Parallax Cepheids With LKH}
\tablehead{ 
\colhead{Cepheid}  & \colhead{$\log P$}  & \colhead{$m_{[3.6]}$}  &  \colhead{$m_{[4.5]}$} & \colhead{LKH} & \colhead{$(m-M)_o$} & \colhead{$A_{[3.6]}$} & \colhead{$A_{[4.5]}$} &  \colhead{$M_{[3.6]}$} & \colhead{$M_{[4.5]}$}    \\
\colhead{} & \colhead{[days]}   & \colhead{[mag]} &  \colhead{[mag]}  & \colhead{[mag]}   &  \colhead{[mag]}  &  \colhead{[mag]}  & \colhead{[mag]}   &  \colhead{ [mag]} &  \colhead{ [mag]} 
}
\startdata
    $\ell$ Car&$  1.551$ & $  0.925\pm 0.004$ & $  1.047\pm 0.004$ & $  -0.08 $ & $   8.56\pm  0.22$ & 	$  0.031\pm 0.004$ & $  0.024\pm 0.003$ & $  -7.67\pm  0.22$ & $  -7.54\pm  0.22$ \\ 
   $\zeta$ Gem&$  1.007$ & $  2.025\pm 0.002$ & $  2.037\pm 0.003$ & $  -0.03 $ & $   7.81\pm  0.14$ & 	$  0.004\pm 0.001$ & $  0.003\pm 0.001$ & $  -5.79\pm  0.14$ & $  -5.78\pm  0.14$ \\ 
   $\beta$ Dor&$  0.993$ & $  1.858\pm 0.003$ & $  1.871\pm 0.003$ & $  -0.02 $ & $   7.54\pm  0.11$ & 	$  0.012\pm 0.002$ & $  0.009\pm 0.002$ & $  -5.69\pm  0.11$ & $  -5.67\pm  0.11$ \\ 
         W Sgr&$  0.881$ & $  2.721\pm 0.003$ & $  2.719\pm 0.004$ & $  -0.06 $ & $   8.27\pm  0.19$ & 	$  0.022\pm 0.003$ & $  0.017\pm 0.002$ & $  -5.57\pm  0.19$ & $  -5.57\pm  0.19$ \\ 
         X Sgr&$  0.846$ & $  2.423\pm 0.003$ & $  2.409\pm 0.003$ & $  -0.03 $ & $   7.64\pm  0.13$ & 	$  0.046\pm 0.005$ & $  0.035\pm 0.004$ & $  -5.27\pm  0.13$ & $  -5.27\pm  0.13$ \\ 
         Y Sgr&$  0.761$ & $  3.486\pm 0.003$ & $  3.483\pm 0.003$ & $  -0.15 $ & $   8.51\pm  0.30$ & 	$  0.041\pm 0.004$ & $  0.032\pm 0.003$ & $  -5.06\pm  0.30$ & $  -5.06\pm  0.30$ \\ 
  $\delta$ Cep&$  0.730$ & $  2.221\pm 0.003$ & $  2.217\pm 0.003$ & $  -0.01 $ & $   7.19\pm  0.09$ & 	$  0.015\pm 0.002$ & $  0.011\pm 0.002$ & $  -4.99\pm  0.09$ & $  -4.99\pm  0.09$ \\ 
        FF Aql&$  0.650$ & $  3.378\pm 0.001$ & $  3.353\pm 0.001$ & $  -0.03 $ & $   7.79\pm  0.14$ & 	$  0.041\pm 0.004$ & $  0.032\pm 0.003$ & $  -4.45\pm  0.14$ & $  -4.47\pm  0.14$ \\
\enddata

\tablecomments{The LKH factor has already been included the reported distance modulus, thus the absolute magnitude is:  $M = m - (m-M)_o -A$.  See Table \ref{hstcephswo} for distance moduli without LKH.  
The distance moduli for $\beta$ Dor and W Sgr differ from the values tabulated in \cite{Benedict:2007};  the values reported here have been confirmed with Benedict (private communication).    
 }
\label{hstcephs}
\end{deluxetable}



\begin{deluxetable}{lllllrllll}
\rotate
\tabletypesize{\scriptsize}
\tablecolumns{10} 
\tablewidth{0pc} 
\tablecaption{HST Cepheids Without LKH}
\tablehead{ 
\colhead{Cepheid}  & \colhead{$\log P$}  & \colhead{$m_{[3.6]}$}  &  \colhead{$m_{[4.5]}$} & \colhead{LKH} & \colhead{$(m-M)_o$} & \colhead{$A_{[3.6]}$} & \colhead{$A_{[4.5]}$} &  \colhead{$M_{[3.6]}$} & \colhead{$M_{[4.5]}$}    \\
\colhead{} & \colhead{[days]}   & \colhead{[mag]} &  \colhead{[mag]}  & \colhead{[mag]}   &  \colhead{[mag]}  &  \colhead{[mag]}  & \colhead{[mag]}   &  \colhead{ [mag]} &  \colhead{ [mag]} 

}
\startdata
    $\ell$ Car&$  1.551$ & $  0.925\pm 0.004$ & $  1.047\pm 0.004$ & $   0.00 $ & $   8.48\pm  0.22$ & 	$  0.031\pm 0.004$ & $  0.024\pm 0.003$ & $  -7.59\pm  0.22$ & $  -7.46\pm  0.22$ \\ 
   $\zeta$ Gem&$  1.007$ & $  2.025\pm 0.002$ & $  2.037\pm 0.003$ & $   0.00 $ & $   7.78\pm  0.14$ & 	$  0.004\pm 0.001$ & $  0.003\pm 0.001$ & $  -5.76\pm  0.14$ & $  -5.75\pm  0.14$ \\ 
   $\beta$ Dor&$  0.993$ & $  1.858\pm 0.003$ & $  1.871\pm 0.003$ & $   0.00 $ & $   7.52\pm  0.11$ & 	$  0.012\pm 0.002$ & $  0.009\pm 0.002$ & $  -5.67\pm  0.11$ & $  -5.65\pm  0.11$ \\ 
         W Sgr&$  0.881$ & $  2.721\pm 0.003$ & $  2.719\pm 0.004$ & $   0.00 $ & $   8.21\pm  0.19$ & 	$  0.022\pm 0.003$ & $  0.017\pm 0.002$ & $  -5.51\pm  0.19$ & $  -5.51\pm  0.19$ \\ 
         X Sgr&$  0.846$ & $  2.423\pm 0.003$ & $  2.409\pm 0.003$ & $   0.00 $ & $   7.61\pm  0.13$ & 	$  0.046\pm 0.005$ & $  0.035\pm 0.004$ & $  -5.24\pm  0.13$ & $  -5.24\pm  0.13$ \\ 
         Y Sgr&$  0.761$ & $  3.486\pm 0.003$ & $  3.483\pm 0.003$ & $   0.00 $ & $   8.36\pm  0.30$ & 	$  0.041\pm 0.004$ & $  0.032\pm 0.003$ & $  -4.91\pm  0.30$ & $  -4.91\pm  0.30$ \\ 
  $\delta$ Cep&$  0.730$ & $  2.221\pm 0.003$ & $  2.217\pm 0.003$ & $   0.00 $ & $   7.18\pm  0.09$ & 	$  0.015\pm 0.002$ & $  0.011\pm 0.002$ & $  -4.98\pm  0.09$ & $  -4.98\pm  0.09$ \\ 
        FF Aql&$  0.650$ & $  3.378\pm 0.001$ & $  3.353\pm 0.001$ & $   0.00 $ & $   7.76\pm  0.14$ & 	$  0.041\pm 0.004$ & $  0.032\pm 0.003$ & $  -4.42\pm  0.14$ & $  -4.44\pm  0.14$ \\
\enddata
\label{hstcephswo}
\end{deluxetable}



\begin{deluxetable}{llllrllll}
\rotate
\tabletypesize{\scriptsize}
\tablecolumns{9} 
\tablewidth{0pc} 
\tablecaption{Cluster Cepheids}
\tablehead{ 
\colhead{Cepheid}  & \colhead{$\log P$}  & \colhead{$m_{[3.6]}$}  &  \colhead{$m_{[4.5]}$} & \colhead{$(m-M)_o$} & \colhead{$A_{[3.6]}$} & \colhead{$A_{[4.5]}$} &  \colhead{$M_{[3.6]}$} & \colhead{$M_{[4.5]}$}    \\
\colhead{} & \colhead{[days]}   & \colhead{[mag]} &  \colhead{[mag]}  & \colhead{[mag]}    &  \colhead{[mag]}  & \colhead{[mag]}   &  \colhead{ [mag]} &  \colhead{ [mag]} 

}
\startdata
         S Vul&$  1.835$ & $  4.358\pm 0.003$ & $  4.394\pm 0.003$ & $  12.47\pm  0.27$ & 	$  0.188\pm 0.019$ & $  0.144\pm 0.015$ & $  -8.30\pm  0.27$ & $  -8.22\pm  0.27$ \\ 
        SV Vul&$  1.653$ & $  3.711\pm 0.005$ & $  3.788\pm 0.005$ & $  11.26\pm  0.12$ & 	$  0.103\pm 0.011$ & $  0.079\pm 0.008$ & $  -7.65\pm  0.12$ & $  -7.55\pm  0.12$ \\ 
        WZ Sgr&$  1.339$ & $  4.364\pm 0.006$ & $  4.443\pm 0.005$ & $  11.26\pm  0.12$ & 	$  0.102\pm 0.010$ & $  0.078\pm 0.008$ & $  -7.00\pm  0.12$ & $  -6.89\pm  0.12$ \\ 
        RU Sct&$  1.294$ & $  4.856\pm 0.005$ & $  4.873\pm 0.005$ & $  11.11\pm  0.11$ & 	$  0.189\pm 0.019$ & $  0.145\pm 0.015$ & $  -6.44\pm  0.11$ & $  -6.38\pm  0.11$ \\ 
         X Cyg&$  1.214$ & $  3.678\pm 0.005$ & $  3.728\pm 0.005$ & $  10.43\pm  0.05$ & 	$  0.049\pm 0.005$ & $  0.038\pm 0.004$ & $  -6.80\pm  0.05$ & $  -6.74\pm  0.05$ \\ 
      V340 Nor&$  1.053$ & $  5.453\pm 0.002$ & $  5.480\pm 0.002$ & $  11.18\pm  0.15$ & 	$  0.064\pm 0.007$ & $  0.049\pm 0.005$ & $  -5.79\pm  0.15$ & $  -5.75\pm  0.15$ \\ 
        TW Nor&$  1.033$ & $  6.152\pm 0.003$ & $  6.160\pm 0.004$ & $  11.42\pm  0.13$ & 	$  0.238\pm 0.024$ & $  0.183\pm 0.018$ & $  -5.51\pm  0.13$ & $  -5.44\pm  0.13$ \\ 
   $\zeta$ Gem&$  1.007$ & $  2.025\pm 0.002$ & $  2.037\pm 0.003$ & $   7.75\pm  0.09$ & 	$  0.004\pm 0.001$ & $  0.003\pm 0.001$ & $  -5.73\pm  0.09$ & $  -5.71\pm  0.09$ \\
\enddata
\label{cluscephs}
\end{deluxetable}



\begin{deluxetable}{llllrllll}
\rotate
\tabletypesize{\scriptsize}
\tablecolumns{9} 
\tablewidth{0pc} 
\tablecaption{IRSB Cepheids}
\tablehead{ 
\colhead{Cepheid}  & \colhead{$\log P$}  & \colhead{$m_{[3.6]}$}  &  \colhead{$m_{[4.5]}$} & \colhead{$(m-M)_o$} & \colhead{$A_{[3.6]}$} & \colhead{$A_{[4.5]}$} &  \colhead{$M_{[3.6]}$} & \colhead{$M_{[4.5]}$}    \\
\colhead{} & \colhead{[days]}   & \colhead{[mag]} &  \colhead{[mag]}  & \colhead{[mag]}    &  \colhead{[mag]}  & \colhead{[mag]}   &  \colhead{ [mag]} &  \colhead{ [mag]} 

}
\startdata
          S Vul&$  1.835$ & $  4.358\pm 0.003$ & $  4.394\pm 0.003$ & $  12.84\pm  0.08$ & 	$  0.188\pm 0.019$ & $  0.144\pm 0.015$ & $  -8.67\pm  0.08$ & $  -8.59\pm  0.08$ \\ 
        GY Sge&$  1.708$ & $  4.311\pm 0.003$ & $  4.345\pm 0.004$ & $  12.28\pm  0.06$ & 	$  0.241\pm 0.042$ & $  0.185\pm 0.032$ & $  -8.21\pm  0.08$ & $  -8.12\pm  0.07$ \\ 
        SV Vul&$  1.653$ & $  3.711\pm 0.005$ & $  3.788\pm 0.005$ & $  11.38\pm  0.04$ & 	$  0.103\pm 0.011$ & $  0.079\pm 0.008$ & $  -7.77\pm  0.04$ & $  -7.67\pm  0.04$ \\ 
         U Car&$  1.589$ & $  3.357\pm 0.005$ & $  3.415\pm 0.006$ & $  10.74\pm  0.03$ & 	$  0.054\pm 0.006$ & $  0.041\pm 0.004$ & $  -7.44\pm  0.03$ & $  -7.37\pm  0.03$ \\ 
    $\ell$ Car&$  1.551$ & $  0.925\pm 0.004$ & $  1.047\pm 0.004$ & $   8.57\pm  0.02$ & 	$  0.031\pm 0.004$ & $  0.024\pm 0.003$ & $  -7.68\pm  0.02$ & $  -7.55\pm  0.02$ \\ 
         T Mon&$  1.432$ & $  3.359\pm 0.006$ & $  3.425\pm 0.005$ & $  10.60\pm  0.03$ & 	$  0.037\pm 0.004$ & $  0.028\pm 0.003$ & $  -7.27\pm  0.03$ & $  -7.20\pm  0.03$ \\ 
        WZ Sgr&$  1.339$ & $  4.364\pm 0.006$ & $  4.443\pm 0.005$ & $  11.26\pm  0.04$ & 	$  0.102\pm 0.010$ & $  0.078\pm 0.008$ & $  -7.00\pm  0.04$ & $  -6.89\pm  0.04$ \\ 
        RU Sct&$  1.294$ & $  4.856\pm 0.005$ & $  4.873\pm 0.005$ & $  11.38\pm  0.04$ & 	$  0.189\pm 0.019$ & $  0.145\pm 0.015$ & $  -6.71\pm  0.05$ & $  -6.65\pm  0.04$ \\
\enddata
\label{irsbcephs}
\end{deluxetable}


\clearpage

\def\crit{$ \rm 6 \leq P \leq  60$}
\begin{deluxetable}{lllrrclrclrcl}
\rotate
\tabletypesize{\tiny}
\tablecolumns{13} 
\tablewidth{0pc} 
\tablecaption{Period-Luminosity (Leavitt) Laws}
\tablehead{ 
\colhead{}              & \colhead{}                 & \colhead{}          &  \colhead{}     & \multicolumn{3}{c}{weights= $\frac{1}{0.1^{2}}$ }  & \multicolumn{3}{c}{weights = $\frac{1}{\sigma_{M}^{2}}$ }  & \multicolumn{3}{c}{weights = $\frac{1}{\sigma_{M}^{2}+0.1^{2}}$ }    \\
\colhead{Galaxy} & \colhead{Sample}   & \colhead{Band} &  \colhead{N}  & \colhead{(slope) $a$}   &  \colhead{(intercept) $b$}  &  \colhead{$\sigma$}  & \colhead{(slope) $a$}   &  \colhead{(intercept) $b$}  &  \colhead{$\sigma$}  & \colhead{(slope) $a$}   &  \colhead{(intercept) $b$}  &  \colhead{$\sigma$} 
}
\startdata
MW &   $\rm\pi_{HST}$                               & [3.6]  &    10 &$   -3.40\pm   0.12 $ &$   -5.81\pm   0.04 $ & 0.09 &$   -3.35\pm   0.22 $ &$   -5.80\pm   0.06 $ &  0.10 & $   -3.39\pm   0.26 $ & $   -5.80\pm   0.07 $ &  0.09\\ 
MW &   $\rm \pi_{HST}$  \crit                      & [3.6]  &      5 &$   -3.33\pm   0.18 $ &$   -5.82\pm   0.05 $ & 0.09 &$   -3.33\pm   0.35 $ &$   -5.79\pm   0.06 $ &  0.08 & $   -3.34\pm   0.39 $ & $   -5.80\pm   0.08 $ &  0.08\\ 
MW &   $\rm\pi_{HST}$  no LKH               & [3.6]  &    10 &$   -3.40\pm   0.12 $ &$   -5.75\pm   0.04 $ & 0.09 &$   -3.33\pm   0.22 $ &$   -5.76\pm   0.06 $ &  0.10 & $   -3.37\pm   0.26 $ & $   -5.76\pm   0.07 $ &  0.10\\ 
MW &   $\rm \pi_{HST}$  no LKH  \crit      & [3.6]  &      5 &$   -3.27\pm   0.18 $ &$   -5.78\pm   0.05 $ & 0.07 &$   -3.27\pm   0.35 $ &$   -5.75\pm   0.06 $ &  0.06 & $   -3.28\pm   0.39 $ & $   -5.76\pm   0.08 $ &  0.07\\ 
MW  &  $\rm \pi_{MS}$                                 & [3.6]  &   18 &$   -3.00\pm   0.07 $ &$   -5.76\pm   0.02 $ & 0.19 &$   -3.43\pm   0.08 $ &$   -5.77\pm   0.02 $ &  0.17 & $   -3.11\pm   0.12 $ & $   -5.75\pm   0.03 $ &  0.19\\ 
MW  &  $\rm \pi_{MS}$  \crit                         & [3.6]  &   11 &$   -3.30\pm   0.12 $ &$   -5.68\pm   0.03 $ & 0.19 &$   -3.70\pm   0.10 $ &$   -5.74\pm   0.02 $ &  0.17 & $   -3.37\pm   0.17 $ & $   -5.70\pm   0.04 $ &  0.19\\ 
MW &  $\rm \pi_{IRSB}$                              & [3.6]  &    34 &$   -3.42\pm   0.05 $ &$   -5.73\pm   0.02 $ & 0.13 &$   -3.39\pm   0.02 $ &$   -5.74\pm   0.01 $ &  0.13 & $   -3.41\pm   0.06 $ & $   -5.73\pm   0.02 $ &  0.13\\ 
MW &  $\rm \pi_{IRSB}$  \crit                      & [3.6]  &   23 &$   -3.35\pm   0.08 $ &$   -5.75\pm   0.02 $ & 0.13 &$   -3.31\pm   0.03 $ &$   -5.77\pm   0.01 $ &  0.14 & $   -3.33\pm   0.09 $ & $   -5.75\pm   0.03 $ &  0.13\\
\enddata
\tablenotetext{1}{The LMC sample does not contain Cepheids with periods less than 6 days.  The LMC data are discussed in \cite{Scowcroft:2011}.   }
\tablenotetext{2}{Force fit the LMC slope from the un-weighting method to find the zero-point. }
\tablecomments{The form of the PL relation used in these fits is: $M = a(\log P-1) +b$.   The values in bold indicate our adopted PL slope
and zero-points.   To eliminate asymmetric rounding errors when reporting two significant figures in the zero-points a floor rounding
function was used which rounds toward negative infinity rather than away from zero.  }
\label{TPL}
\end{deluxetable}


\clearpage

\begin{deluxetable}{lcccccccccccccc}
\rotate
\tabletypesize{\scriptsize}
\tablecolumns{5} 
\tablewidth{0pc} 
\tablecaption{Multi-Wavelength parameters for Galactic Cepheids.}
\tablehead{ 
\colhead{Cepheid}      & \colhead{$\log P$} & \colhead{$E(B-V)$} & \colhead{$\pi$ [mas]} & \colhead{$(m-M)$} & \colhead{LKH} & \colhead{$m_B$}  & \colhead{$m_V$} & \colhead{$m_I$}   & \colhead{$m_J$}
& \colhead{$m_H$} & \colhead{$m_K$} & \colhead{$m_{[3.6]}$} & \colhead{$m_{[4.5]}$} 
 }
\startdata
    $\ell$ Car & 1.551  & $ 0.154 \pm 0.011 $ & $ 2.01 \pm 0.20  $ & $ 8.48 \pm 0.22 $ & $  -0.08 $ & $ 4.986  $ & $  3.722  $ & $ 2.552  $ & $  1.674  $ & $ 1.198  $ & $  1.076  $ & $ 0.925  $ & $  1.047  $   \\ 
   $\zeta$ Gem & 1.007  & $ 0.018 \pm 0.007 $ & $ 2.78 \pm 0.18  $ & $ 7.78 \pm 0.14 $ & $  -0.03 $ & $ 4.701  $ & $  3.889  $ & $ 3.107  $ & $  2.585  $ & \nodata    & $  2.114  $ & $ 2.025  $ & $  2.037  $   \\ 
   $\beta$ Dor & 0.993  & $ 0.058 \pm 0.009 $ & $ 3.14 \pm 0.16  $ & $ 7.52 \pm 0.11 $ & $  -0.02 $ & $ 4.542  $ & $  3.741  $ & $ 2.939  $ & $  2.376  $ & $ 2.020  $ & $  1.948  $ & $ 1.858  $ & $  1.871  $   \\ 
         W Sgr & 0.881  & $ 0.109 \pm 0.007 $ & $ 2.28 \pm 0.20  $ & $ 8.21 \pm 0.19 $ & $  -0.06 $ & $ 5.412  $ & $  4.665  $ & $ 3.846  $ & $  3.252  $ & $ 2.952  $ & $  2.828  $ & $ 2.721  $ & $  2.719  $   \\ 
         X Sgr & 0.846  & $ 0.227 \pm 0.013 $ & $ 3.00 \pm 0.18  $ & $ 7.61 \pm 0.13 $ & $  -0.03 $ & $ 5.299  $ & $  4.550  $ & $ 3.649  $ & $  2.916  $ & $ 2.626  $ & $  2.636  $ & $ 2.423  $ & $  2.409  $   \\ 
         Y Sgr & 0.761  & $ 0.203 \pm 0.007 $ & $ 2.13 \pm 0.29  $ & $ 8.36 \pm 0.30 $ & $  -0.15 $ & $ 6.595  $ & $  5.738  $ & $ 4.789  $ & $  4.086  $ & $ 3.690  $ & $  3.609  $ & $ 3.486  $ & $  3.483  $   \\ 
  $\delta$ Cep & 0.730  & $ 0.073 \pm 0.007 $ & $ 3.66 \pm 0.15  $ & $ 7.18 \pm 0.09 $ & $  -0.01 $ & $ 4.602  $ & $  3.955  $ & $ 3.168  $ & $  2.689  $ & $ 2.378  $ & $  2.310  $ & $ 2.221  $ & $  2.217  $   \\ 
        FF Aql & 0.650  & $ 0.204 \pm 0.008 $ & $ 2.81 \pm 0.18  $ & $ 7.76 \pm 0.14 $ & $  -0.03 $ & $ 6.135  $ & $  5.372  $ & $ 4.488  $ & $  3.864  $ & $ 3.565  $ & $  3.478  $ & $ 3.378  $ & $  3.353  $   \\
\enddata
\label{nest}
\end{deluxetable}



\clearpage

\begin{deluxetable}{ccrcccrcccc}
\tabletypesize{\scriptsize}
\tablecolumns{5} 
\tablewidth{0pc} 
\tablecaption{Multi-wavelength Period-Luminosity Relations for the LMC and MW.  }
\tablehead{ 
\colhead{}              & \colhead{}                 & \multicolumn{4}{c}{LMC}  & \multicolumn{3}{c}{MW }  &  \colhead{}  & \colhead{}      \\
\colhead{Filter}      & \colhead{$\lambda$ [\micron]} & \colhead{N}  & \colhead{$a$} & \colhead{$b$} & \colhead{$\sigma$}
& \colhead{N}   & \colhead{$b$} & \colhead{$\sigma$} & \colhead{$\mu$} & \colhead{$A_{\lambda}$}
   
 }
\startdata
    B &  0.44 &   384 & $ -2.47 \pm  0.11 $ & $ 15.60 \pm 0.03 $ &   0.42 & 10 & $ -3.37 \pm  0.07 $ &   0.24 & $  18.97 \pm 0.08 $ &  0.51 \\
    V &  0.55 &   388 & $ -2.77 \pm  0.08 $ & $ 14.81 \pm 0.02 $ &   0.31 & 10 & $ -4.08 \pm  0.04 $ &   0.15 & $  18.88 \pm 0.05 $ &  0.39 \\
I$_C$ &  0.80 &   392 & $ -2.97 \pm  0.05 $ & $ 13.93 \pm 0.01 $ &   0.20 & 10 & $ -4.77 \pm  0.03 $ &   0.12 & $  18.70 \pm 0.04 $ &  0.23 \\
    J &  1.24 &    82 & $ -3.15 \pm  0.07 $ & $ 13.26 \pm 0.02 $ &   0.15 & 10 & $ -5.33 \pm  0.03 $ &   0.09 & $  18.59 \pm 0.04 $ &  0.11 \\
    H &  1.66 &    82 & $ -3.23 \pm  0.06 $ & $ 12.90 \pm 0.02 $ &   0.12 & 9 & $ -5.65 \pm  0.03 $ &   0.09 & $  18.54 \pm 0.03 $ &  0.07 \\
K$_s$ &  2.16 &    82 & $ -3.27 \pm  0.05 $ & $ 12.80 \pm 0.02 $ &   0.11 & 10 & $ -5.70 \pm  0.03 $ &   0.10 & $  18.50 \pm 0.03 $ &  0.04 \\
{[3.6]} &  3.55 &    80 & $ -3.31 \pm  0.05 $ & $ 12.70 \pm 0.02 $ &   0.11 & 10 & $ -5.79 \pm  0.03 $ &   0.10 & $  18.50 \pm 0.03 $ &  0.02 \\
{[4.5]} &  4.49 &    80 & $ -3.21 \pm  0.06 $ & $ 12.69 \pm 0.02 $ &   0.11 & 10 & $ -5.77 \pm  0.03 $ &   0.10 & $  18.46 \pm 0.03 $ &  0.01 \\
\enddata
\tablecomments{The form of the PL relation is: $M_{\lambda} = a(\log P-1)+b$.   The LMC slope is used to constrain the MW zero-point.   The apparent distance moduli ($\mu$) are found by differencing the LMC and MW zero-points and the extinction ($A_{\lambda}$) is the solution from Figure \ref{LMC_2}.   }
\label{mpl}
\end{deluxetable}



\clearpage

\begin{deluxetable}{lcccc}
\tabletypesize{\scriptsize}
\tablecolumns{5} 
\tablewidth{0pc} 
\tablecaption{Constants used in the photometric reduction.}
\tablehead{ 
\colhead{Channel Criterion}      & \colhead{$u$} & \colhead{$v$}  & \colhead{$f_{corr}$} & \colhead{$zp$}   
 }
\startdata
Ch 1. non-saturated   &   1.000         &    0.080          &   1.021         &  $280.9\pm4.1$            \\  
Ch 1. saturated       &   1.000         &    0.000          &   1.021         &  $280.9\pm4.1$            \\  
Ch 2. non-saturated   &   1.000         &    0.025          &   1.012         &  $179.7\pm2.6$            \\  
Ch 2. saturated       &   0.965         &    -0.050         &   1.012         &  $179.7\pm2.6$            \\  
\enddata
\tablecomments{A source is considered saturated if 1 or more pixels were masked during the PRF fitting procedure; see text.  }
\label{constants}
\end{deluxetable}




\clearpage

\begin{figure}
\begin{tabular}{ccc}
                \resizebox{50mm}{50mm}{\includegraphics{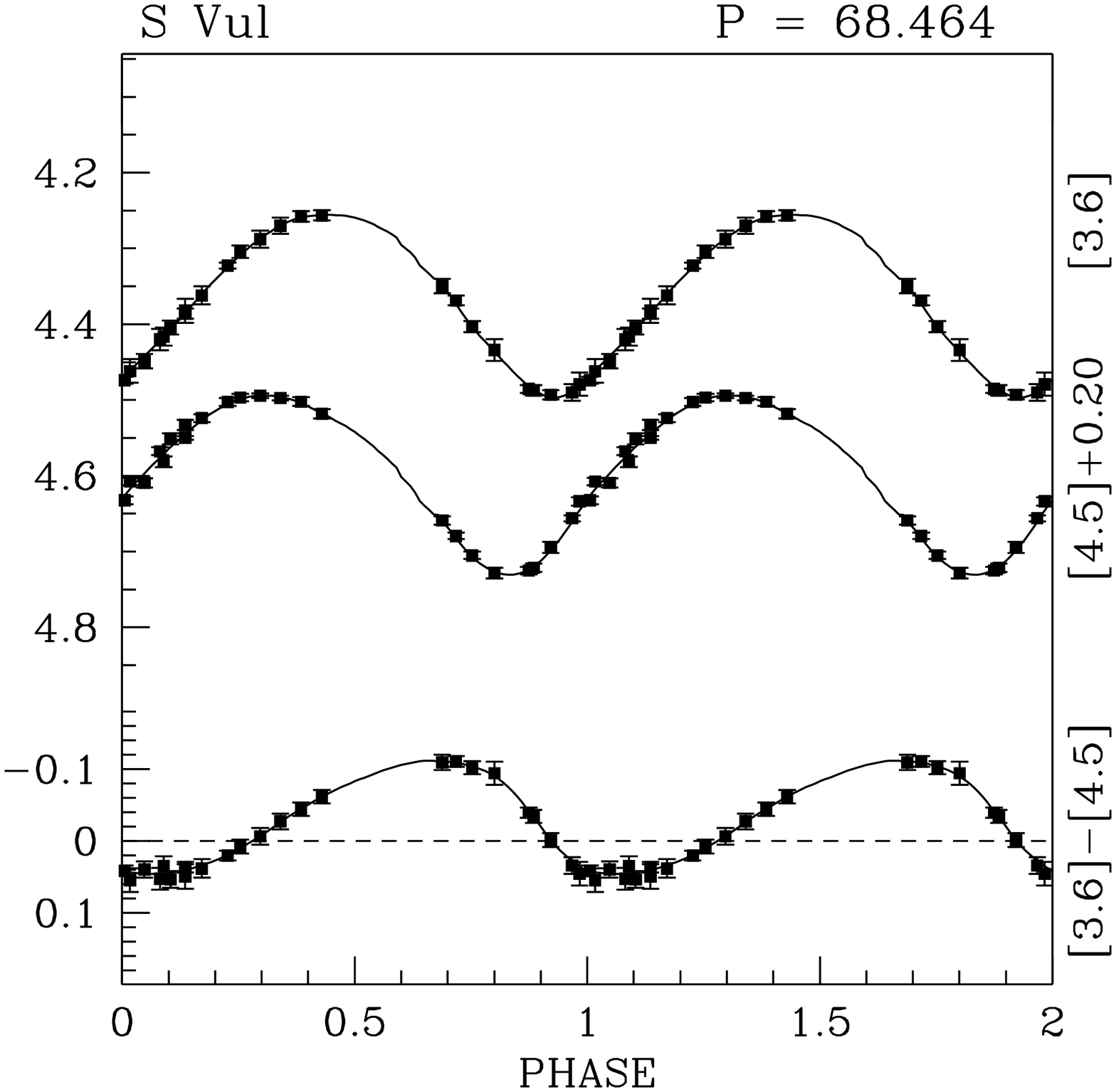}} & 
                \resizebox{50mm}{50mm}{\includegraphics{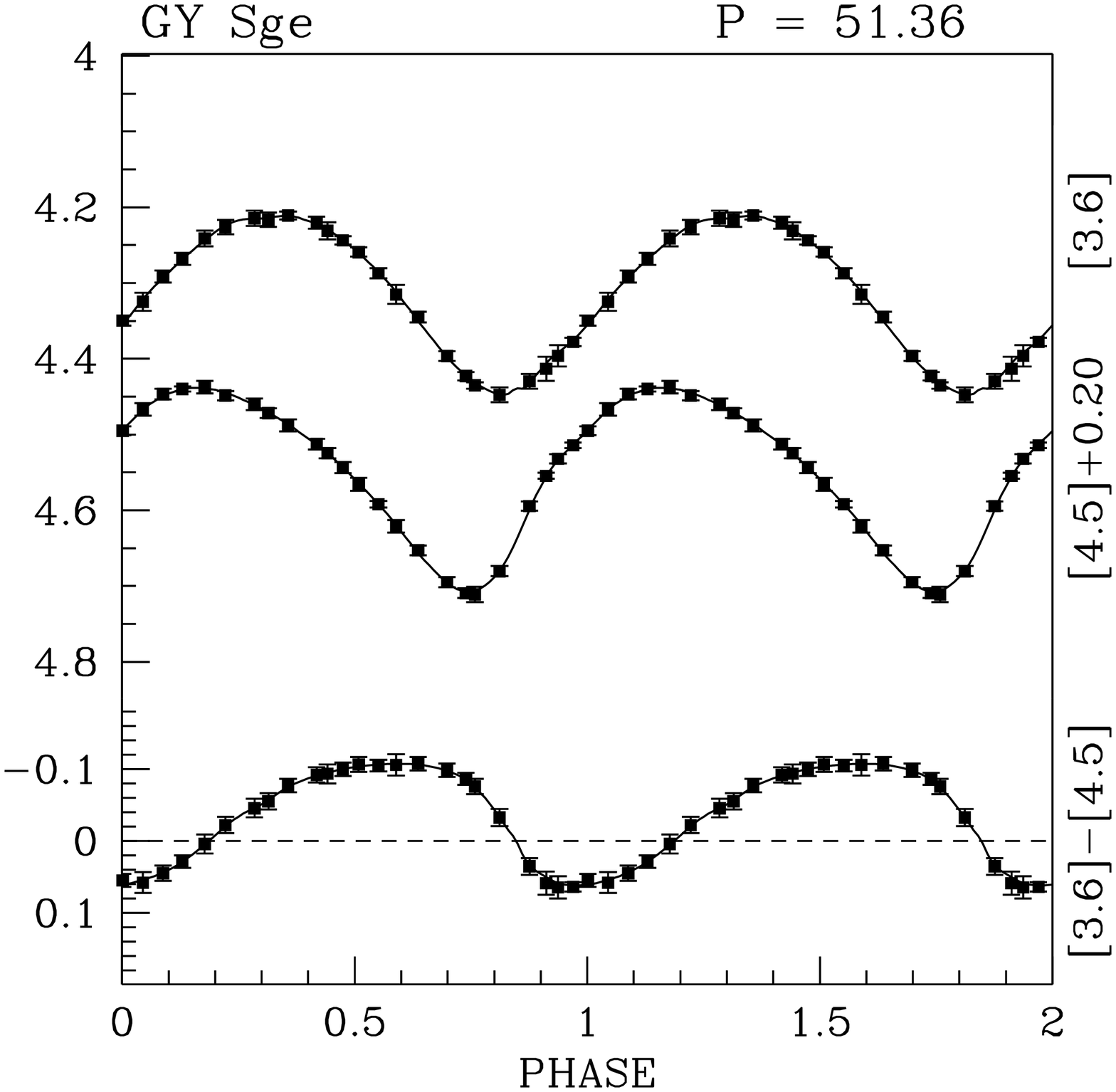}} & 
                \resizebox{50mm}{50mm}{\includegraphics{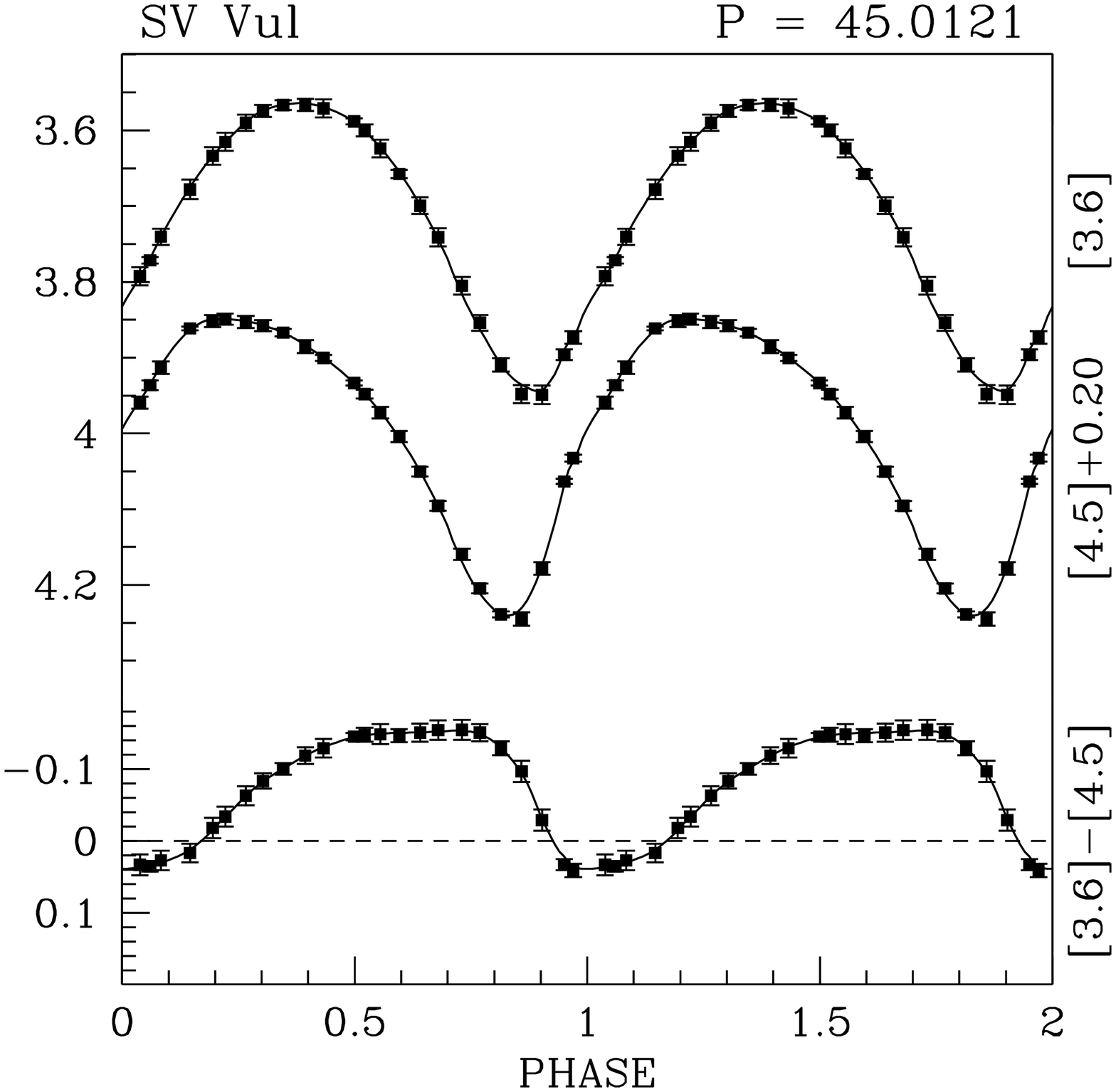}} \\ 
                \resizebox{50mm}{50mm}{\includegraphics{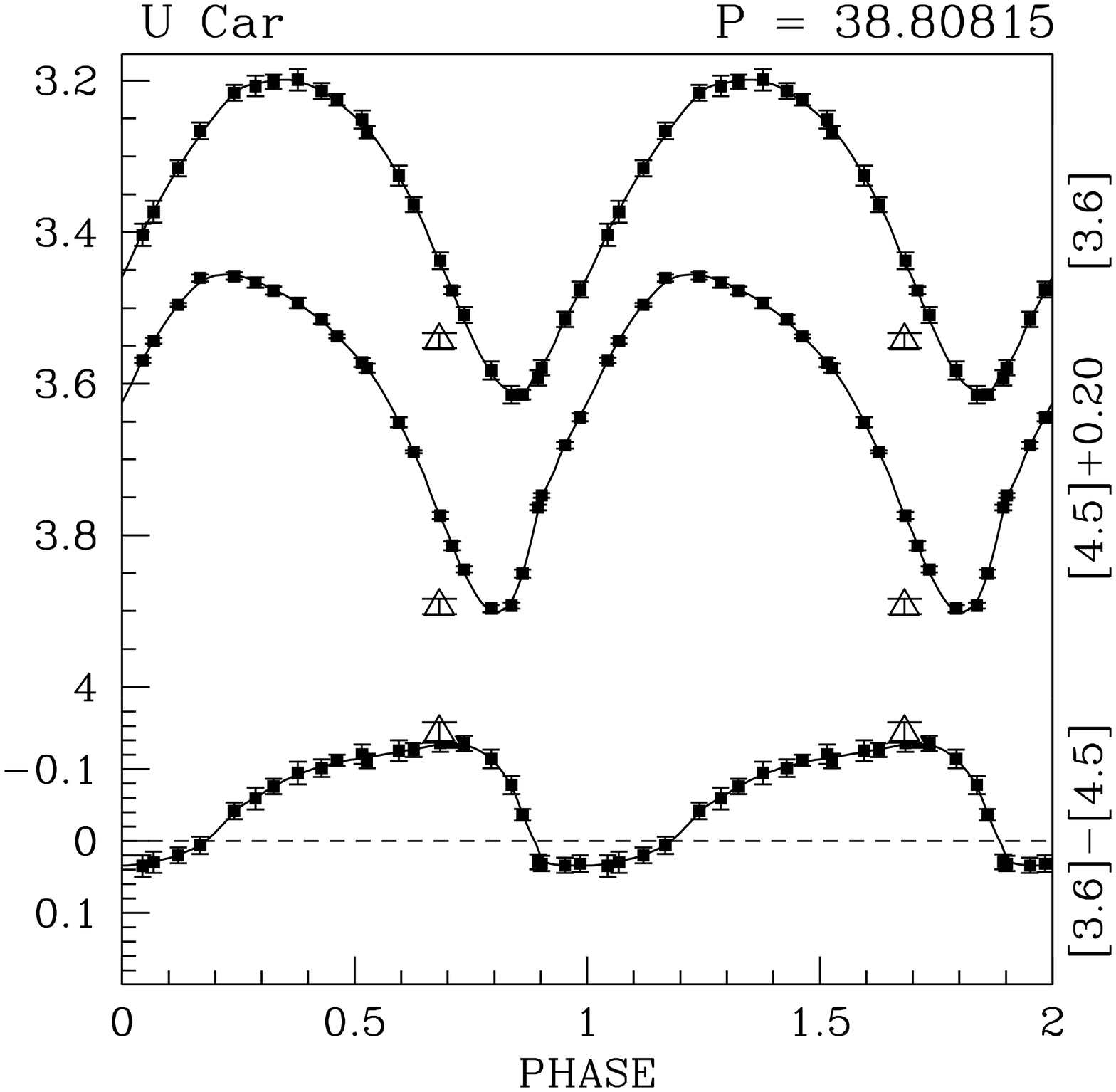}} & 
                \resizebox{50mm}{50mm}{\includegraphics{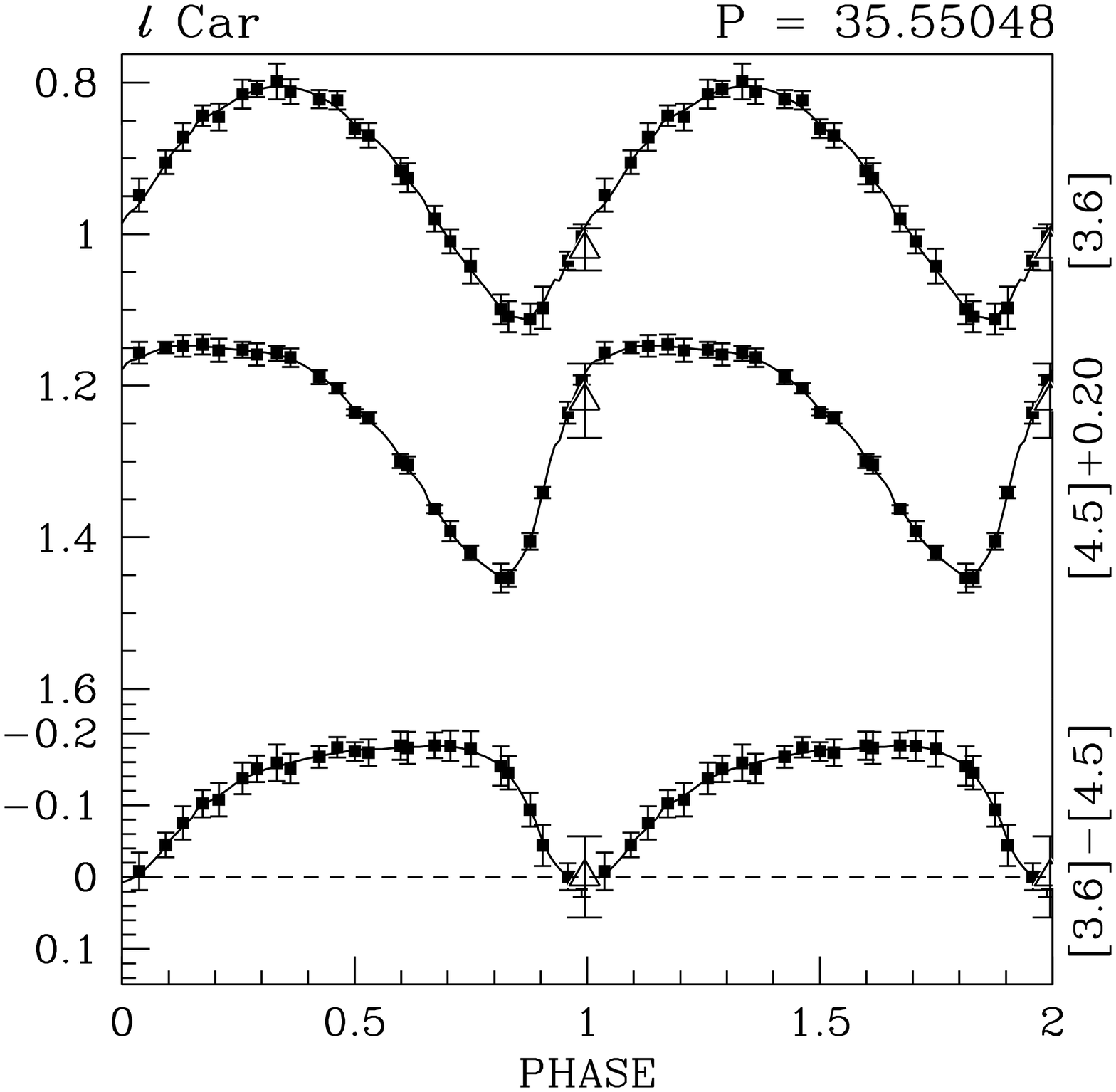}} & 
                \resizebox{50mm}{50mm}{\includegraphics{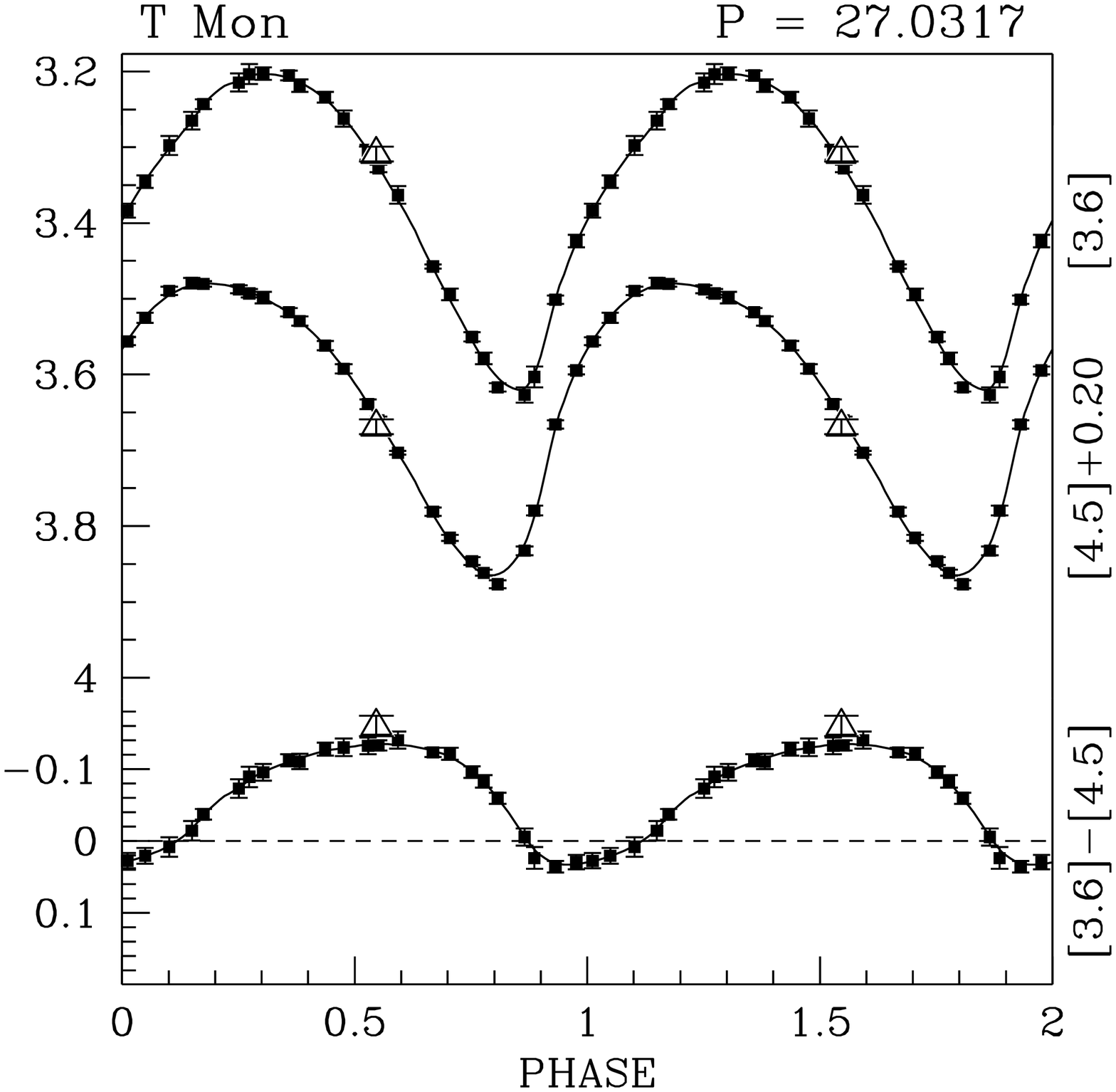}} \\ 
                \resizebox{50mm}{50mm}{\includegraphics{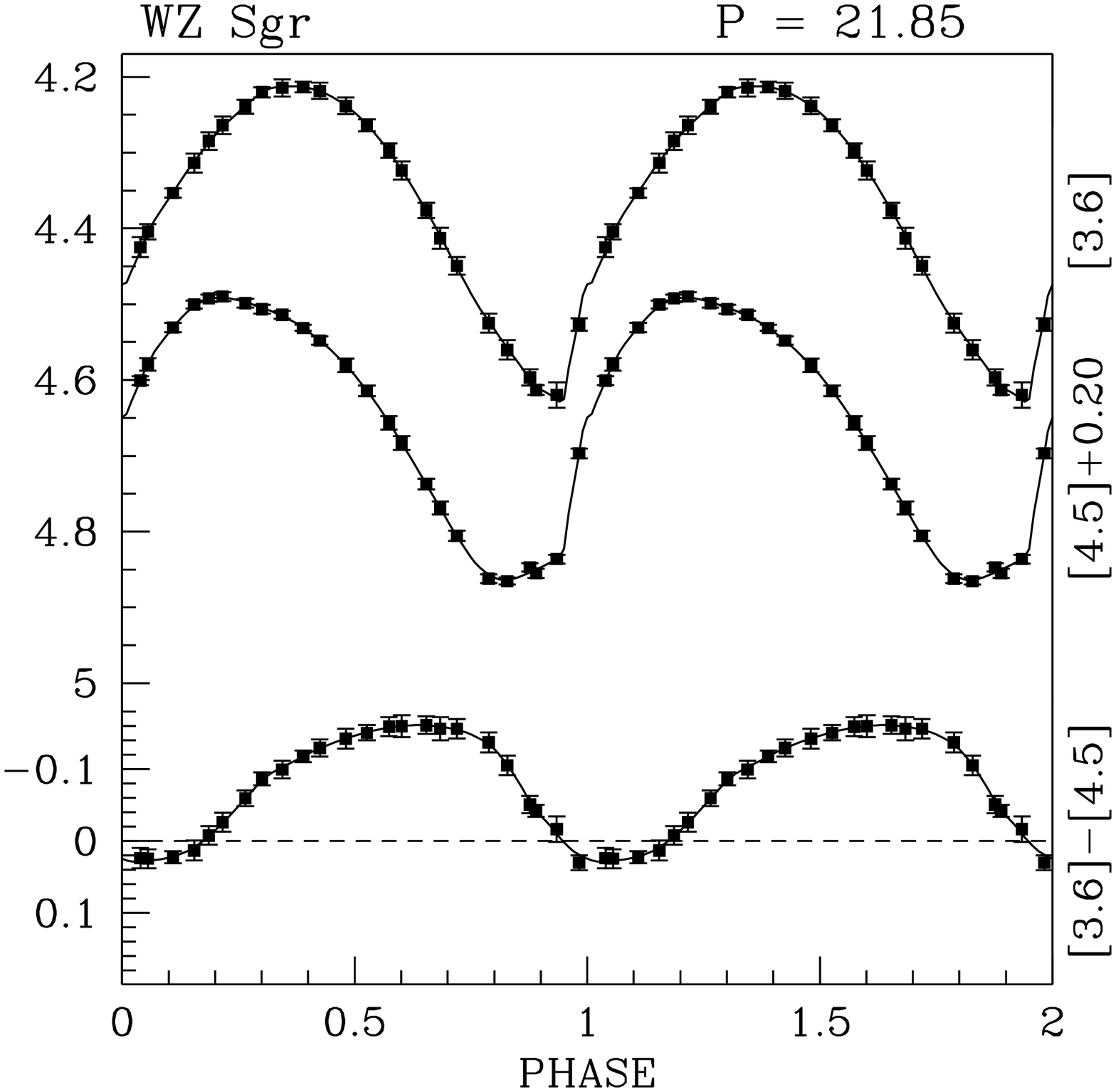}} & 
                \resizebox{50mm}{50mm}{\includegraphics{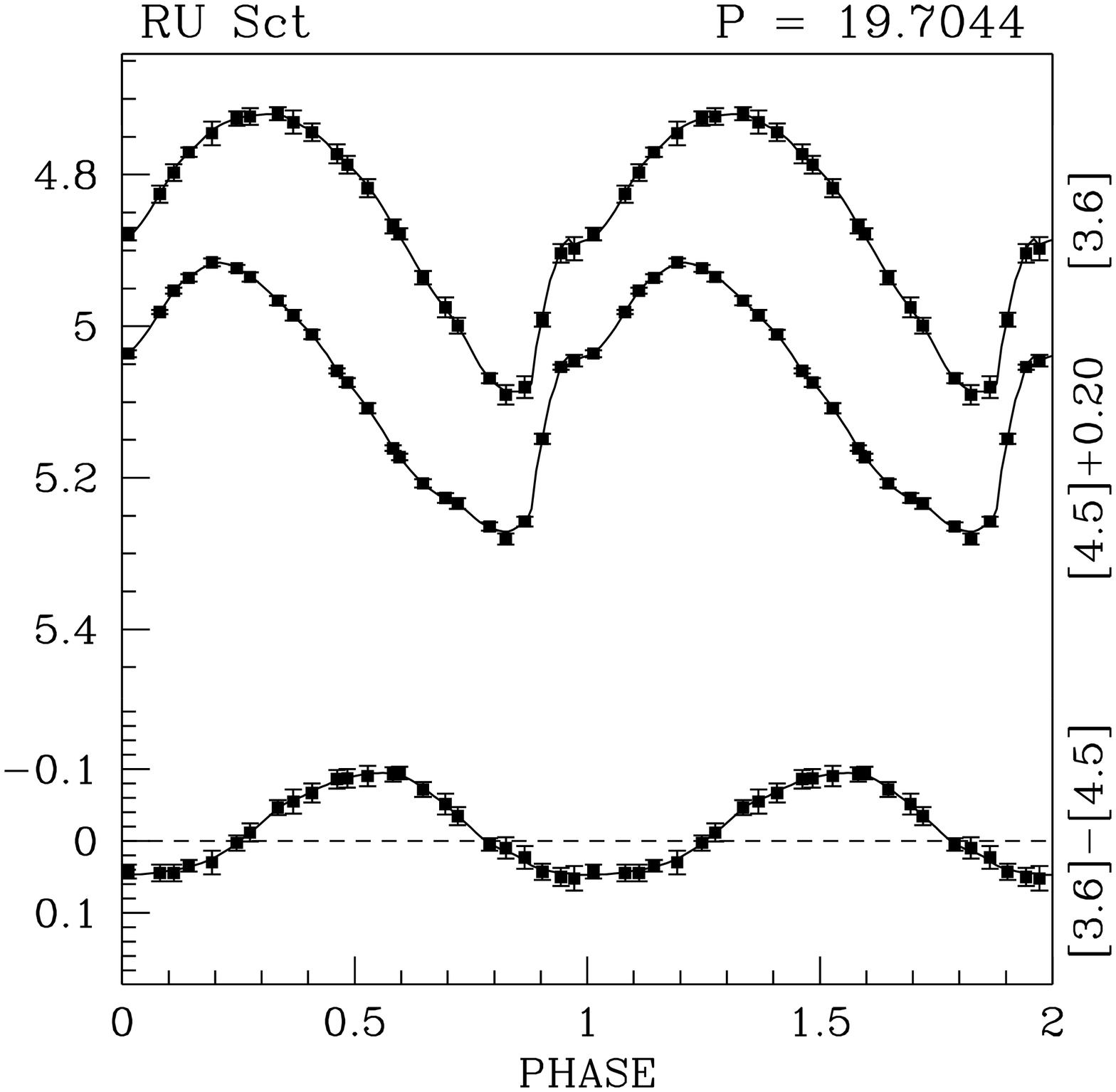}} & 
                \resizebox{50mm}{50mm}{\includegraphics{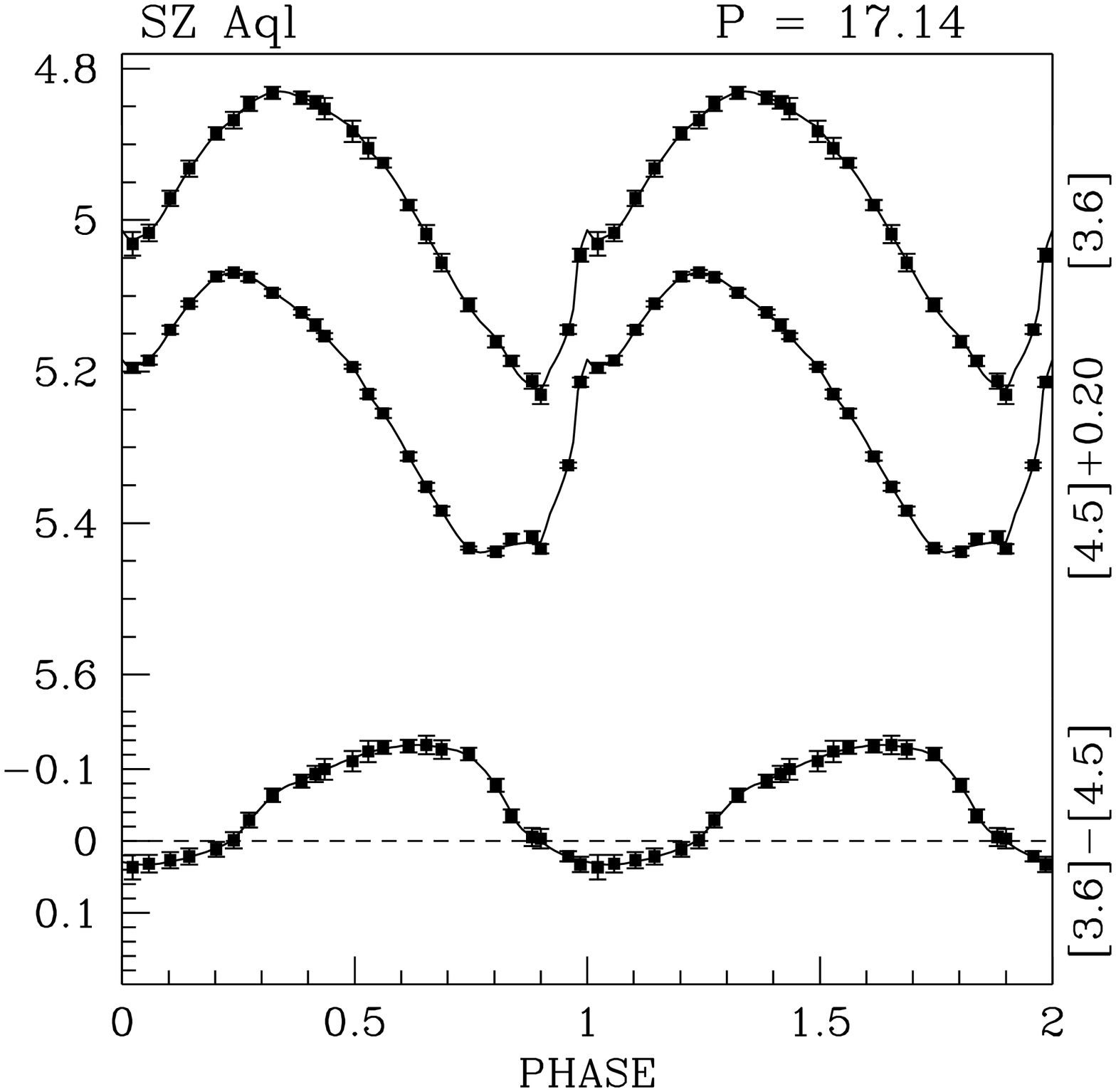}} \\ 
\end{tabular}
\caption{IRAC [3.6] and [4.5] light-curves for 37 Galactic Cepheids.  The error bars correspond to the random photometric error and the solid line is the GLOESS interpolated curve.  Data points from \cite{Marengo:2010} are over-plotted as open triangles when available.   }
\label{LC}
\end{figure}

\addtocounter{figure}{-1}
\begin{figure}
\begin{tabular}{ccc}
                \resizebox{50mm}{50mm}{\includegraphics{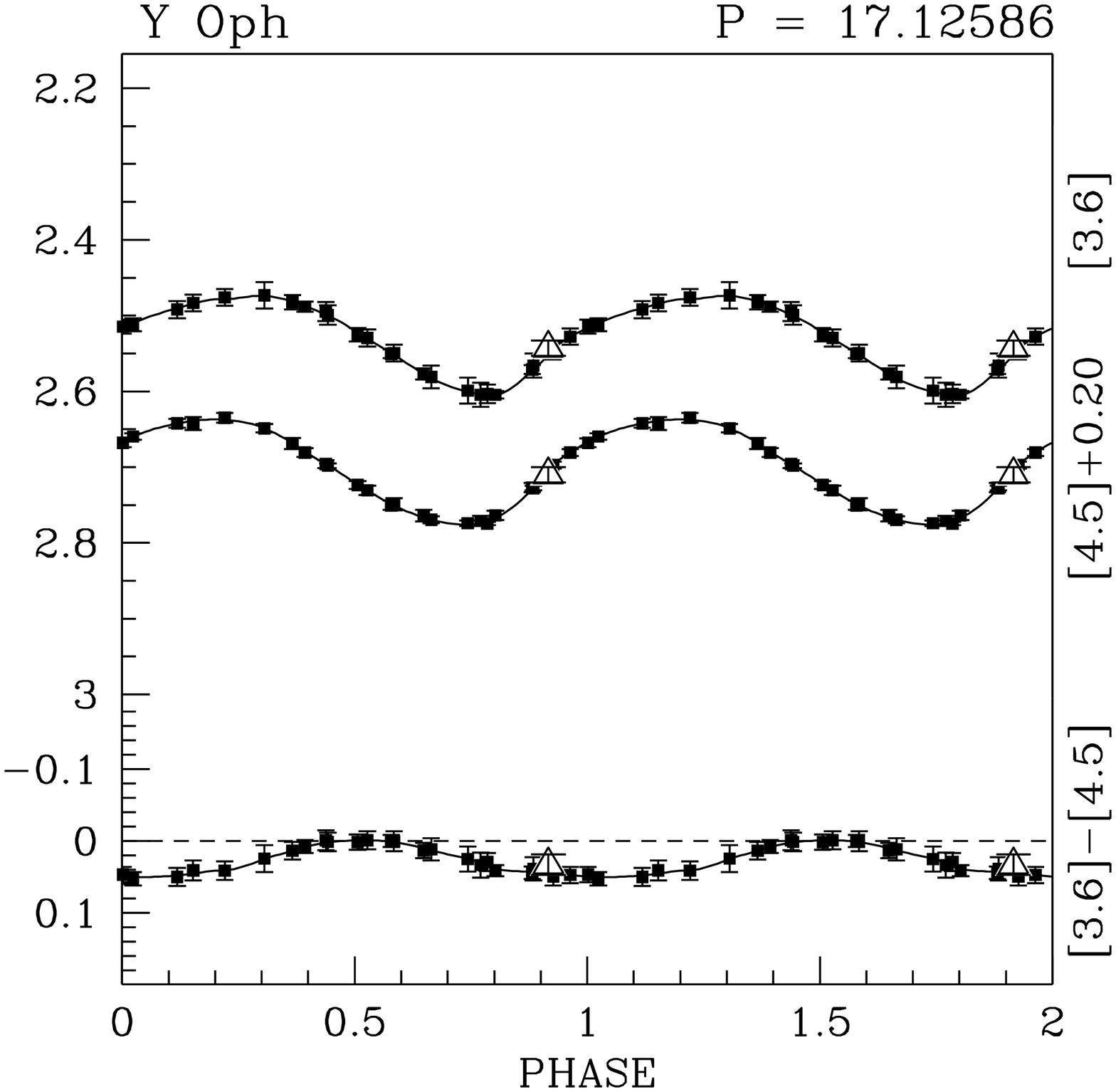}} & 
                \resizebox{50mm}{50mm}{\includegraphics{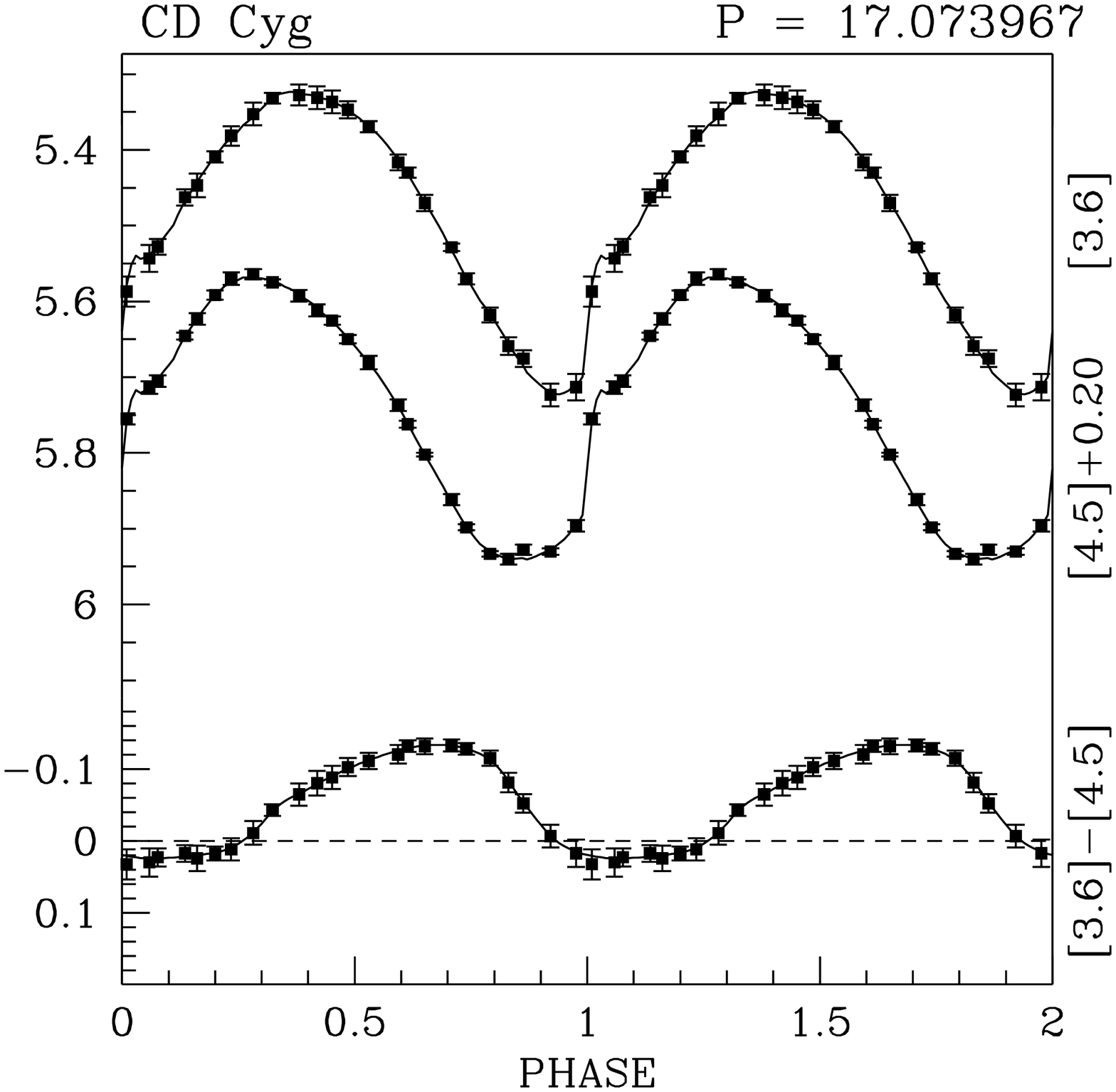}} & 
                \resizebox{50mm}{50mm}{\includegraphics{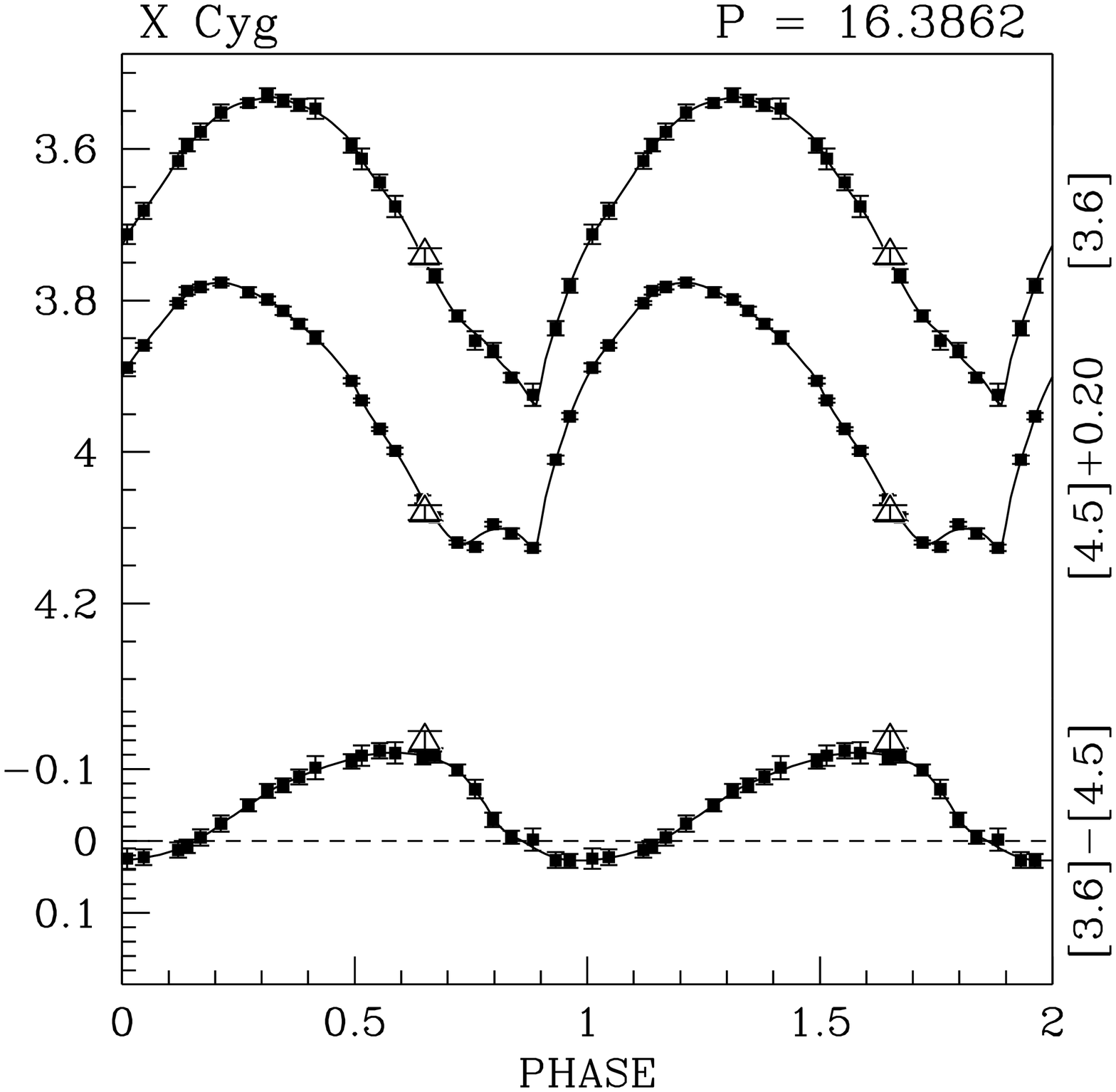}} \\ 
                \resizebox{50mm}{50mm}{\includegraphics{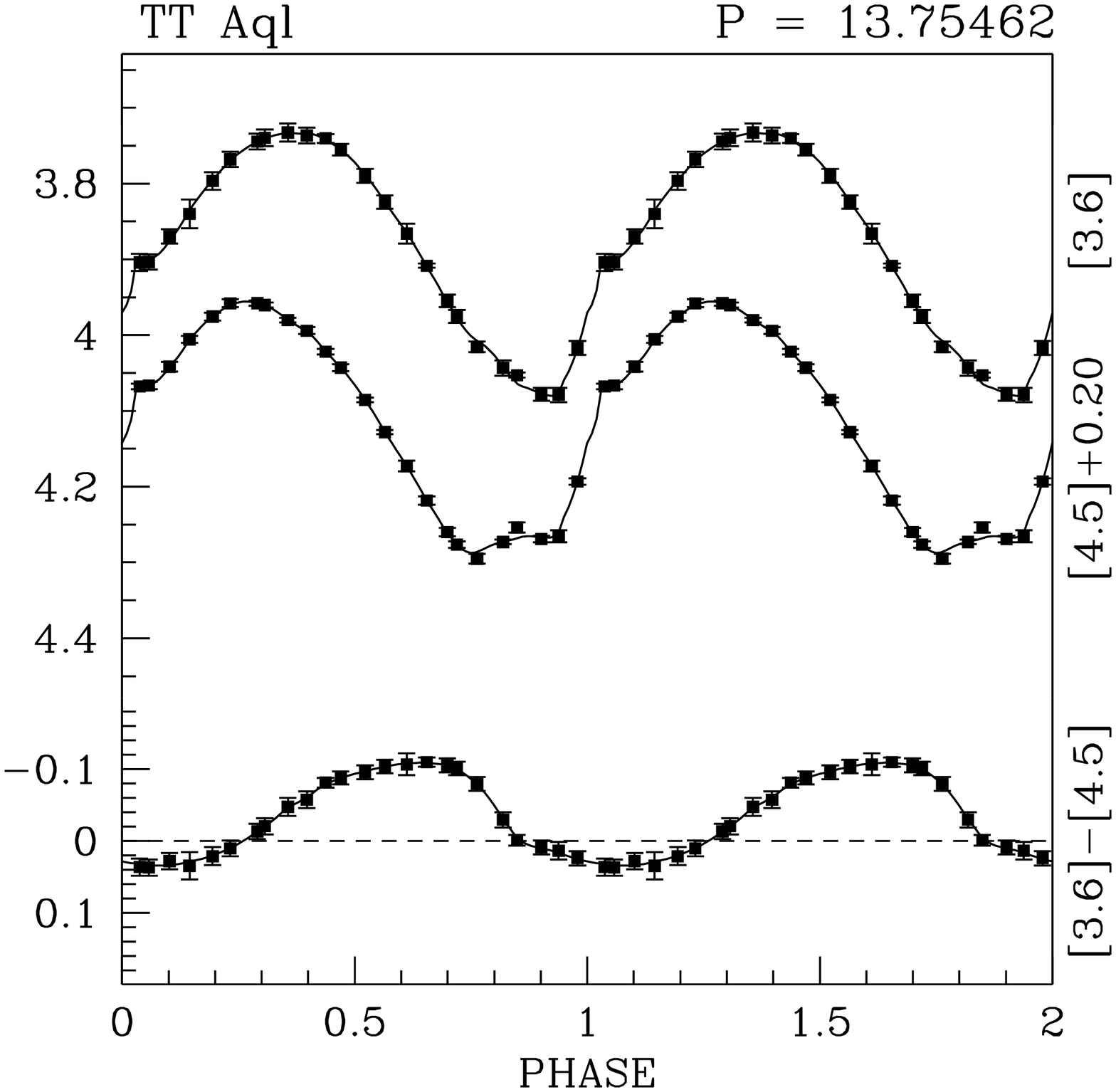}} & 
                \resizebox{50mm}{50mm}{\includegraphics{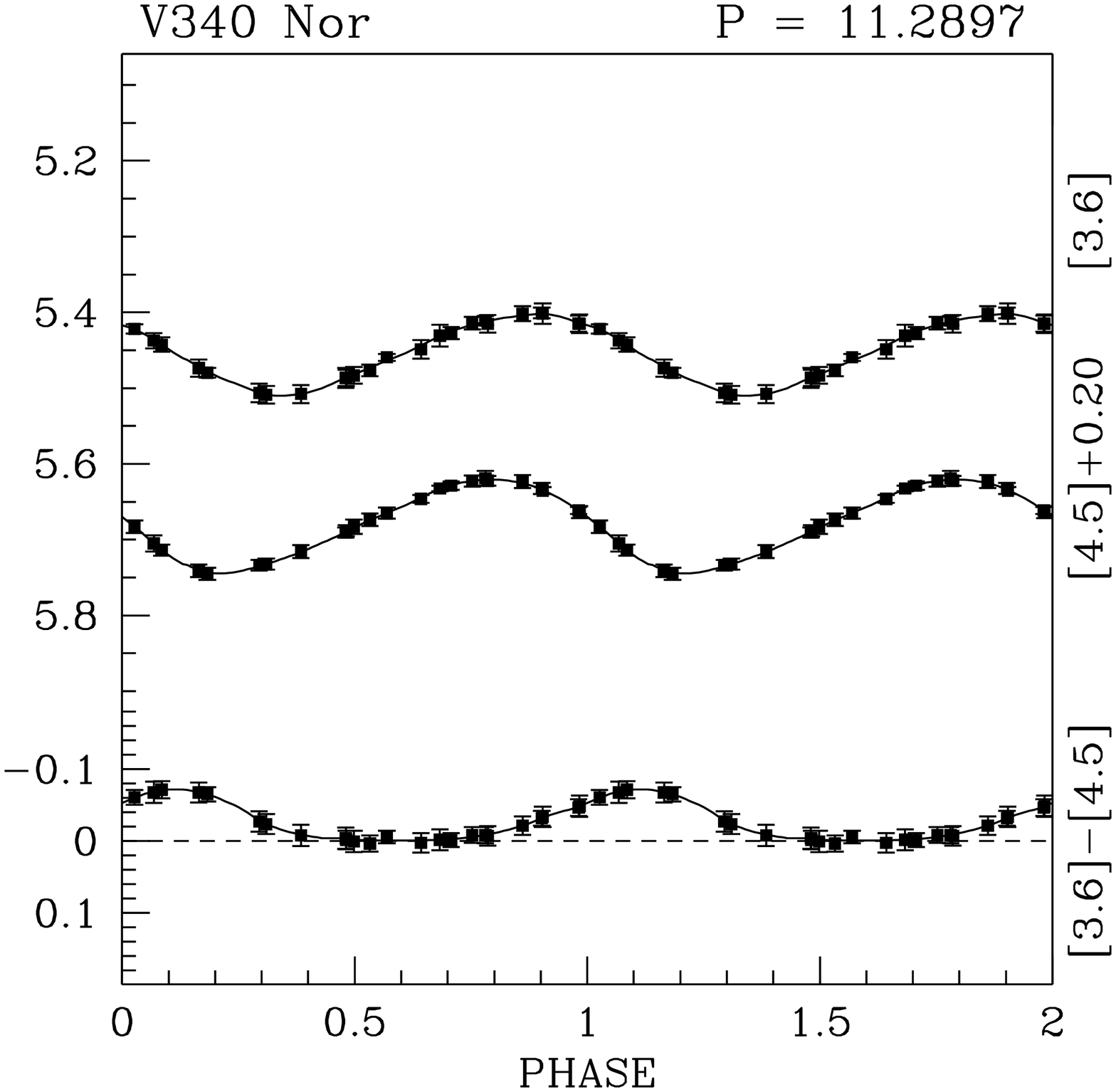}} & 
                \resizebox{50mm}{50mm}{\includegraphics{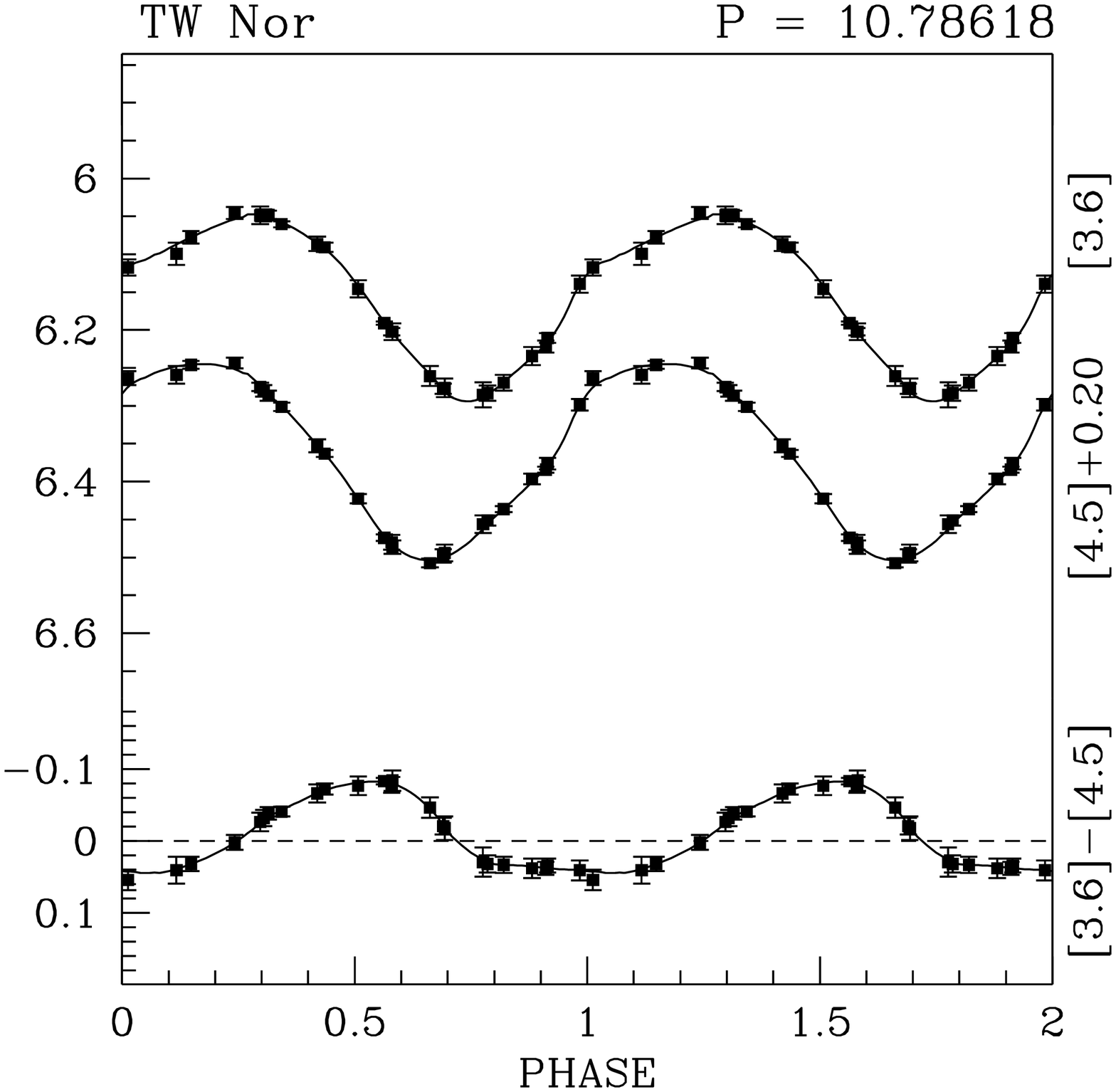}} \\ 
                \resizebox{50mm}{50mm}{\includegraphics{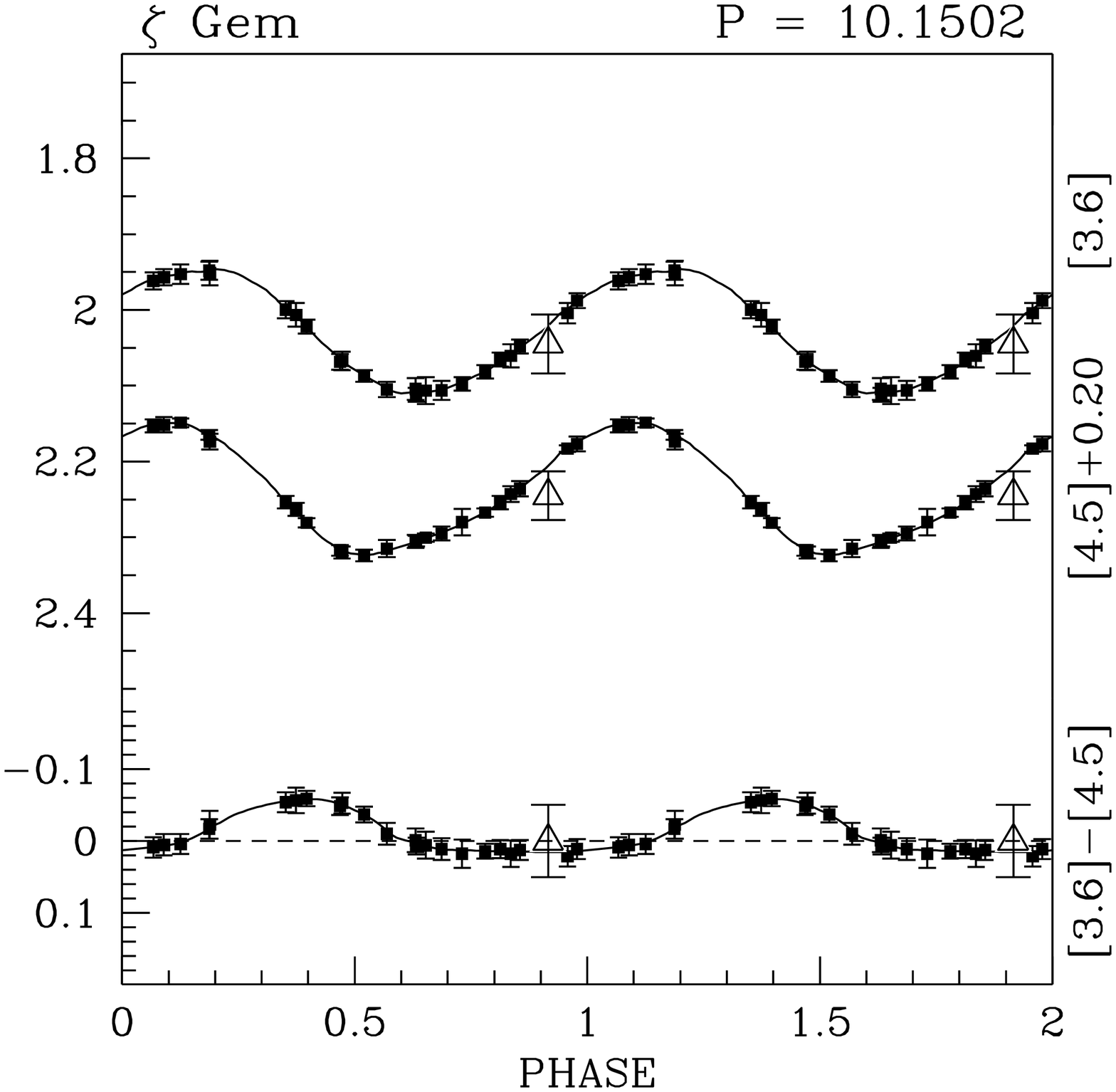}} & 
                \resizebox{50mm}{50mm}{\includegraphics{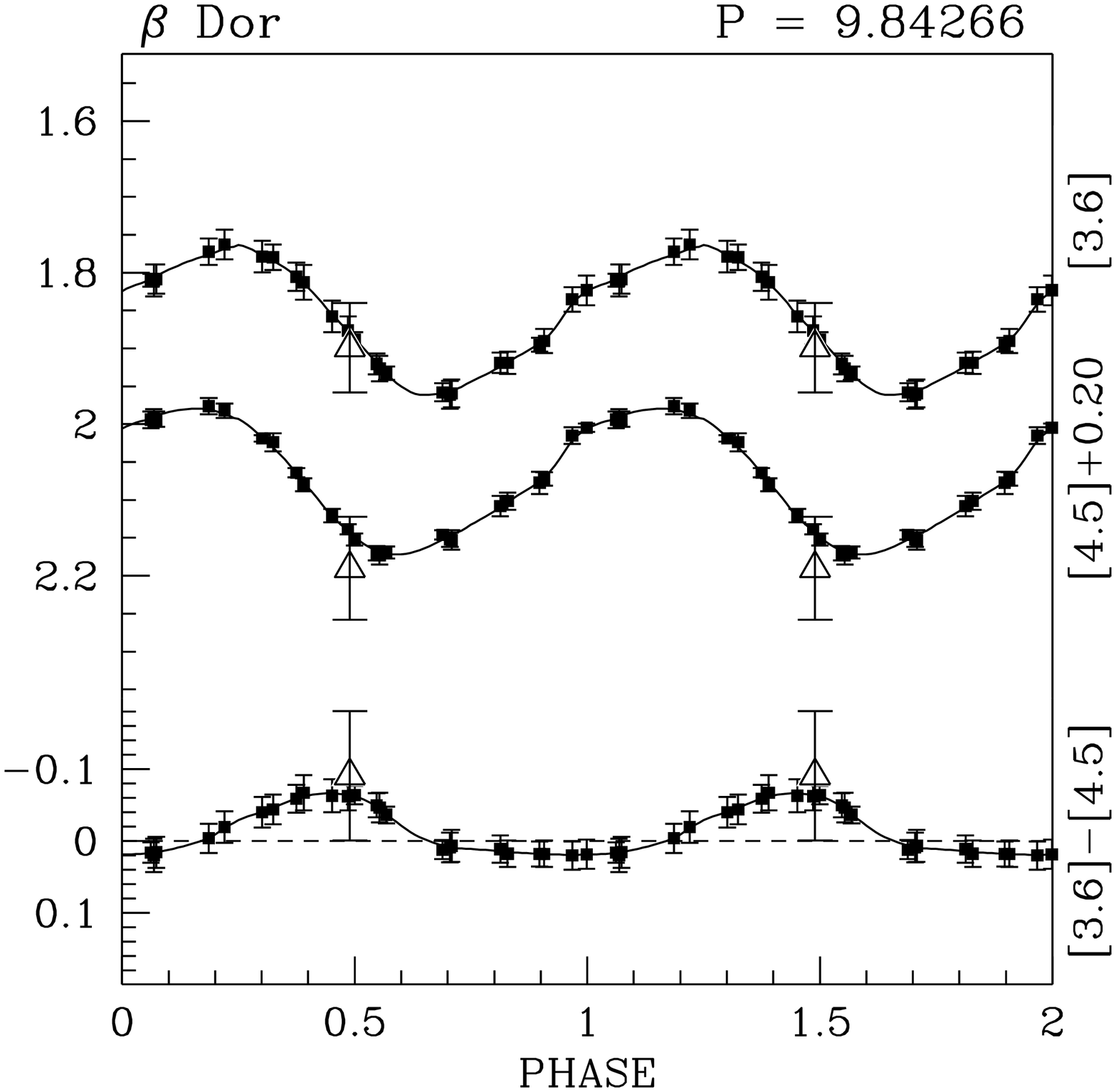}} & 
                \resizebox{50mm}{50mm}{\includegraphics{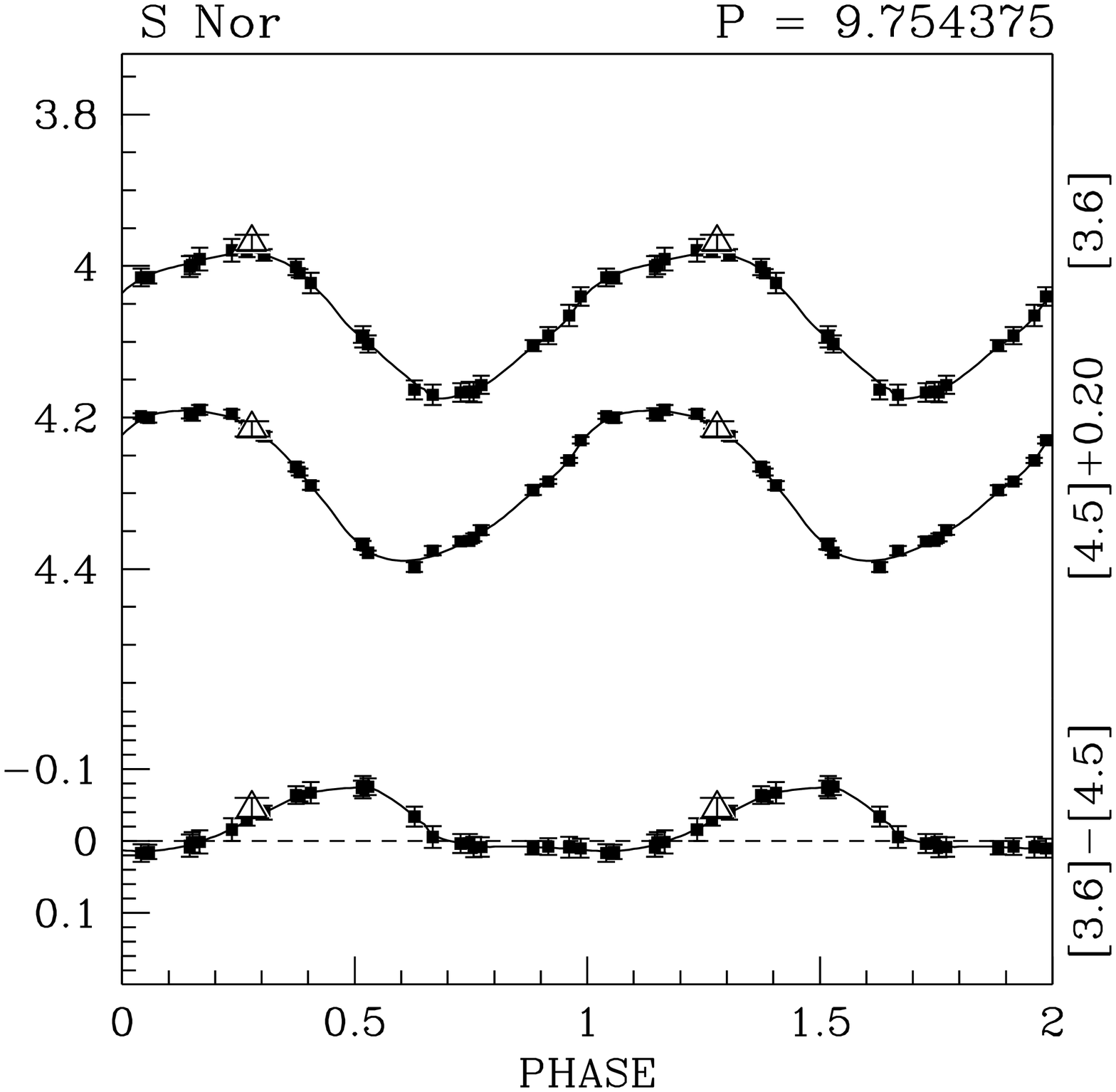}} \\ 
\end{tabular}
\caption{continued.}
\end{figure}

\addtocounter{figure}{-1}
\begin{figure}
\begin{tabular}{ccc}
                \resizebox{50mm}{50mm}{\includegraphics{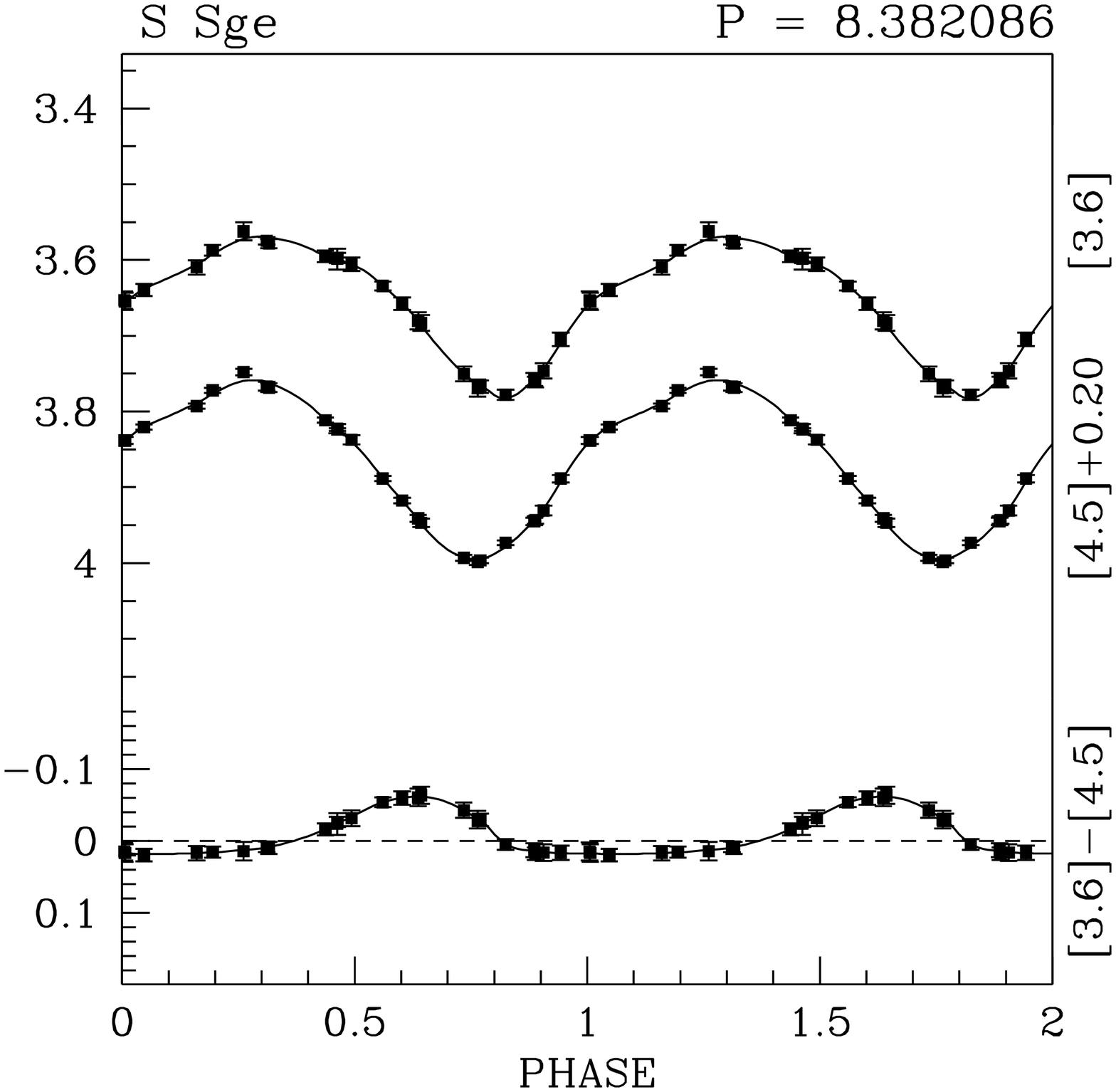}} & 
                \resizebox{50mm}{50mm}{\includegraphics{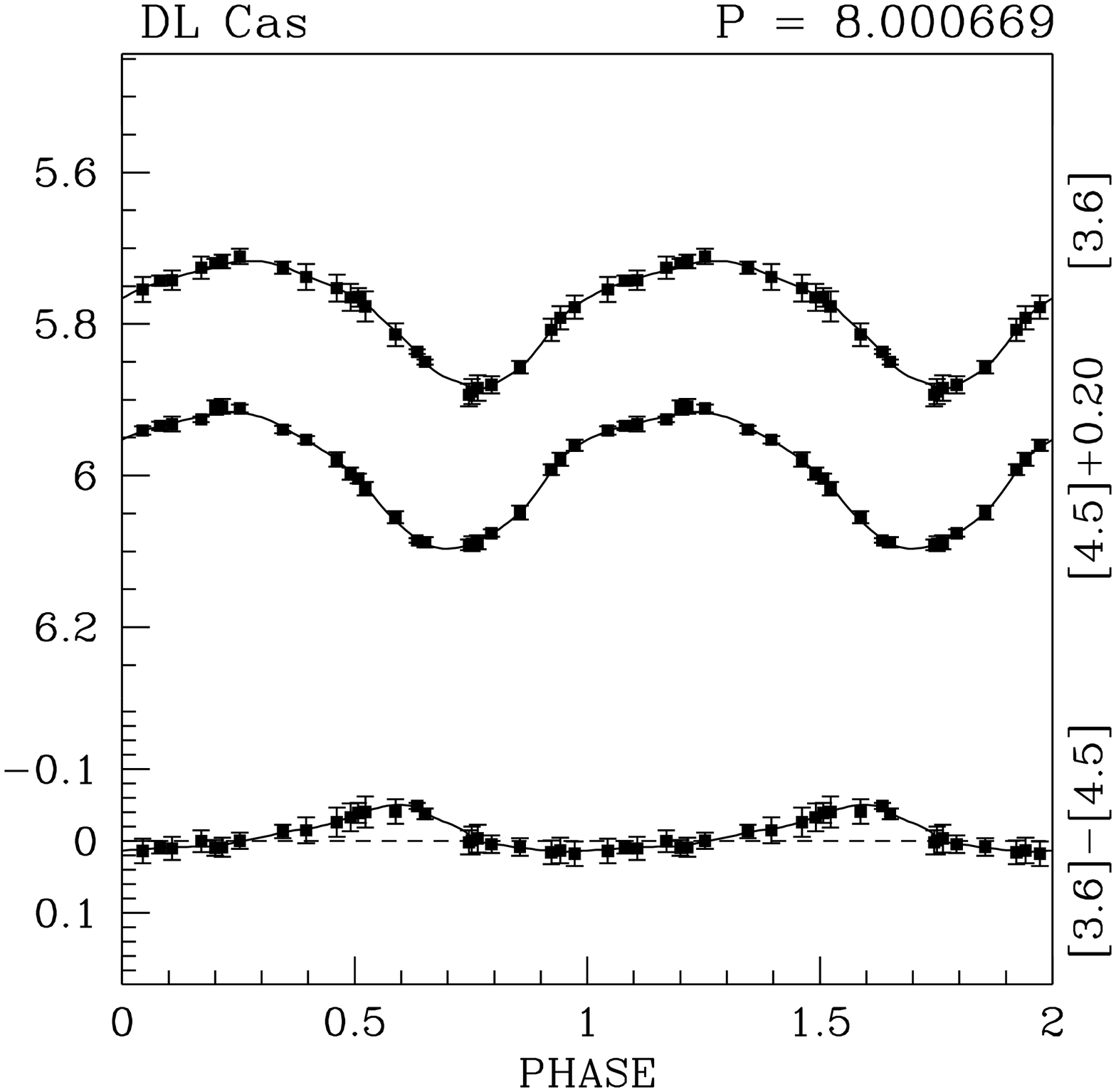}} & 
                \resizebox{50mm}{50mm}{\includegraphics{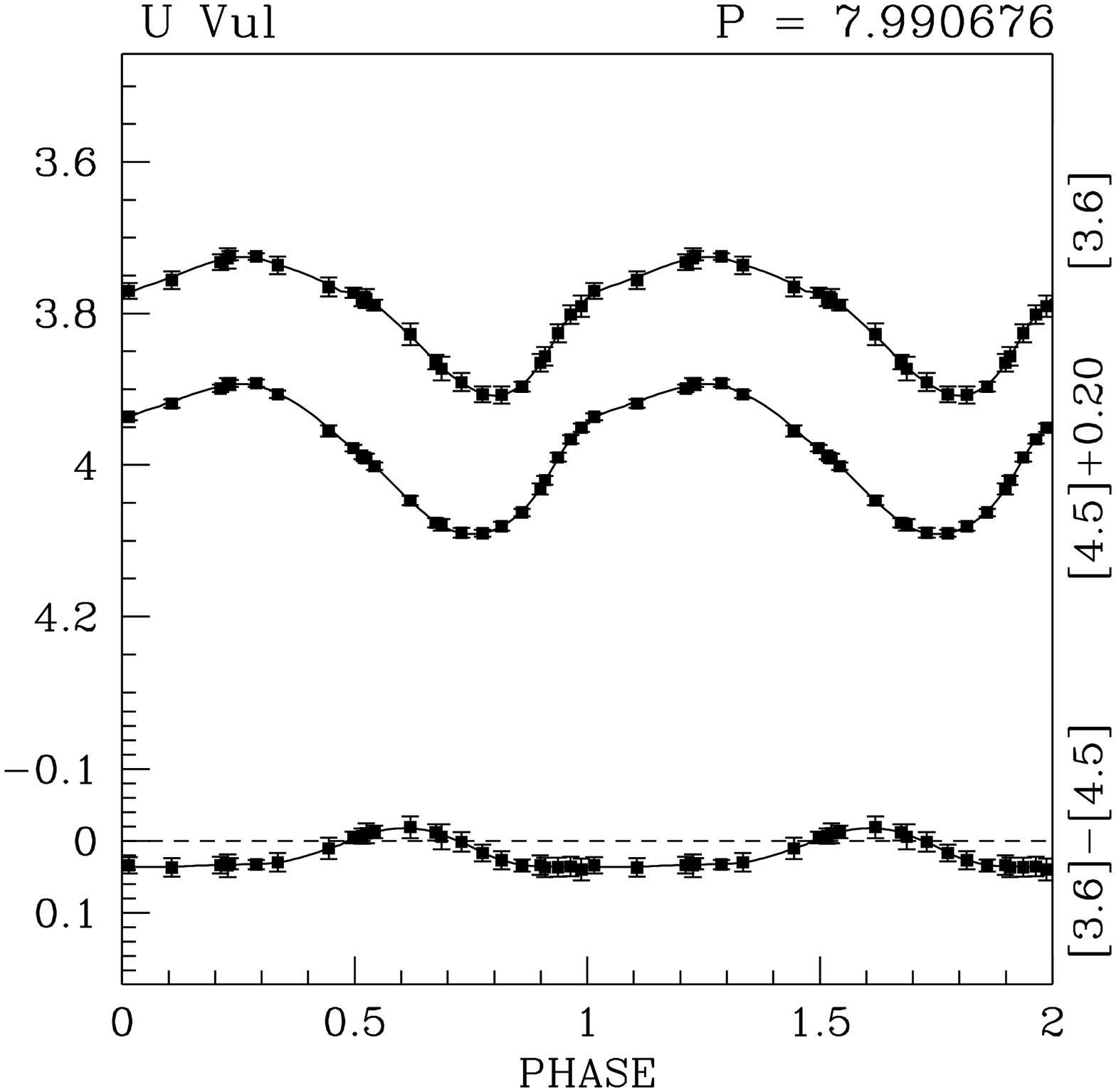}} \\ 
                \resizebox{50mm}{50mm}{\includegraphics{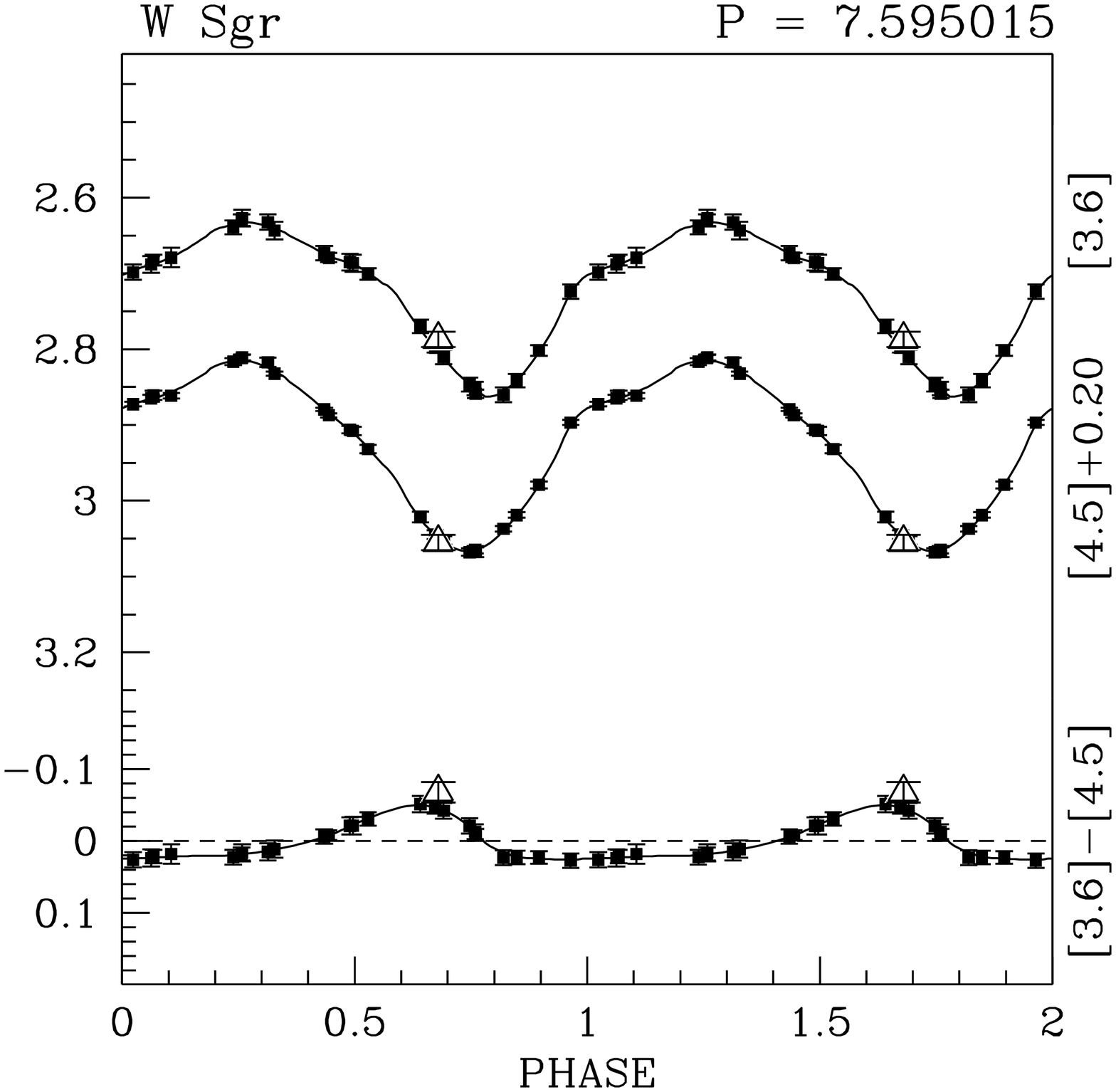}} & 
                \resizebox{50mm}{50mm}{\includegraphics{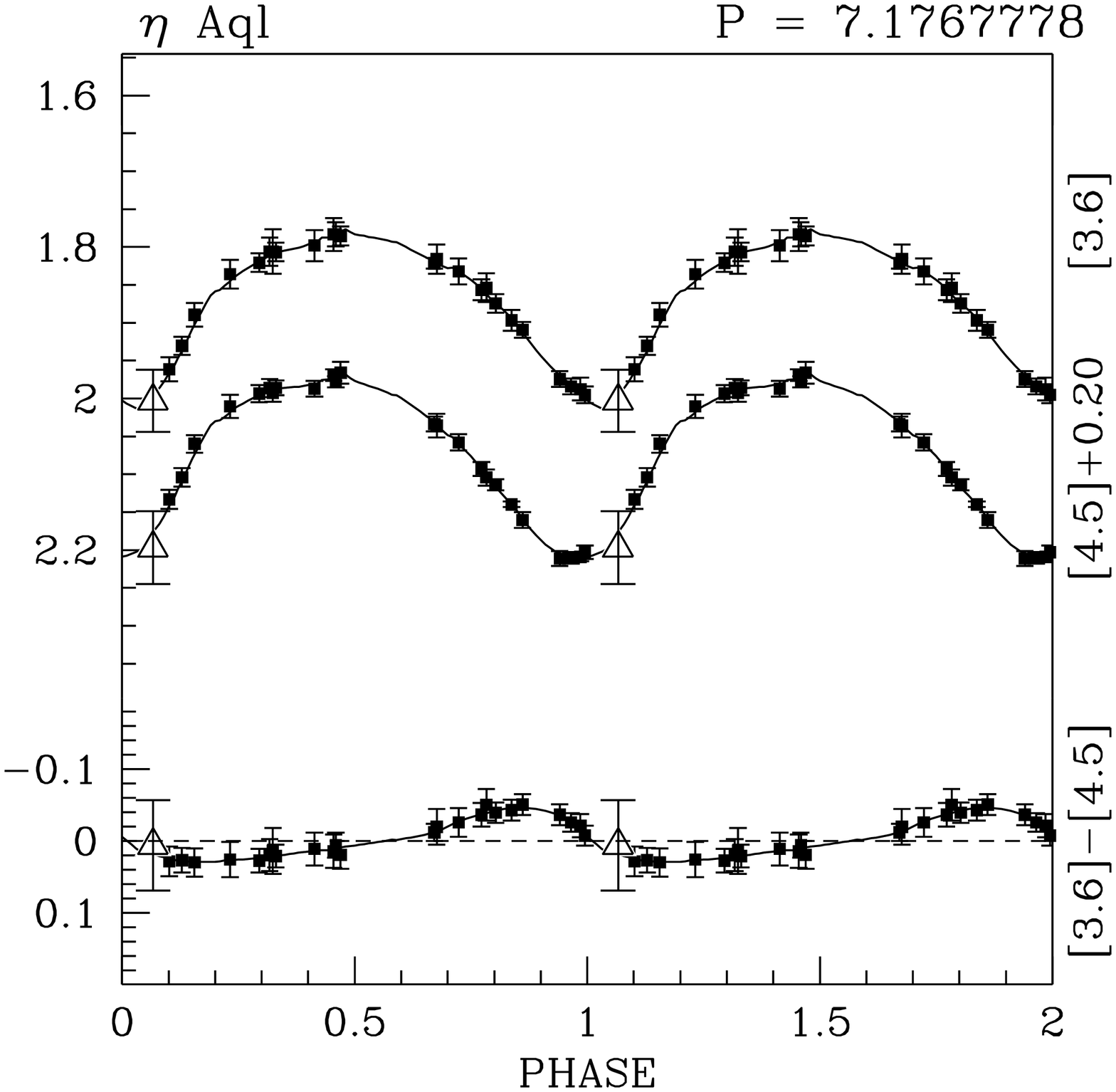}} & 
                \resizebox{50mm}{50mm}{\includegraphics{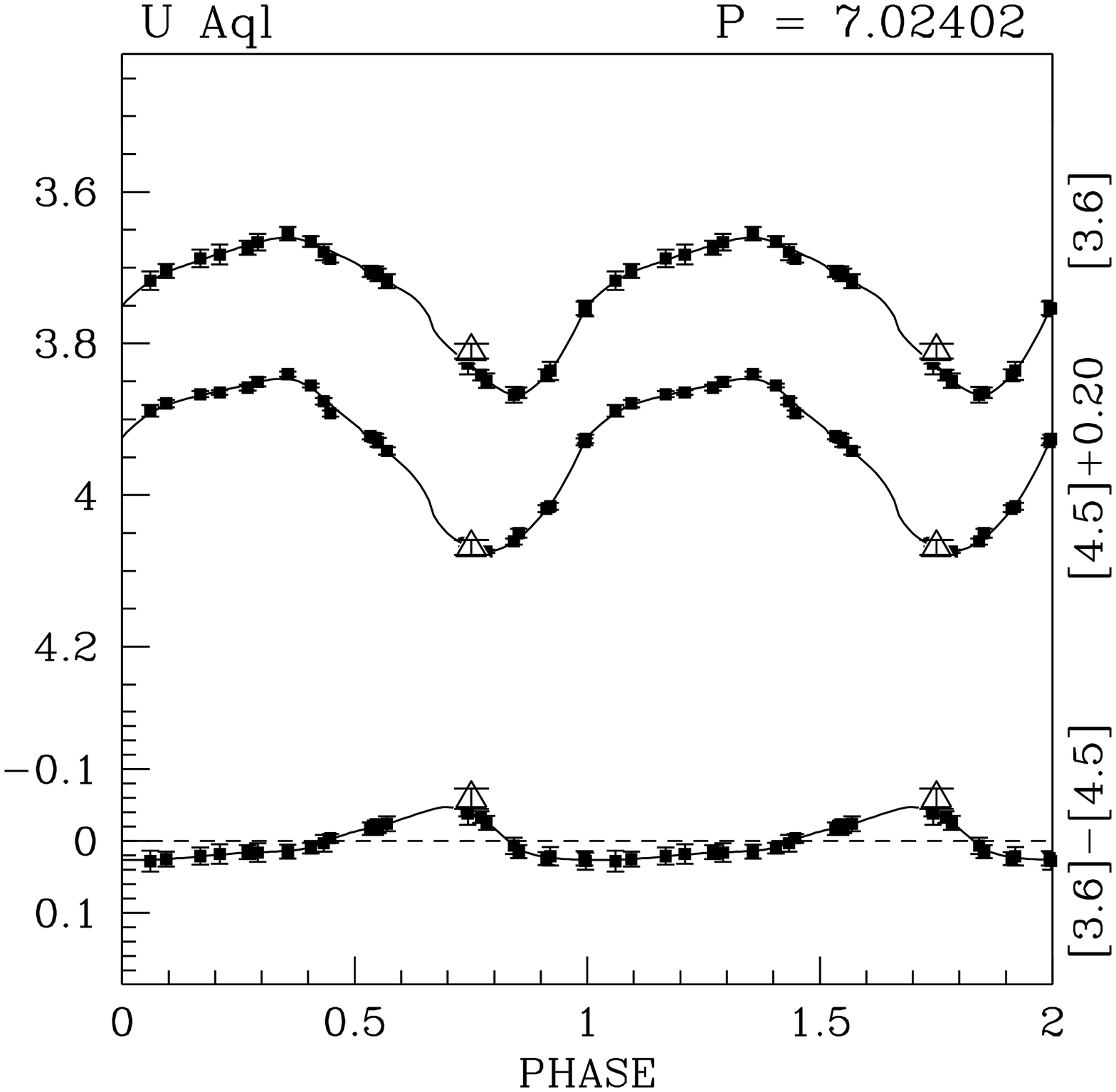}} \\ 
                \resizebox{50mm}{50mm}{\includegraphics{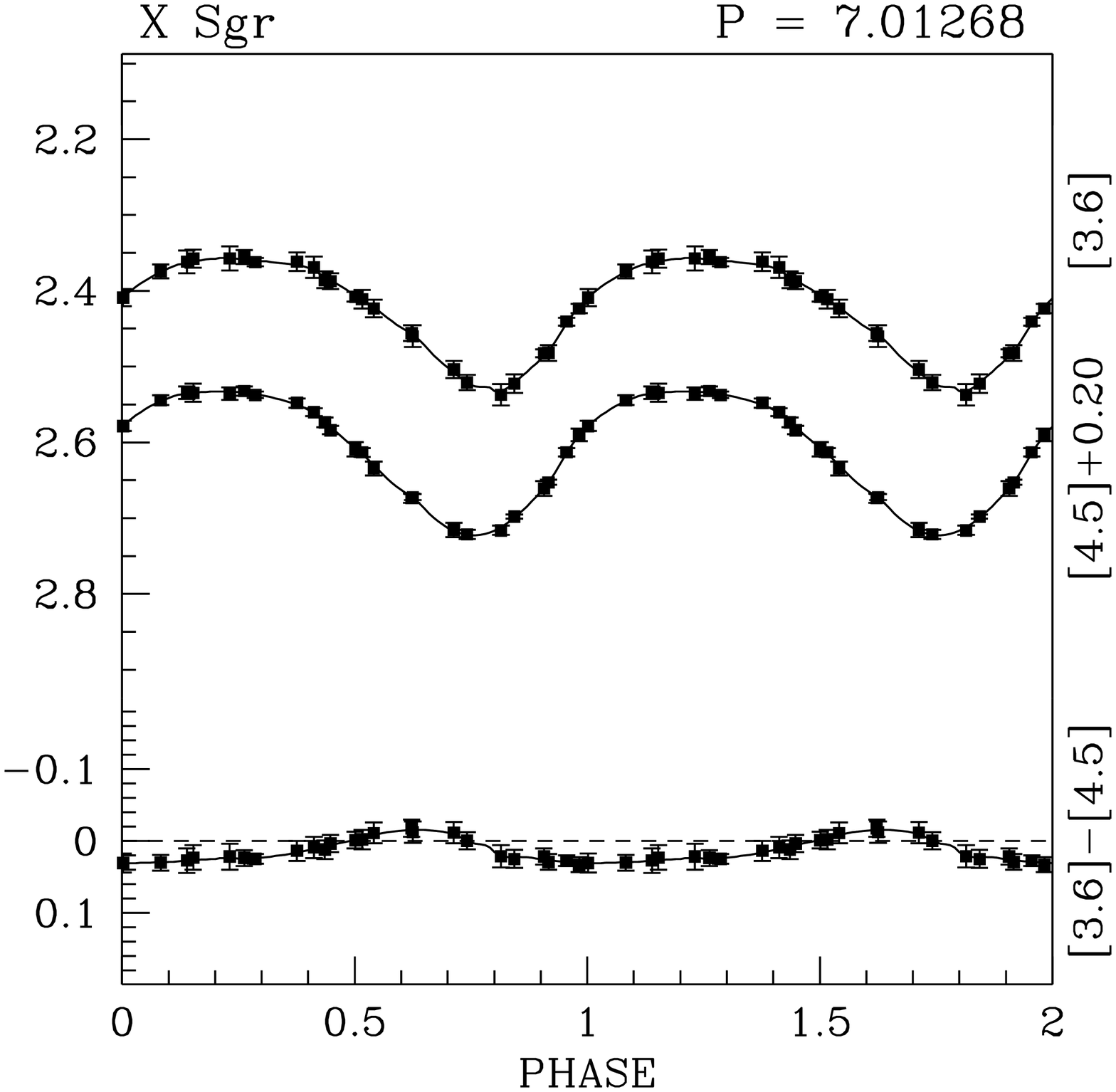}} & 
                \resizebox{50mm}{50mm}{\includegraphics{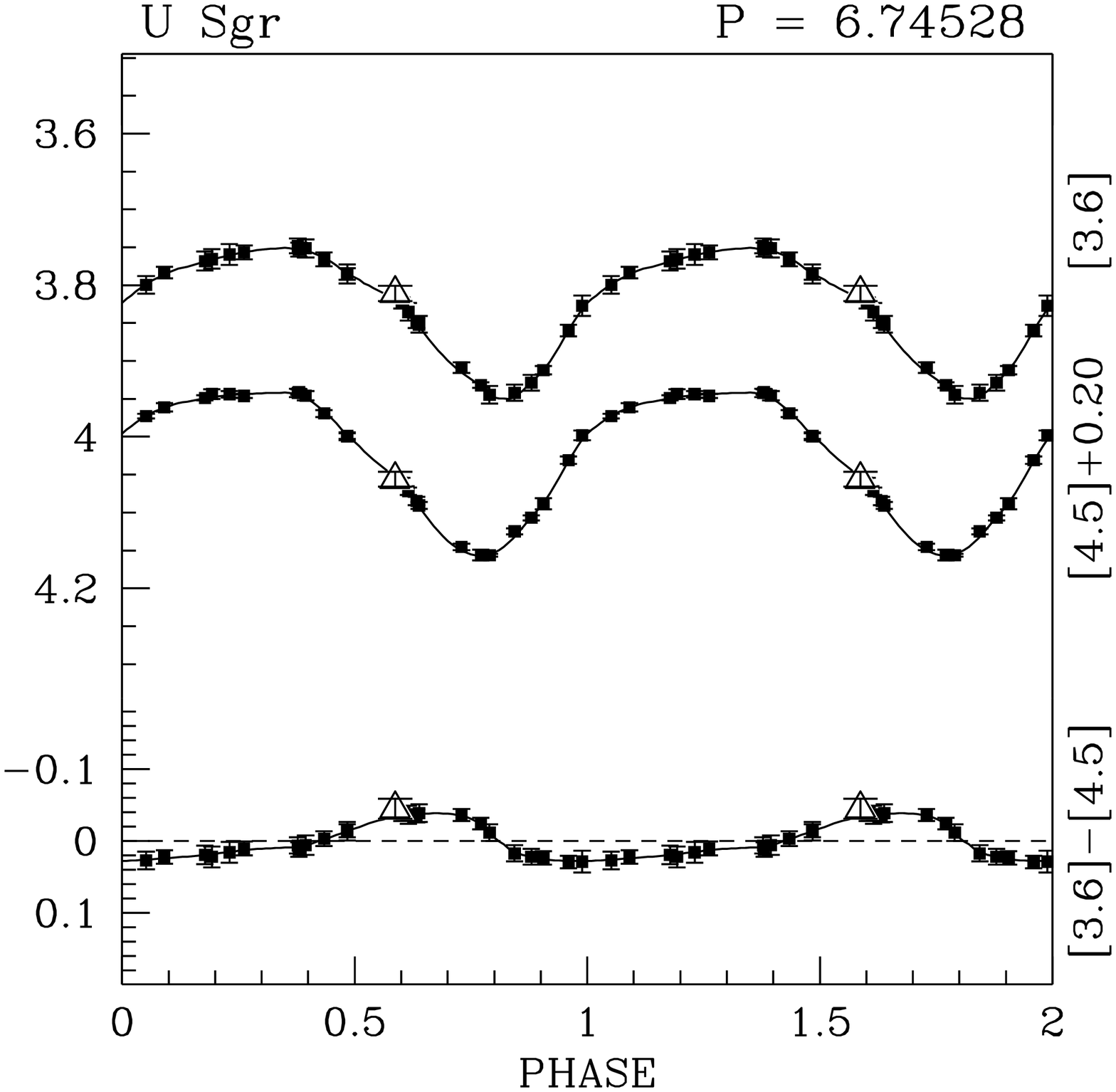}} & 
                \resizebox{50mm}{50mm}{\includegraphics{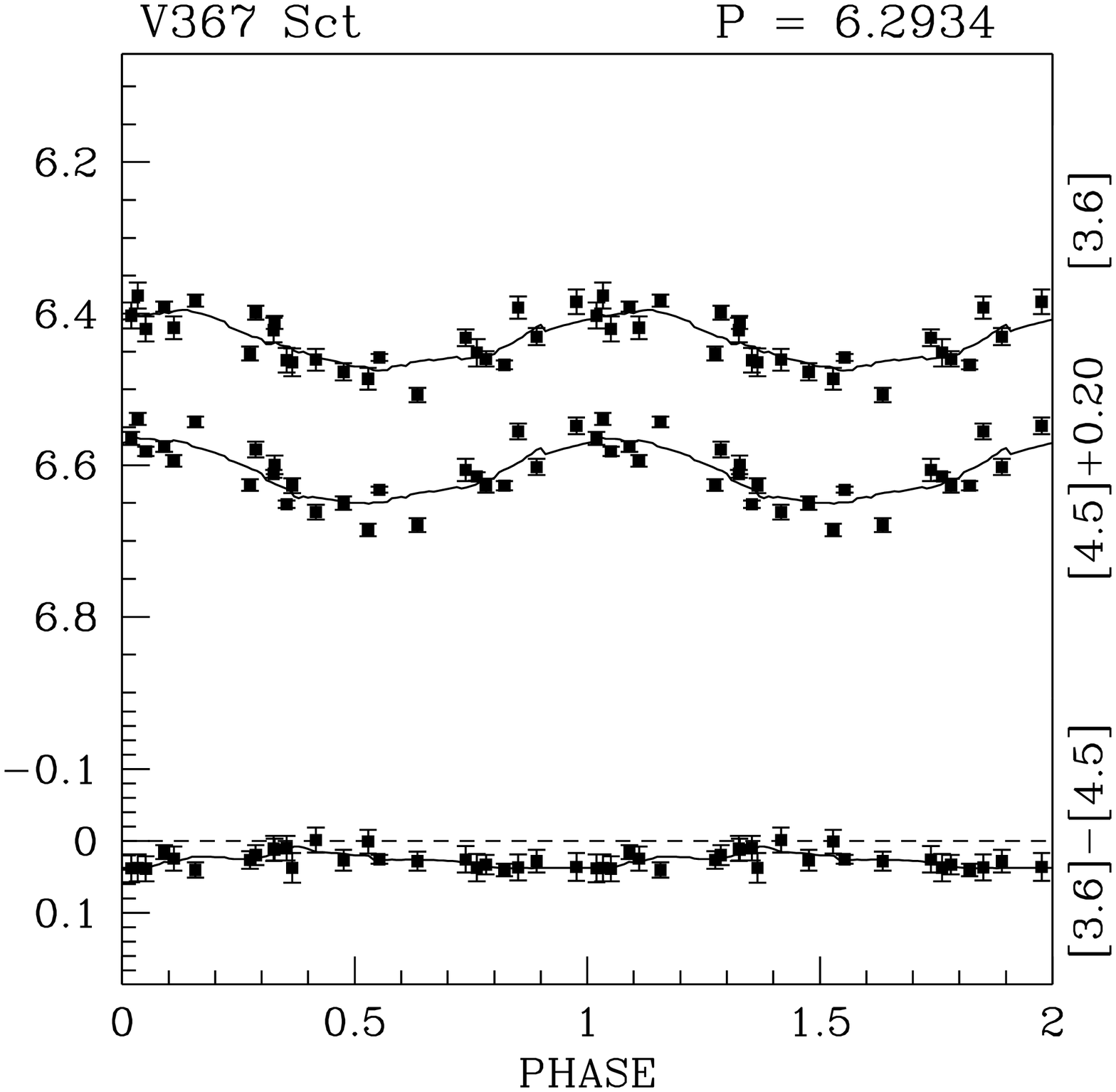}} \\ 
\end{tabular}
\caption{continued.}
\end{figure}

\addtocounter{figure}{-1}
\begin{figure}
\begin{tabular}{ccc}
                \resizebox{50mm}{50mm}{\includegraphics{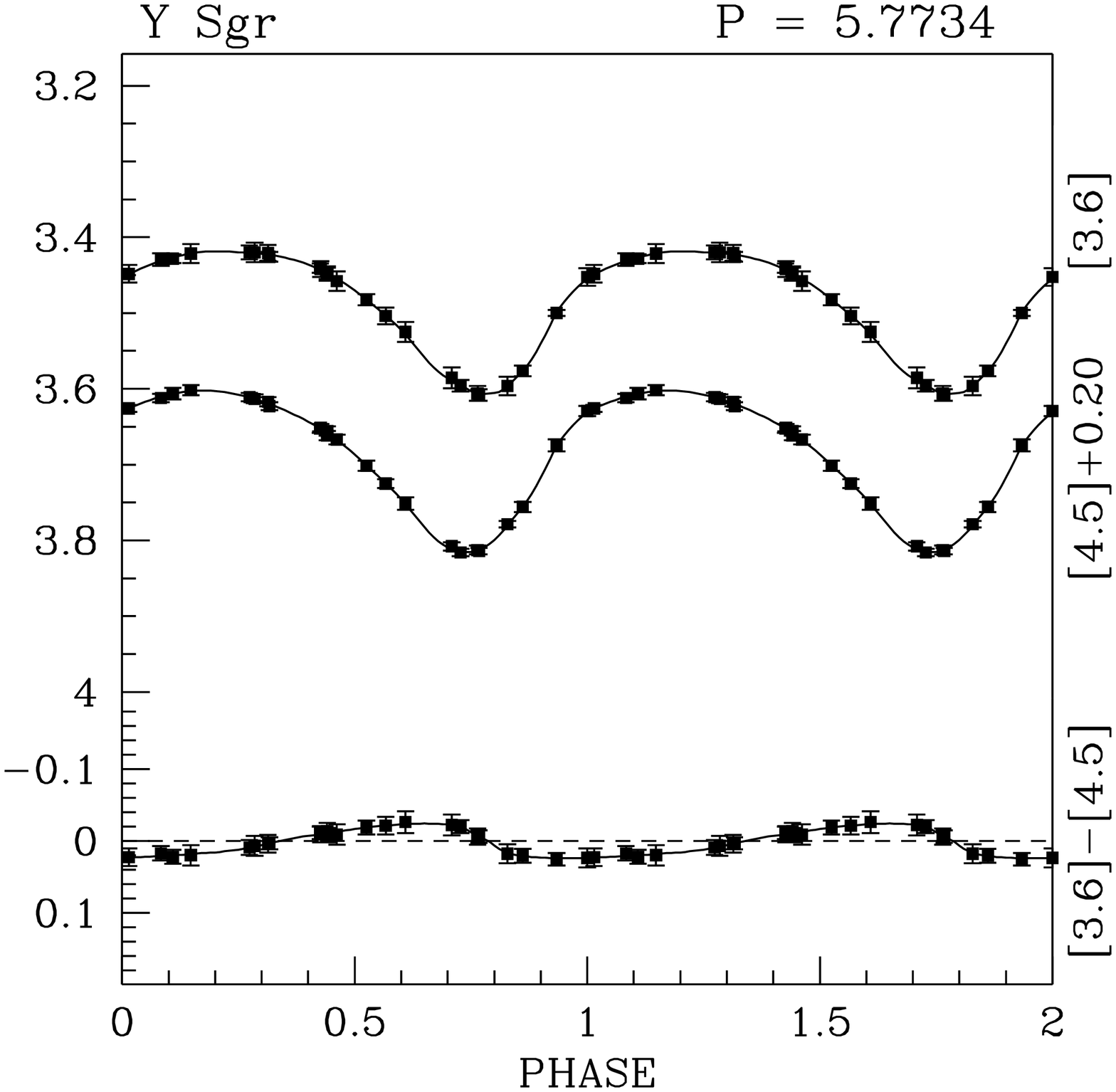}} & 
                \resizebox{50mm}{50mm}{\includegraphics{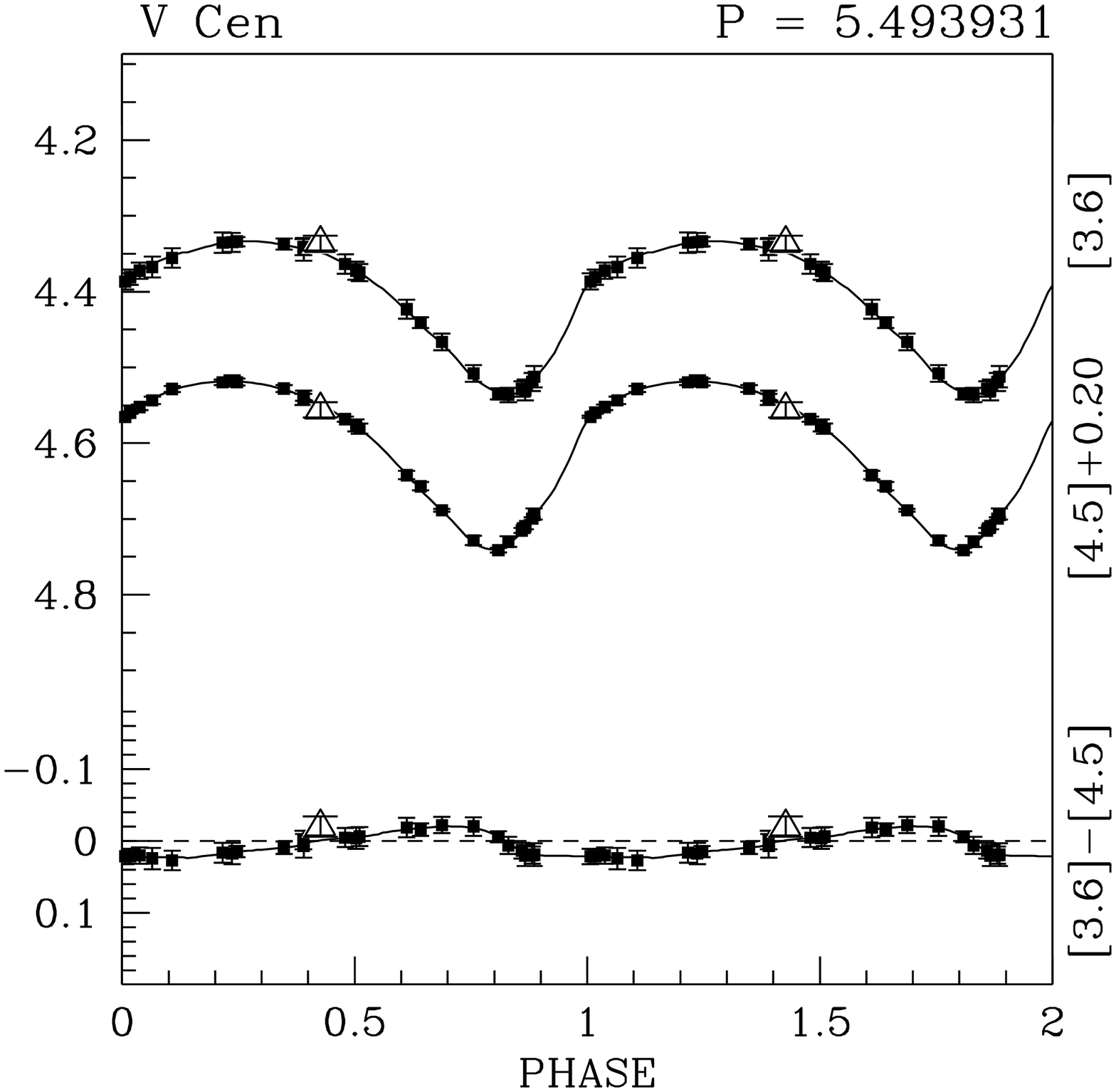}} & 
                \resizebox{50mm}{50mm}{\includegraphics{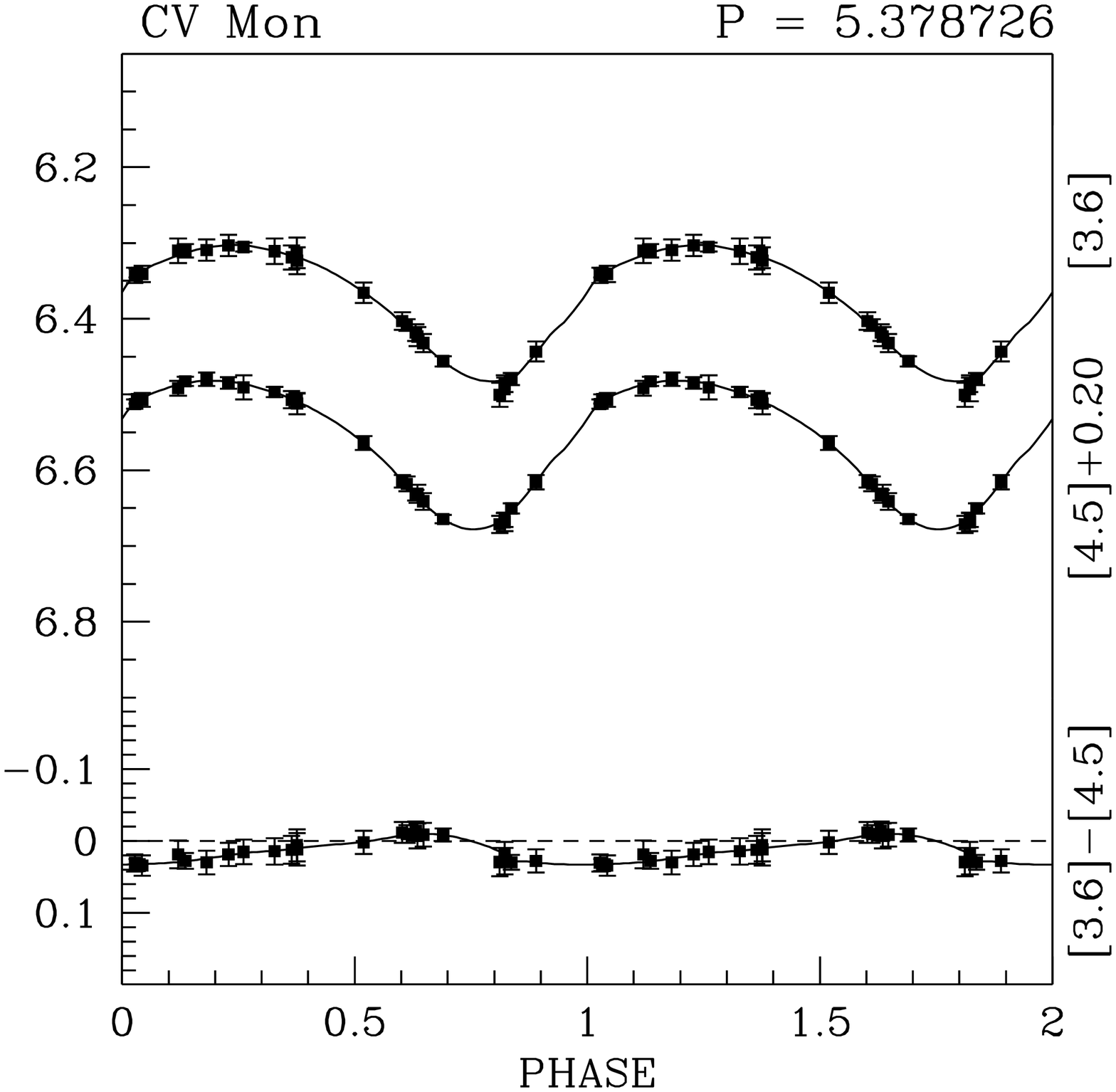}} \\ 
                \resizebox{50mm}{50mm}{\includegraphics{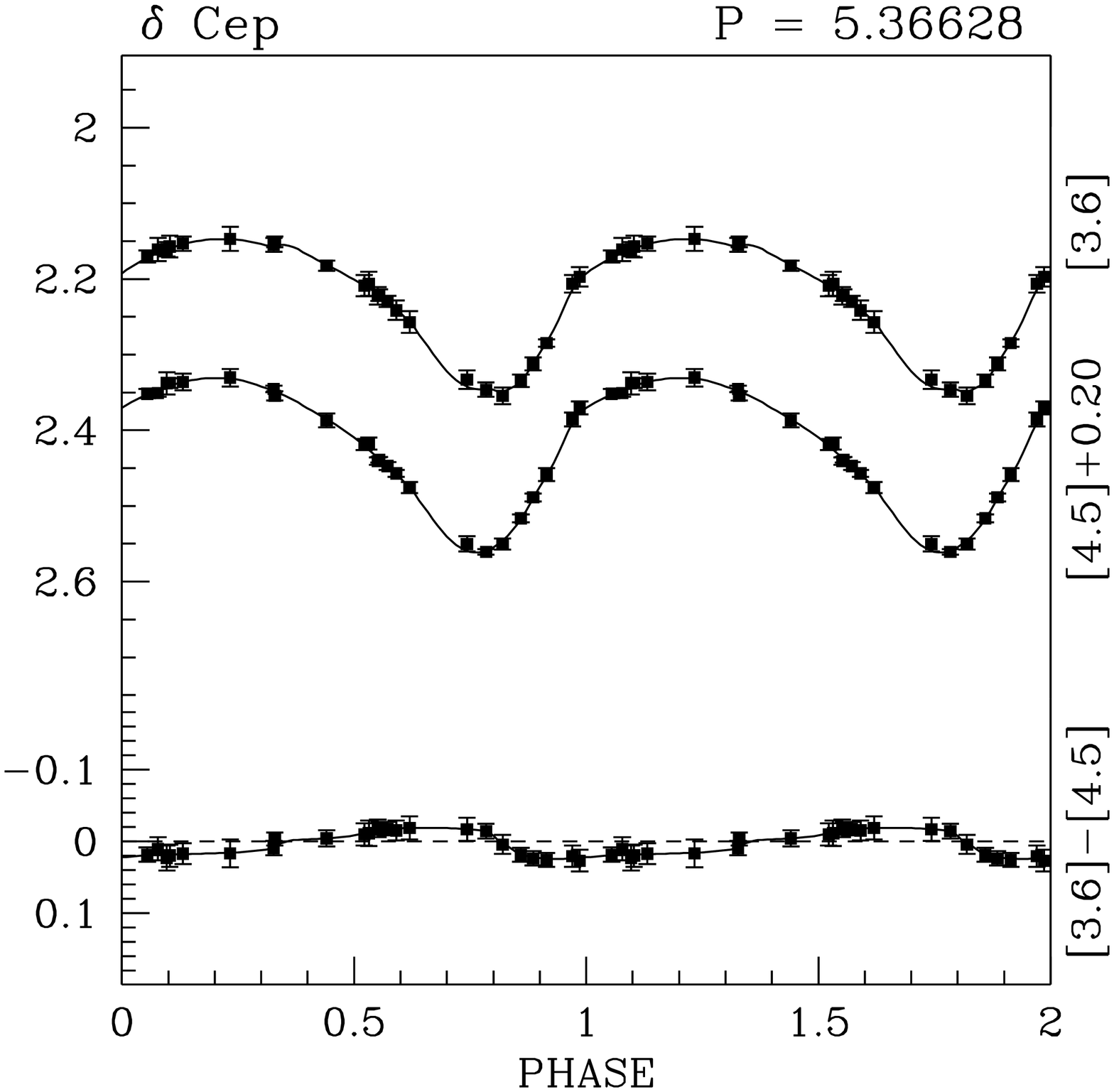}} & 
                \resizebox{50mm}{50mm}{\includegraphics{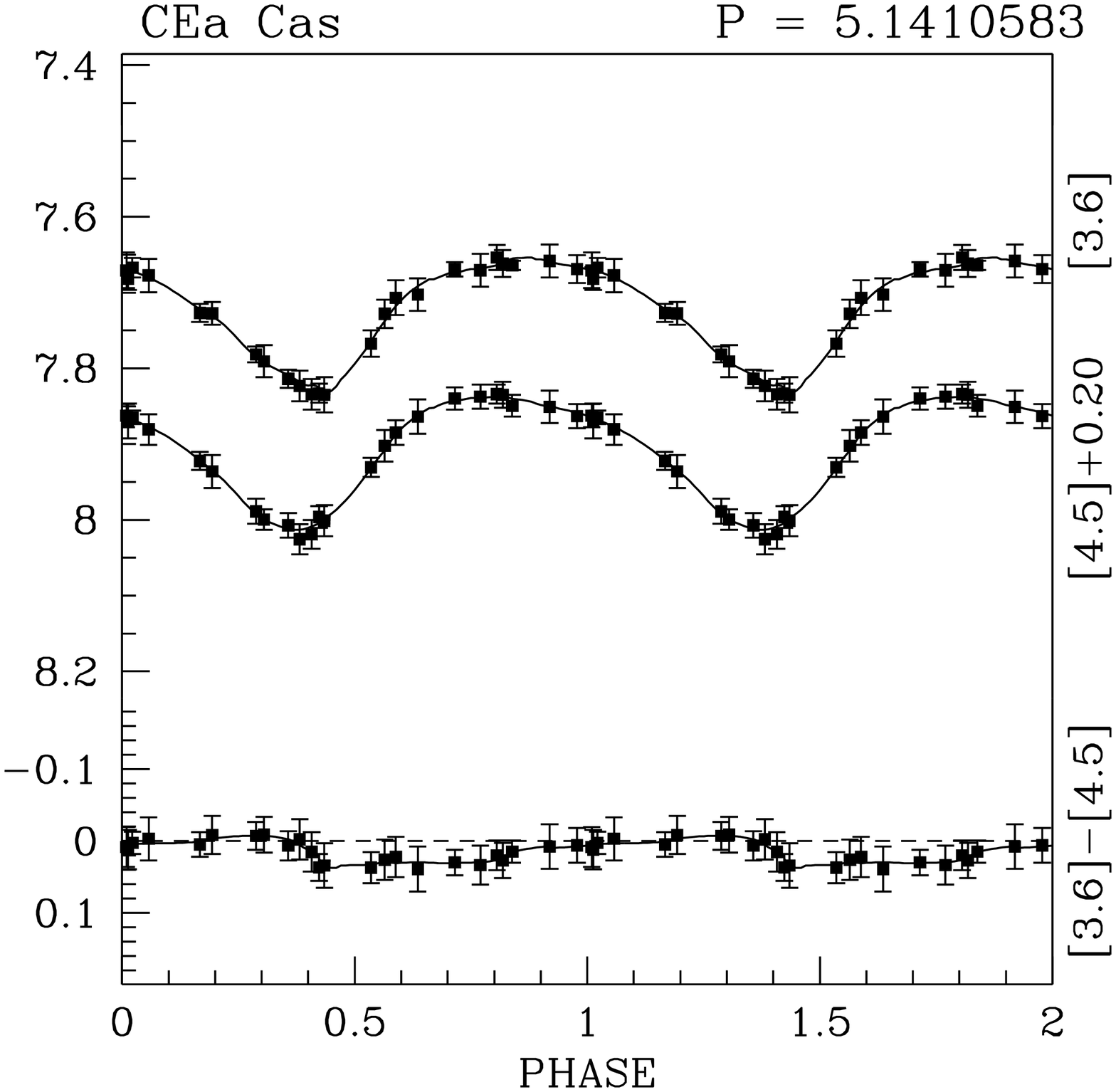}} & 
                \resizebox{50mm}{50mm}{\includegraphics{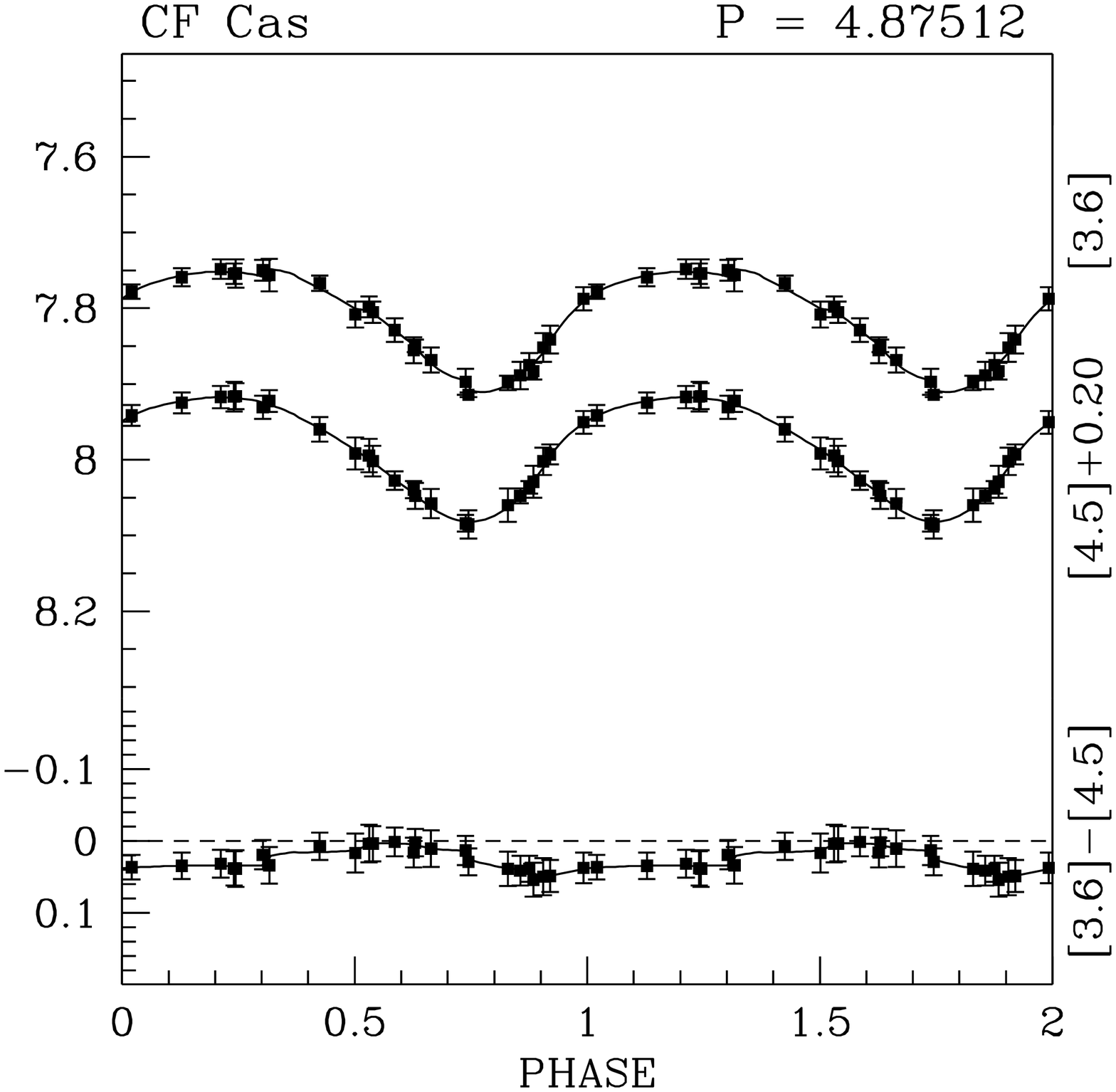}} \\
                \resizebox{50mm}{50mm}{\includegraphics{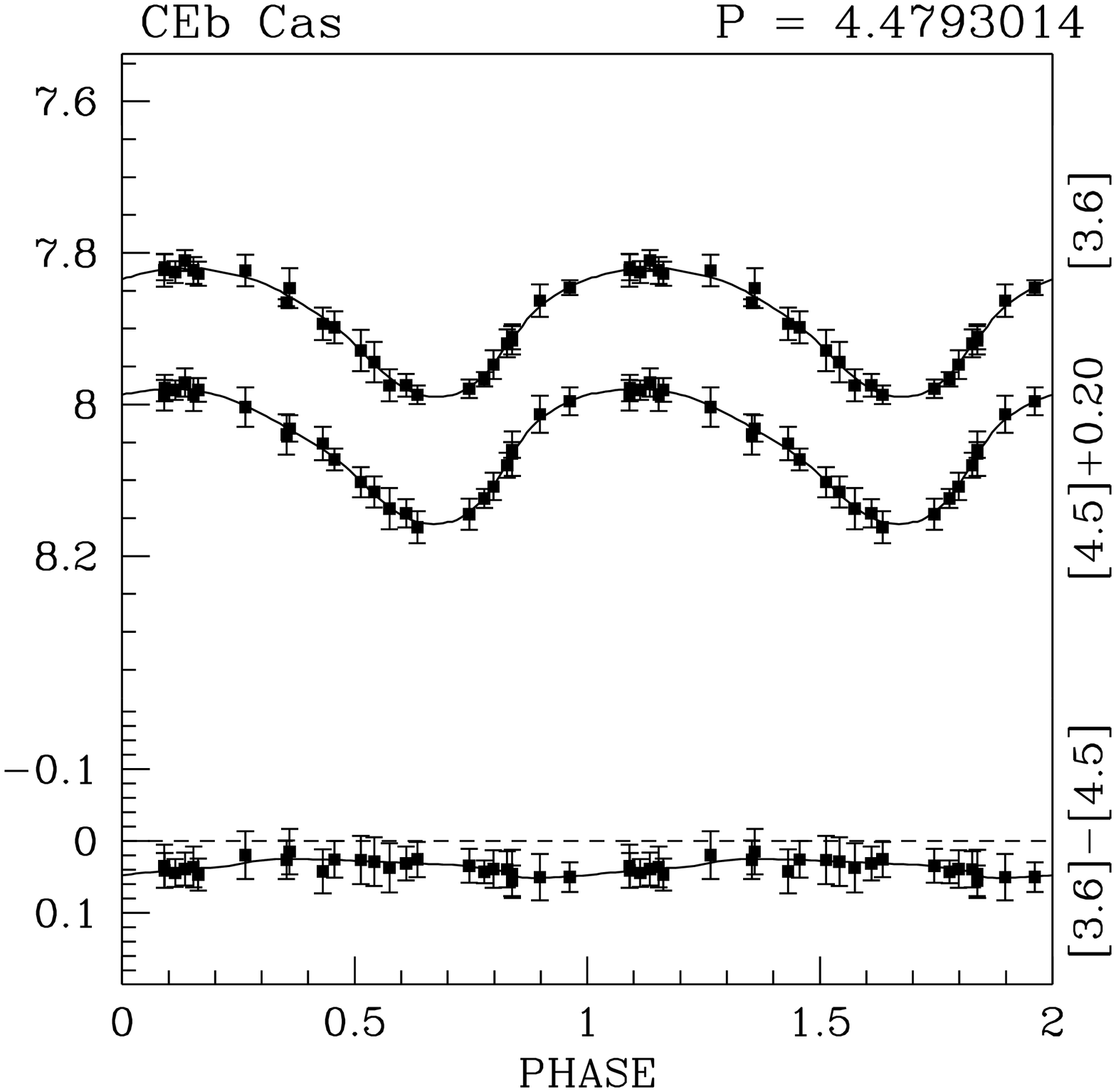}} & 
                \resizebox{50mm}{50mm}{\includegraphics{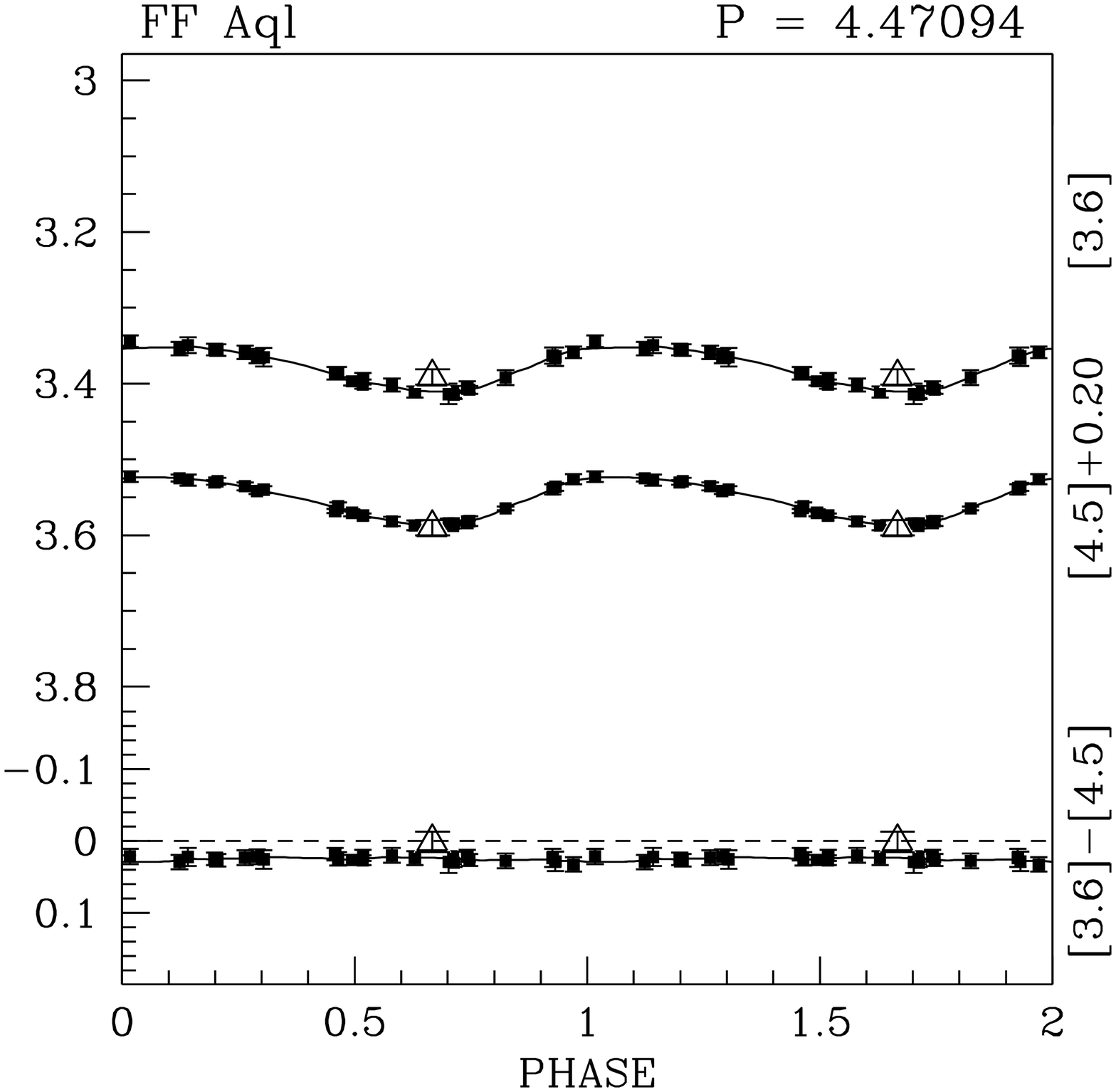}} &
                \resizebox{50mm}{50mm}{\includegraphics{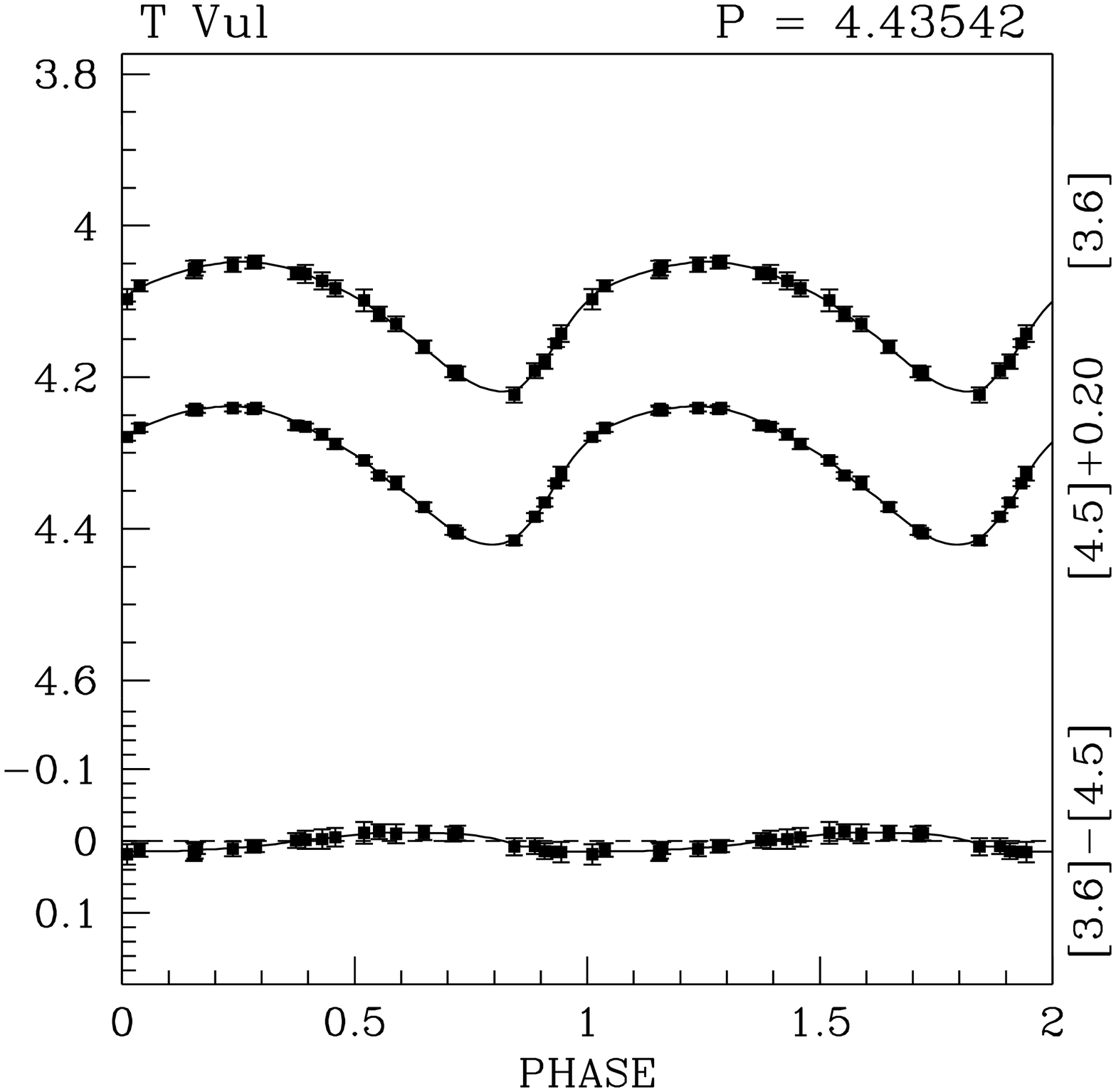}} \\ 

\end{tabular}
\caption{continued.}
\end{figure}

\addtocounter{figure}{-1}
\begin{figure}
\begin{tabular}{ccc}
                \resizebox{50mm}{50mm}{\includegraphics{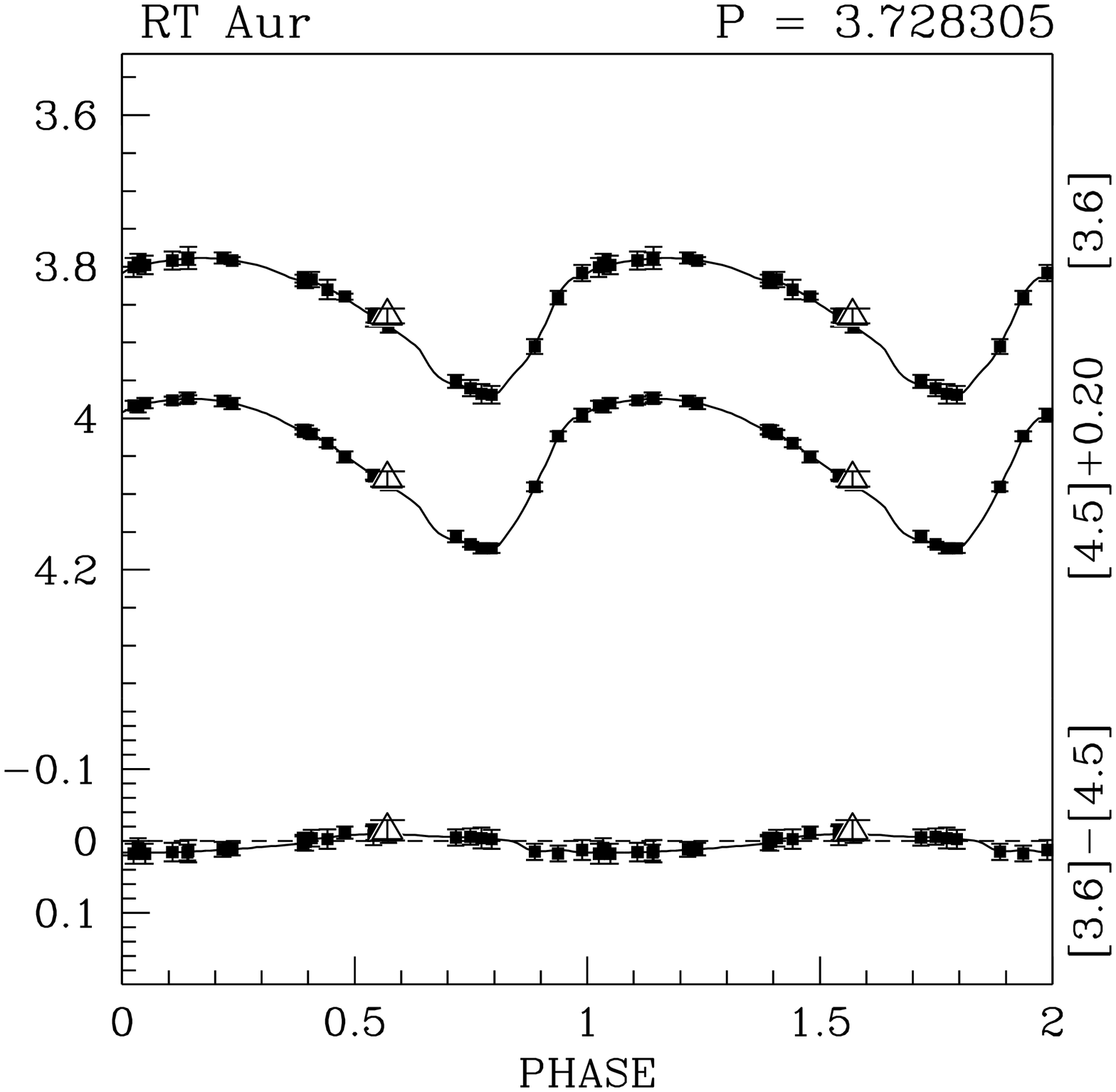}} \\ 
\end{tabular}
\caption{continued.}
\end{figure}


\clearpage

\begin{figure}
\begin{centering}
\resizebox{6in}{6in}{\includegraphics{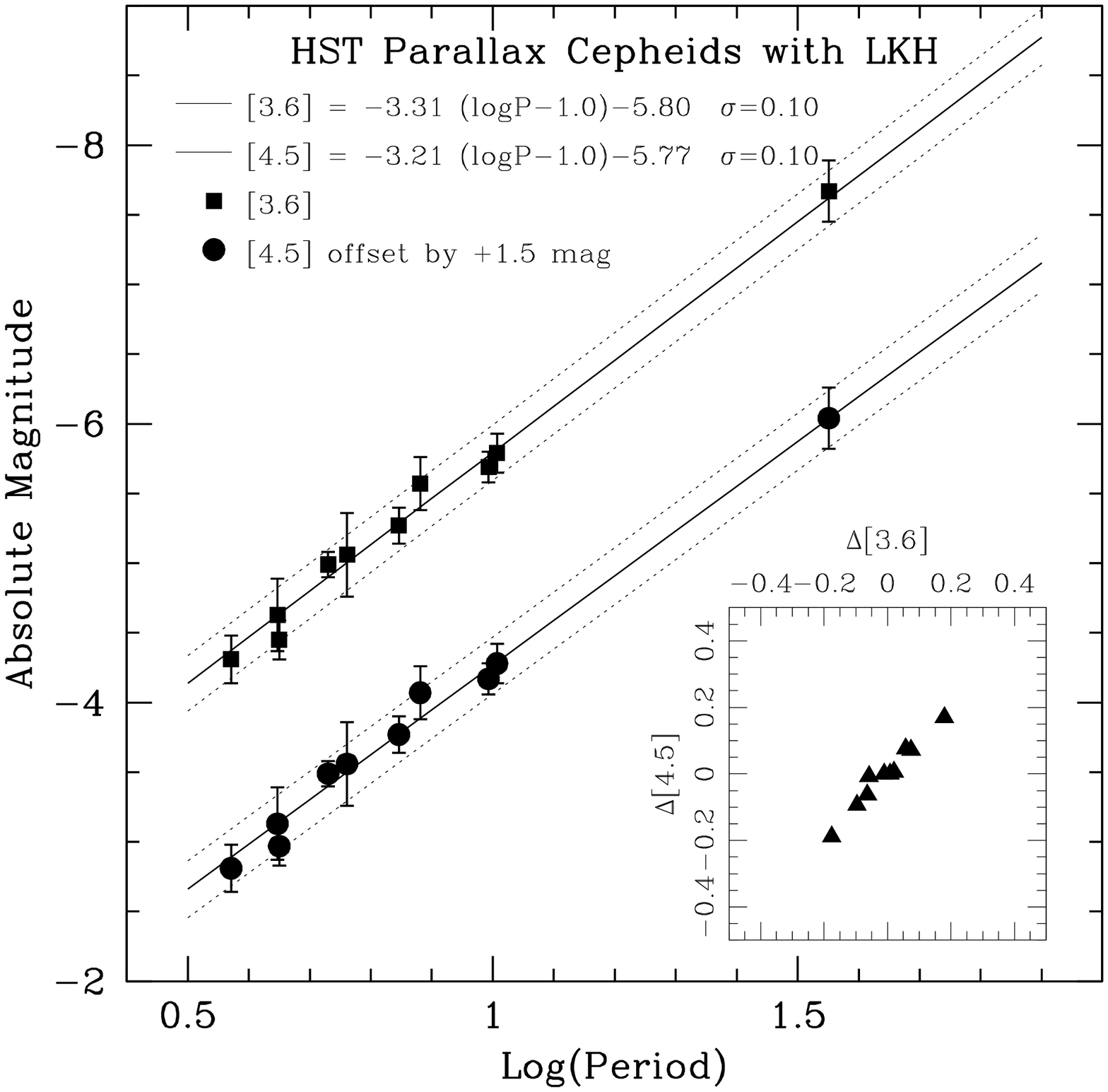}}
\caption{Leavitt PL Relations for the HST Calibrators in the Galaxy at 3.6 (upper plot) and 4.5 $\mu$m (lower plot).  The data have been corrected for Lutz-Kelker-Hanson bias. The relations are shown as solid lines which were determined using a fixed slope found from the LMC data and a zero-point found from the HST parallaxes \citep{Benedict:2007}; the $\pm2\sigma$ boundaries are shown as dotted lines.   The highly correlated magnitude residuals are plotted in the inset, showing that the peak-to-peak width of the IS, as defined by the HST Galactic Calibrators is less than 0.4 mag.  }
\label{plhstw}
\end{centering}
\end{figure}


\clearpage

\begin{figure}
\begin{centering}
\resizebox{6in}{6in}{\includegraphics{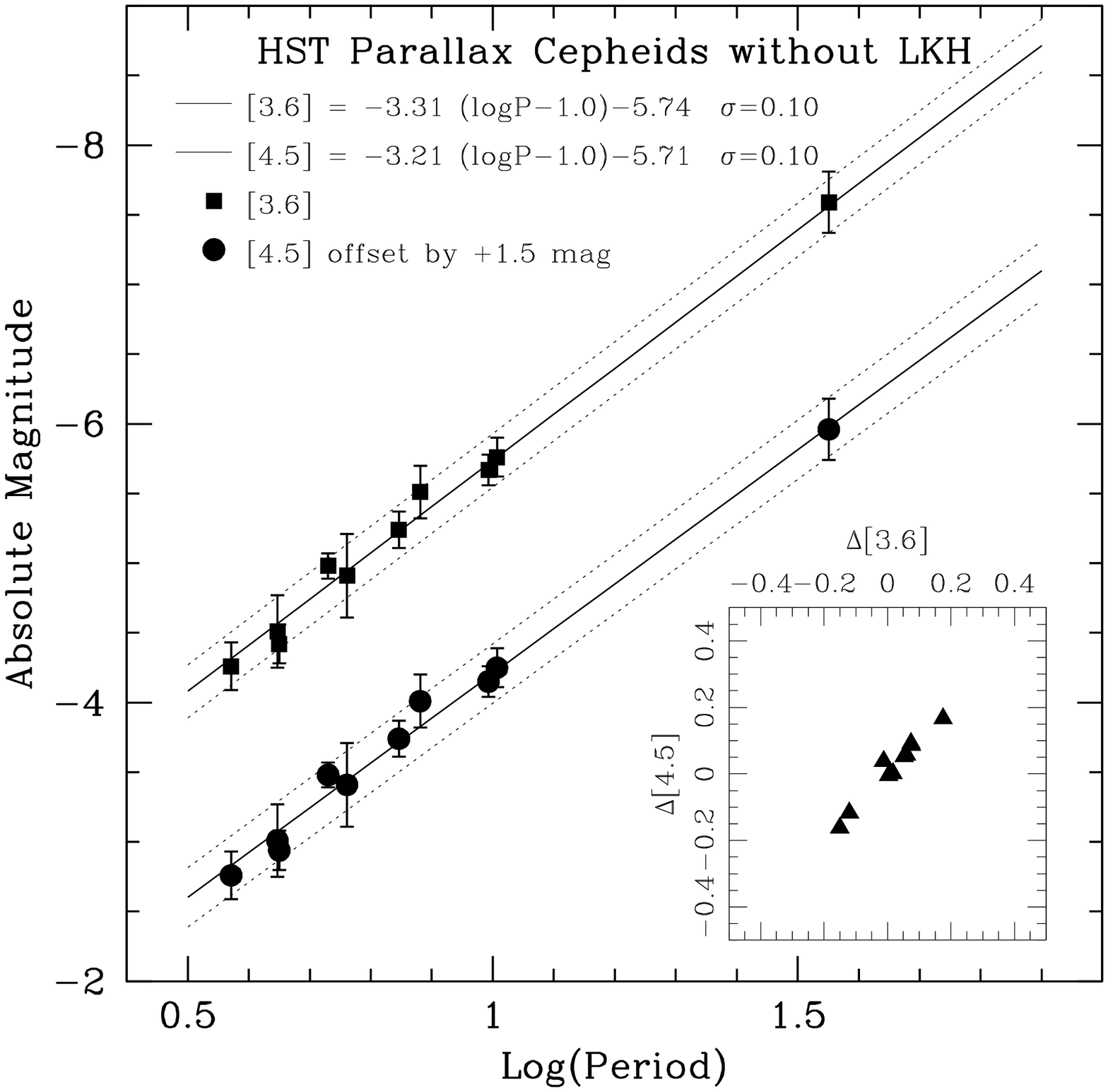}}
\caption{Leavitt PL Relations for the HST Calibrators in the Galaxy at 3.6 (upper plot) and 4.5 $\mu$m (lower plot).  The data have \emph{not} been corrected for Lutz-Kelker-Hanson bias.  The relations are shown as solid lines which were determined using a fixed slope found from the LMC data and a zero-point found from the HST parallaxes \citep{Benedict:2007}; the $\pm2\sigma$ boundaries are shown as dotted lines.   The highly correlated magnitude residuals are plotted in the inset, showing that the peak-to-peak width of the IS, as defined by the HST Galactic Calibrators is less than 0.4 mag.  }
\label{plhstwo}
\end{centering}
\end{figure}

\clearpage

\begin{figure}
\begin{centering}
\resizebox{6in}{6in}{\includegraphics{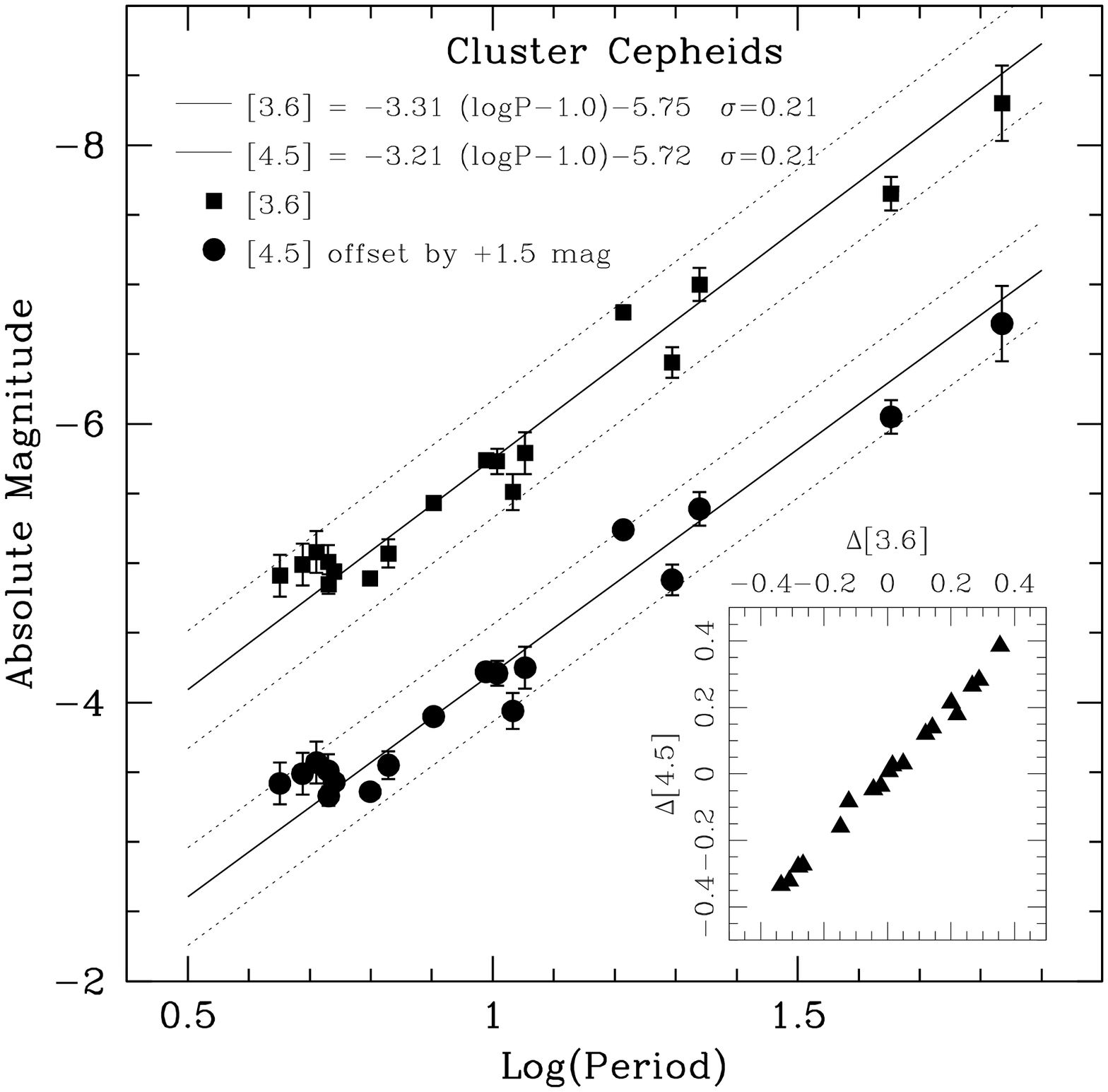}}
\caption{Leavitt PL Relations for the Cluster Cepheids in the Galaxy at 3.6 (upper plot) and 4.5 $\mu$m (lower plot).   The relations are shown as solid lines which were determined using a fixed slope found from the LMC data.   The $\pm2\sigma$ boundaries are shown as dotted lines.   The highly correlated magnitude residuals are plotted in the inset, showing that the peak-to-peak width of the IS, as defined by the Cluster Galactic Calibrators is less than 0.8 mag.       }
\label{plclus}
\end{centering}
\end{figure}

\clearpage

\begin{figure}
\begin{centering}
\resizebox{6in}{6in}{\includegraphics{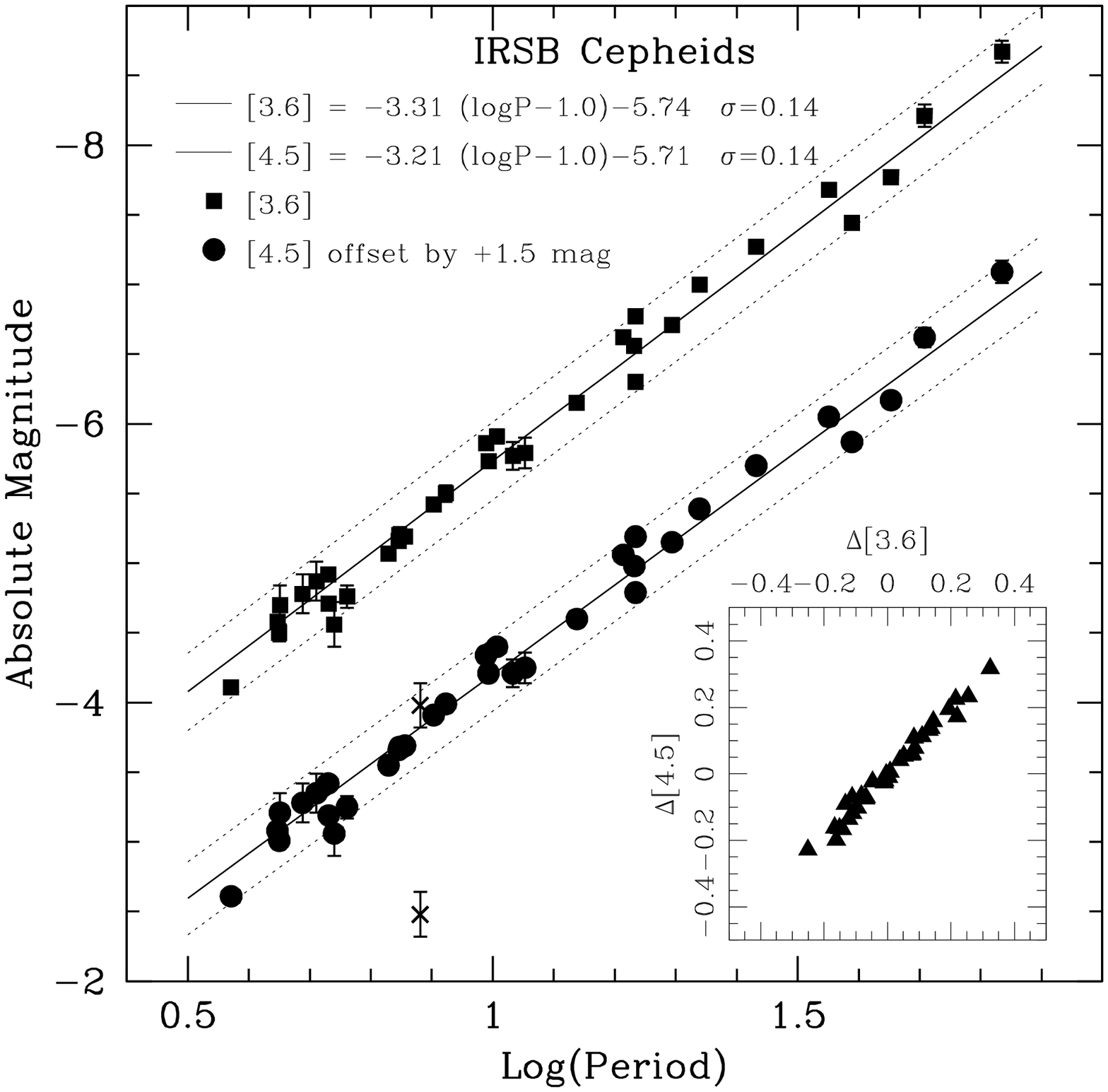}}
\caption{Leavitt PL Relations for the IRSB Cepheids in the Galaxy at 3.6 (upper plot) and 4.5 $\mu$m (lower plot).   The relations are shown as solid lines which were determined using a fixed slope found from the LMC data. The $\pm2\sigma$ boundaries are shown as dotted lines.   The highly correlated magnitude residuals are plotted in the inset, showing that the peak-to-peak width of the IS, as defined by the IRSB Galactic Calibrators is less than 0.6 mag.  }
\label{plirsb}
\end{centering}
\end{figure}


\clearpage

\begin{figure}
\begin{centering}
\resizebox{6in}{6in}{\includegraphics{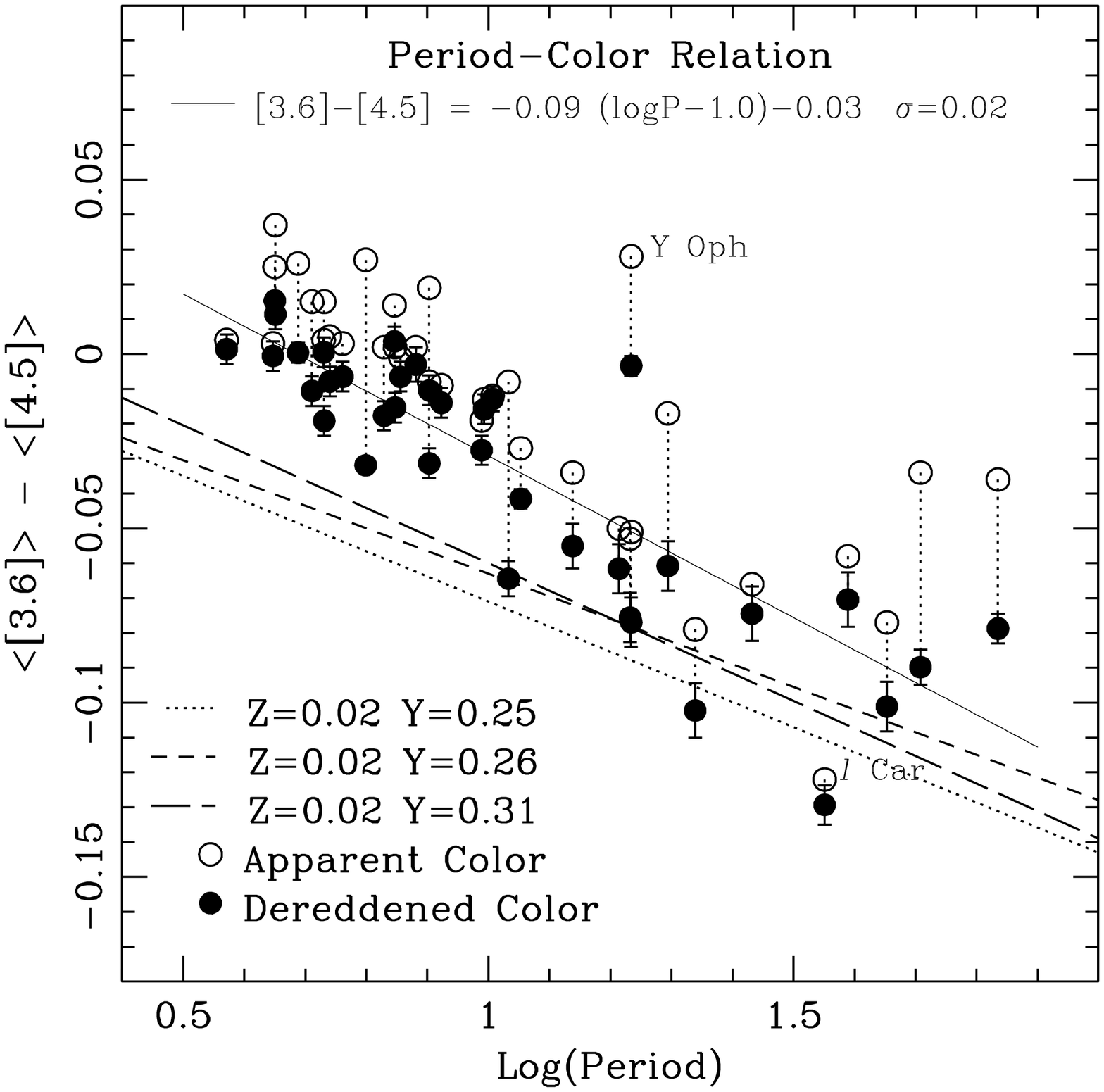}}
\caption{The Galactic Period-Color Relation.  The solid circles are the de-reddened colors for the 37 Cepheids in this study; see Tables \ref{target} \& \ref{galcephs}.  The solid line is the fit to all the data except Y Oph.  The broken lines are theoretical trends from \cite{Ngeow:2012}. }

\label{PLC}
\end{centering}
\end{figure}


\clearpage

\begin{figure}
\begin{centering}
\resizebox{6in}{!}{\includegraphics{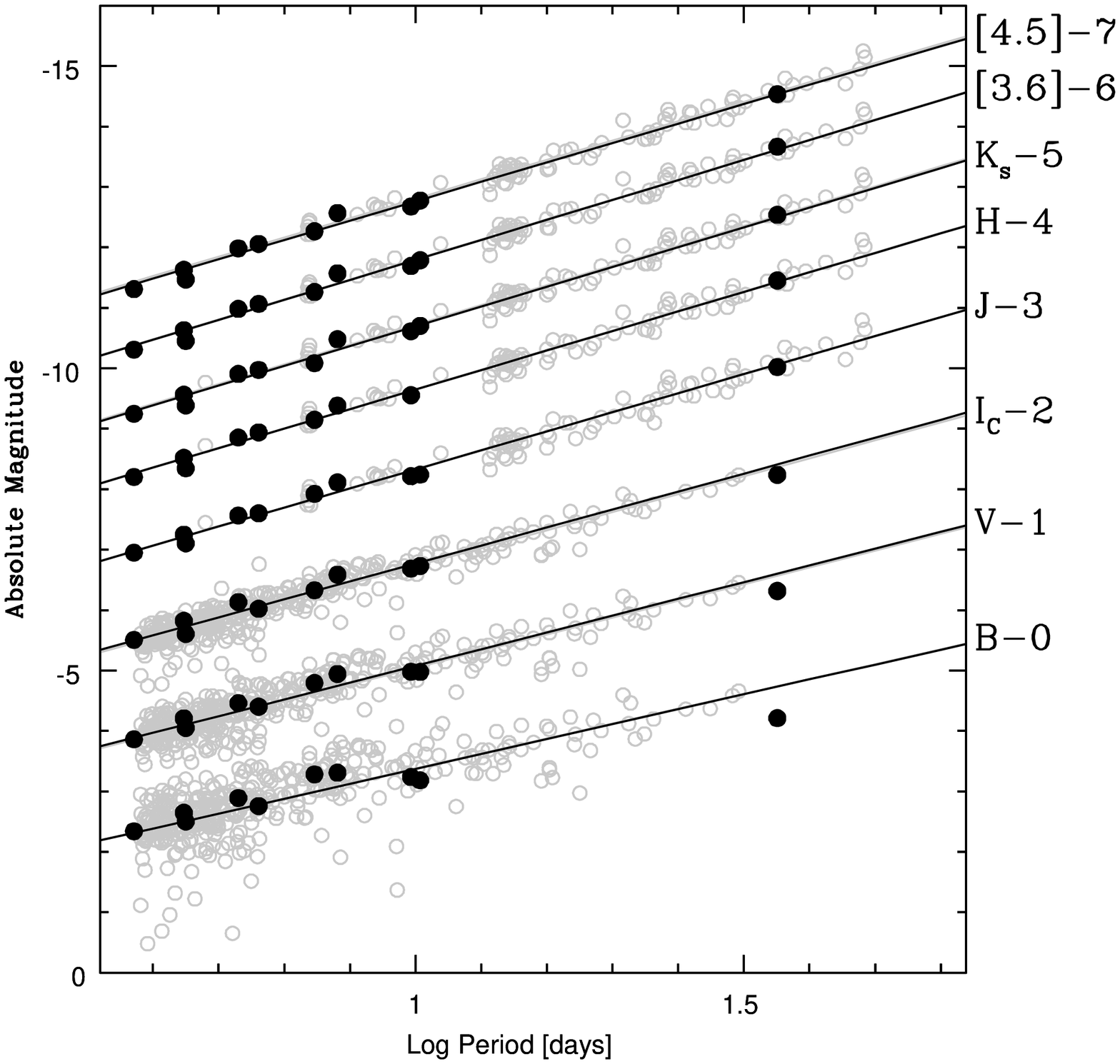}}
\caption{The Period Luminosity laws for the B, V, I$_C$, J, H, K$_s$, [3.6] and [4.5] bands found for LMC data (open gray circles) and the Galaxy (filled black dots).    Solid lines are the PL relations for each band corrected for LMC distance and extinction.   }

\label{LMC_1}
\end{centering}
\end{figure}


\clearpage

\begin{figure}
\begin{centering}
\resizebox{6in}{!}{\includegraphics{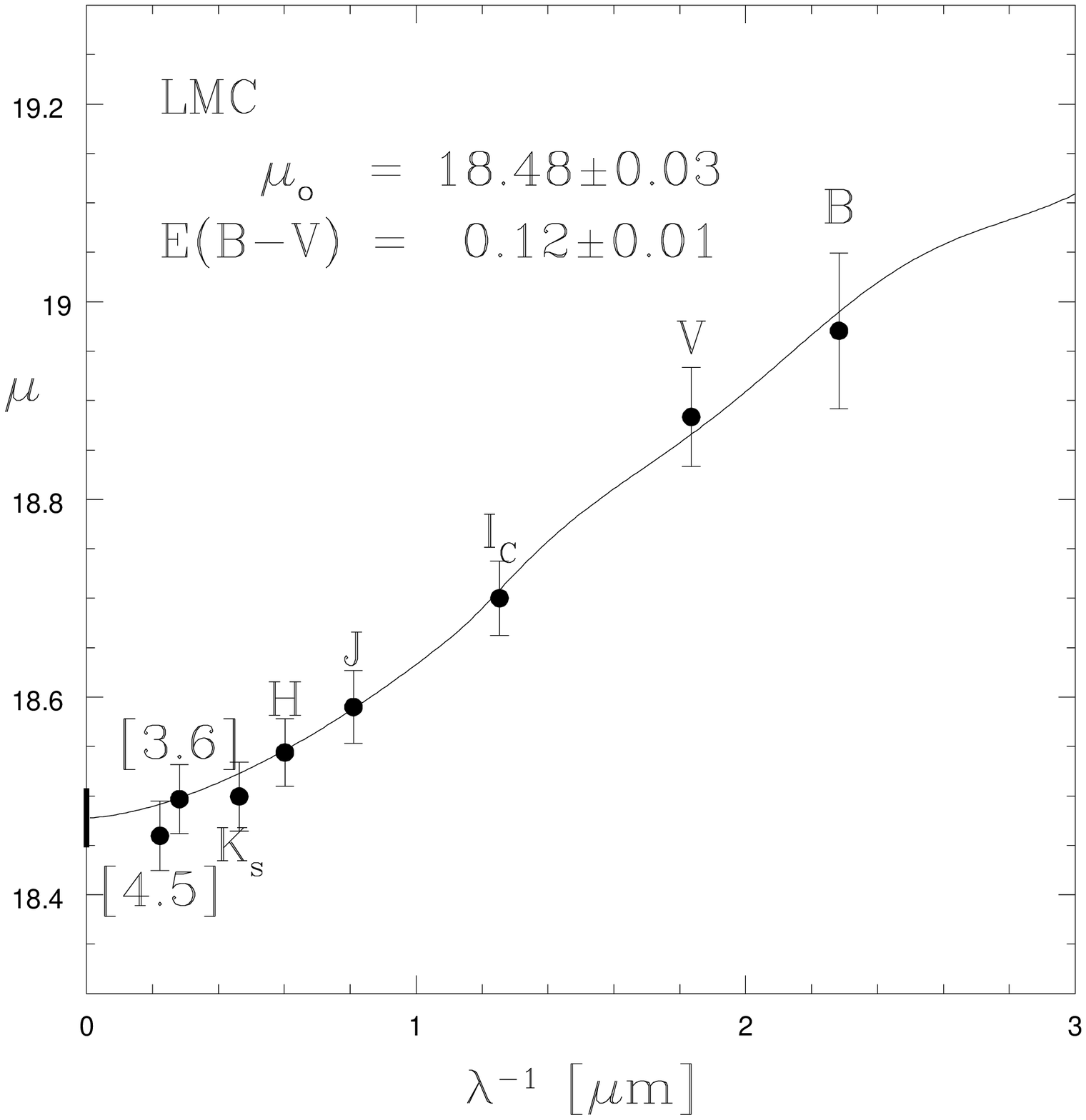}}
\caption{Standard extinction curve fit to multi-wavelength apparent distance moduli to the LMC.  The fitted extinction model yields a true distance modulus (intercept) of  $18.48\pm0.03$ and an average color-excess (slope) of $E(B-V)=0.12\pm0.01$ mag.  The K$_s$ and [4.5] bands were not included in the fit due to the effect of CO in those bands.  }

\label{LMC_2}
\end{centering}
\end{figure}


\clearpage

\begin{figure}
\begin{centering}
\resizebox{6in}{6in}{\includegraphics{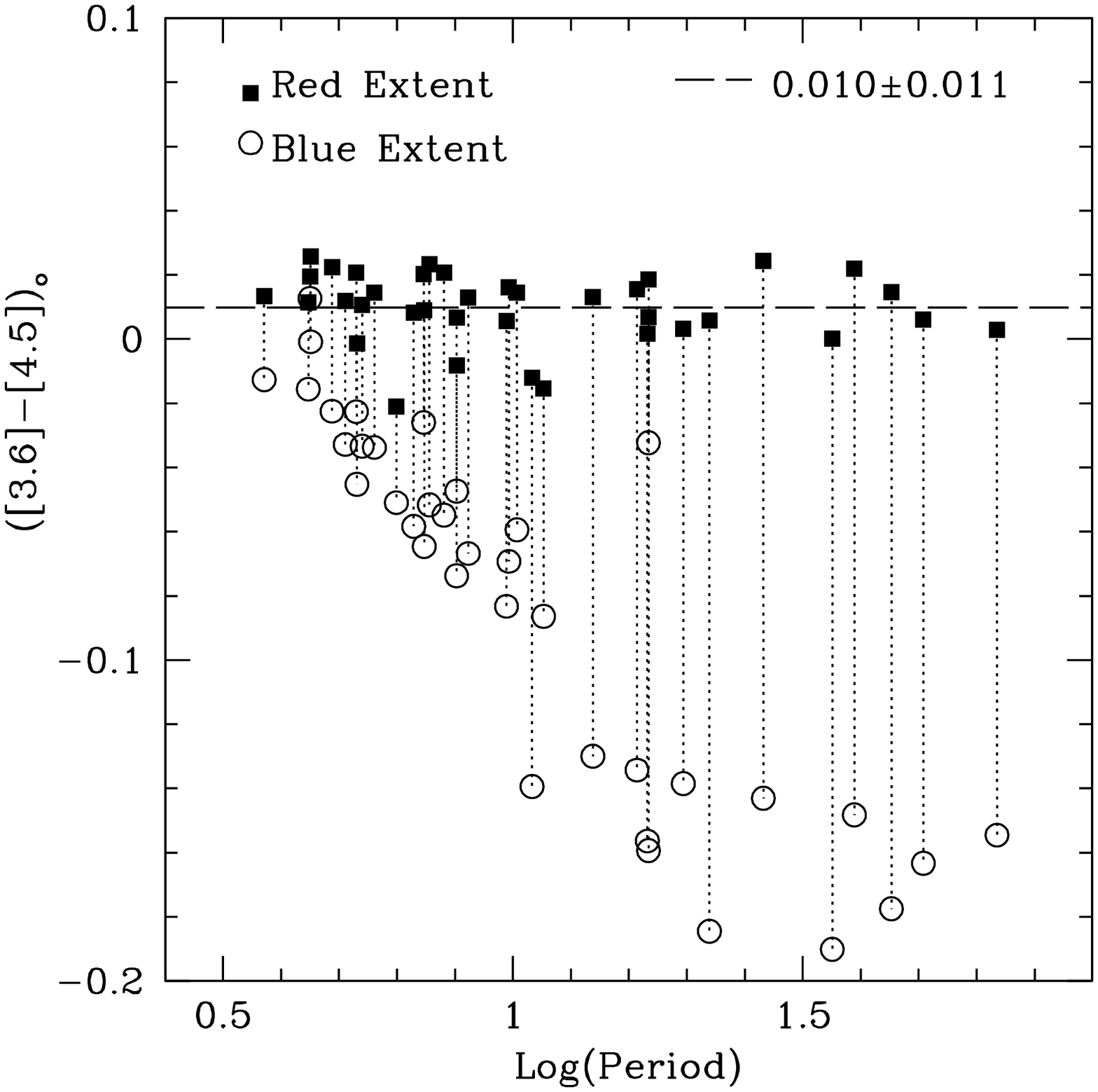}} 
\caption{De-reddened color versus Period.  The red extent of each Cepheid is well defined near 0.01 mag.  The blue extent increases with period due to CO molecular absorption in Cepheids that reach intrinsically cooler temperatures.  }
\label{extent}
\end{centering}
\end{figure}


\clearpage

\begin{figure}
\begin{centering}
\resizebox{6in}{3in}{\includegraphics{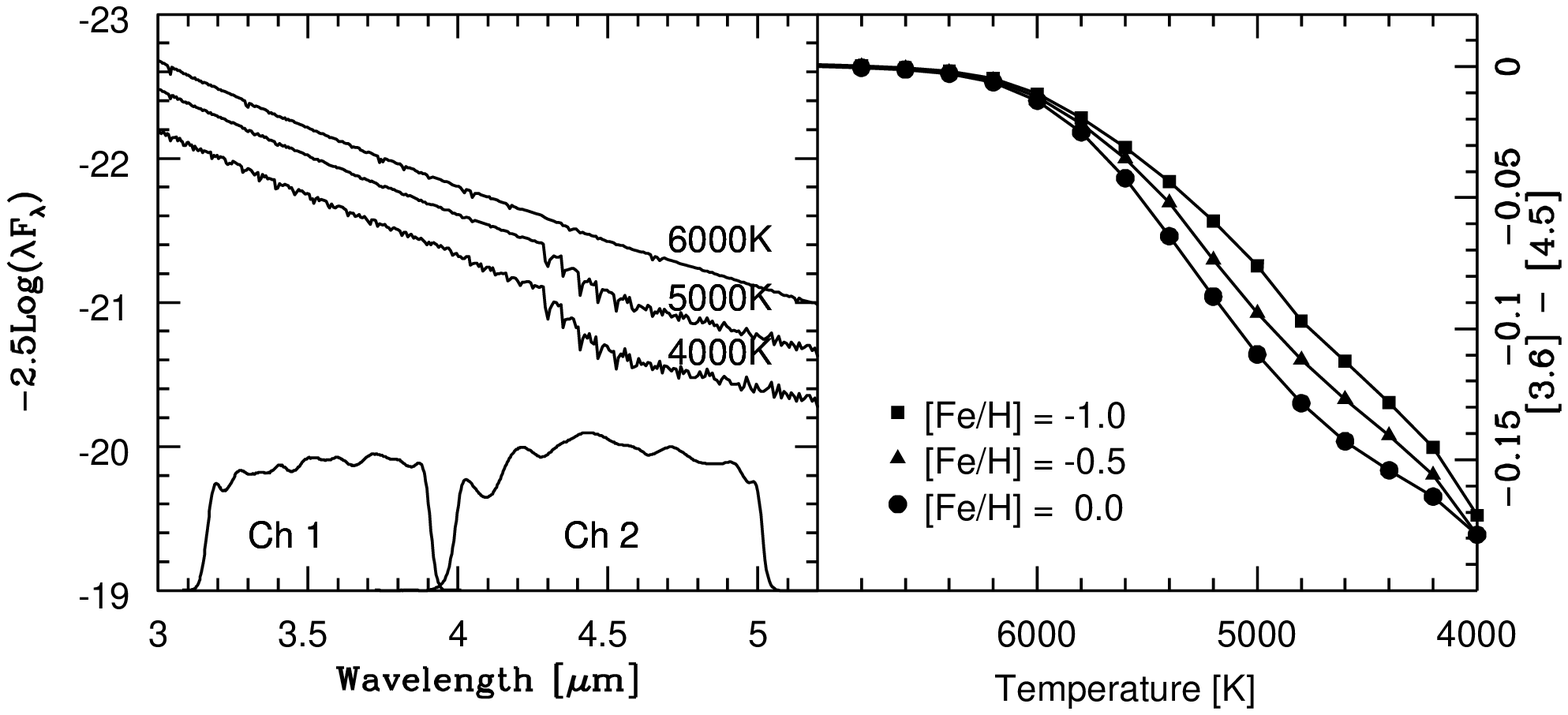}}
\caption{Left: Synthetic spectra showing the onset of CO molecular absorption in the IRAC Channel 2 bandpass.  Right:  The synthetic color, normalized to zero at 7000K, for the generated model spectra.  The color trends to the blue at cooler temperatures, a result of the CO molecular absorption.   }
\label{kurucz}
\end{centering}
\end{figure}


\clearpage

\begin{figure}
\begin{centering}
\resizebox{6in}{6in}{\includegraphics{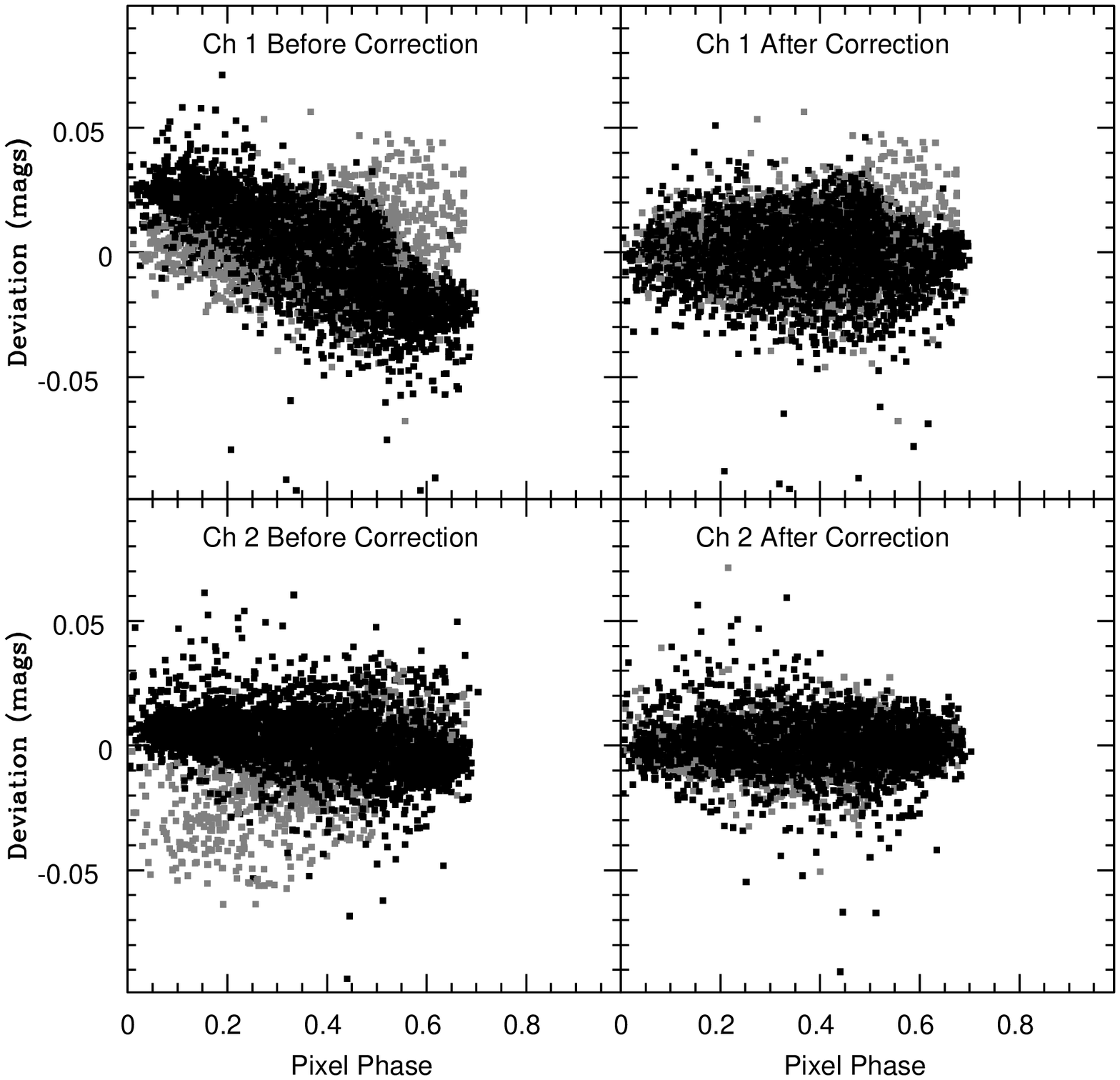}}
\caption{The pixel phase correction (PPC) as a function of pixel phase for a sample of Galactic Cepheids.  The black points are for non-saturated sources and the gray points are for saturated sources; see text.  Top left:  The Channel 1  data as output from the MOPEX PRF fitting algorithm.  Top right:  The Channel 1 data after the pixel phase correction has been applied.   Bottom panels: same as top, but for Channel 2 data.     }
\label{apexppc}
\end{centering}
\end{figure}


\clearpage

\begin{figure}
\begin{centering}
\resizebox{6in}{6in}{\includegraphics{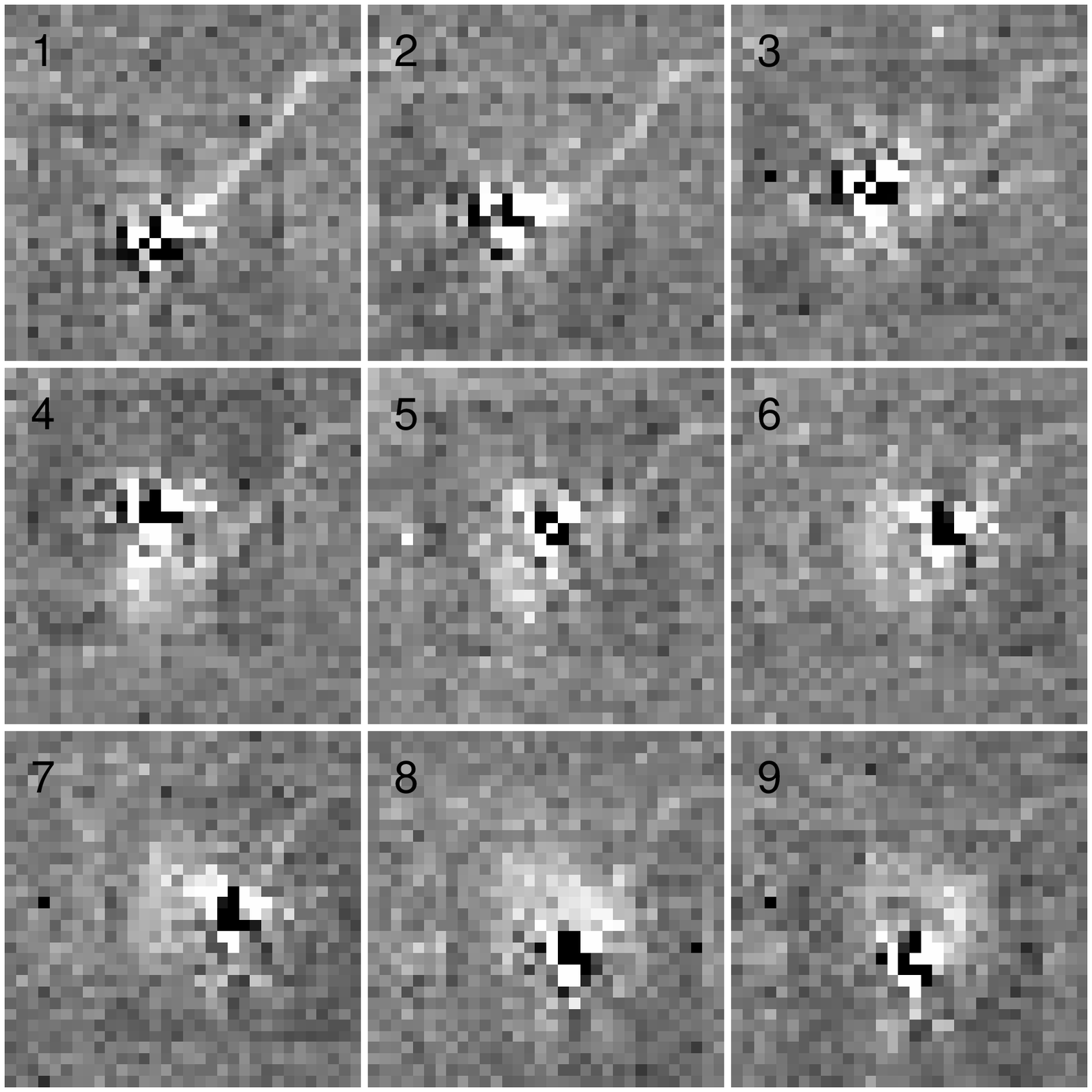}} \\
\caption{Dither time-series of PSF subtracted images for one epoch of RT Aur showing the 9 dither positions (Channel 1).  RT Aur has been fit using MOPEX and the model fit was subtracted leaving the residual.  In these frames (ordered 1-9) it is possible to see the latent image of the star trail as the telescope slewed into position at time-stamp 1.  The trail is nearly dissipated by time-stamp 9; but new latent images are formed at the location of each the previous dithers.  By time-stamp 9 the previous 8 latent images reveal the full Reuleaux dither pattern as a faint after glowing latent triangle.   }
\label{persist}
\end{centering}
\end{figure}


\clearpage

\begin{figure}
\begin{centering}
\resizebox{6in}{4in}{\includegraphics{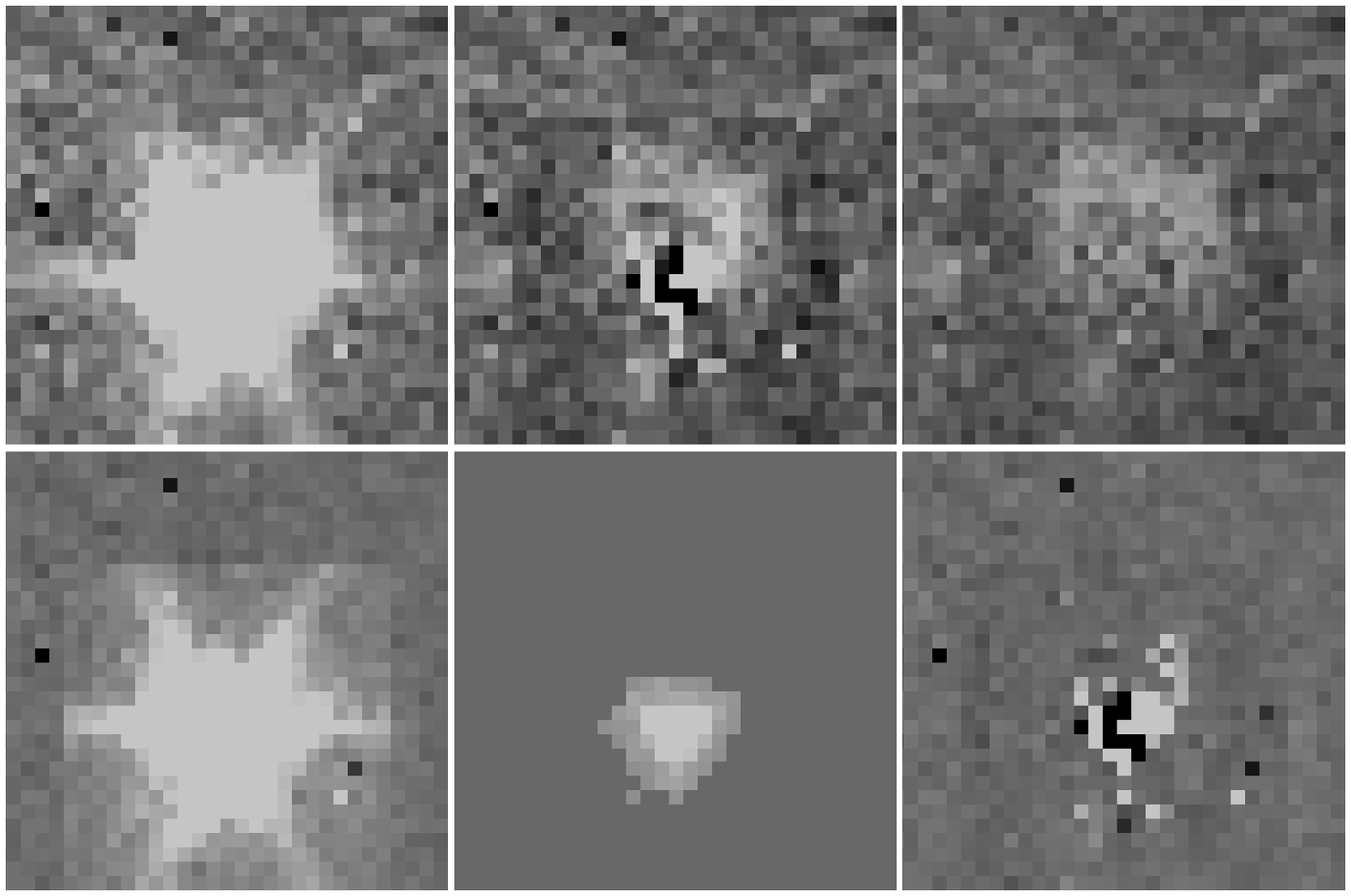}}
\caption{Image persistence mitigation for Channel 1.  Top left: Original BCD image of the Galactic Cepheid RT Aur.  Top middle: Residual image after preliminary PRF fitting.  Top right: Persistence map created from weighted average of nine dithered residual images.  Bottom left: Original BCD with background/persistence map subtracted.  Bottom middle: The uncertainty map with background pixels masked and only the core region used for PRF fitting.   Bottom right: Final residual image after persistence map was subtracted.  The same procedure was followed for Channel 2 although the latency effect was negligible.  }
\label{phot}
\end{centering}
\end{figure}


\clearpage


\end{document}